\documentclass[11pt, draftclsnofoot, onecolumn]{IEEEtran}

\usepackage{cite,graphicx,amsmath,amssymb}
\usepackage{bbm}

\usepackage{fancyhdr}
\usepackage{mdwmath}
\usepackage{mdwtab}
\usepackage{balance}
\usepackage{xcolor}
\usepackage{bm}
\usepackage{amsthm}

\usepackage{multirow}
\usepackage{flafter}

\usepackage{caption}
\usepackage{subcaption}

\usepackage[english]{babel}

\usepackage{tcolorbox}
\usepackage{framed}
\usepackage{adjustbox}
\usepackage{blindtext}

\usepackage[normalem]{ulem}

\usepackage[ruled,vlined]{algorithm2e}
\allowdisplaybreaks[4]

\usepackage{tikz}
\usetikzlibrary{quantikz}
\usepackage{tkz-tab}

\usepackage{epstopdf}

\usepackage{hyperref}
\hypersetup{
	colorlinks=true,
    allcolors = black
}

\newtheorem{remark}{Remark}
\newtheorem{theorem}{Theorem}

\newtheorem{lemma}{Lemma}

\newtheorem{corollary}{Corollary}

\newtheorem{proposition}{Proposition}

\definecolor{coolgrey}{rgb}{0.55, 0.57, 0.67}
\definecolor{airforceblue}{rgb}{0.36, 0.54, 0.66}
\definecolor{americanrose}{rgb}{1.0, 0.01, 0.24}
\definecolor{amber}{rgb}{1.0, 0.49, 0.0}

\begin{document}

\title{General Hamiltonian Representation of ML Detection  Relying on the Quantum Approximate Optimization Algorithm}

\author{
Jingjing Cui,~\IEEEmembership{Member, IEEE,}
Gui Lu Long, \IEEEmembership{Member, IEEE} and \\
 Lajos Hanzo,~\IEEEmembership{Life Fellow, IEEE}

}

\maketitle

\vspace{-4.9em}

\begin{abstract}

The quantum approximate optimization algorithm (QAOA) conceived for solving combinatorial optimization problems has attracted significant interest since it can be run on the existing noisy intermediate-scale quantum (NISQ) devices.  A primary step of using the QAOA is the efficient Hamiltonian  construction based on  different problem instances. 
Hence, we solve the maximum likelihood (ML) detection problem for  general constellations by appropriately adapting the QAOA, which gives rise to a new paradigm in communication systems.   We first  transform the ML detection problem into a weighted minimum $N$-satisfiability (WMIN-$N$-SAT) problem, where we formulate the objective function of the WMIN-$N$-SAT as a pseudo Boolean function.  Furthermore, we formalize the connection between the degree of  the objective function and the Gray-labelled modulation constellations.  Explicitly, we show a series of results exploring the connection between the coefficients of the monomials and the patterns of the associated constellation points, which substantially simplifies the objective function with respect to  the problem Hamiltonian of  the QAOA.  In particular, for an  M-ary Gray-mapped quadrature amplitude modulation (MQAM) constellation, we show that the specific qubits encoding the in-phase components and those encoding the quadrature components are independent in the quantum system of interest, which allows  the in-phase  and  quadrature components to be detected separately using the QAOA.   Furthermore, we  characterize the degree of the objective function in the WMIN-$N$-SAT problem  corresponding to the ML detection of   multiple-input and multiple-output (MIMO) channels. Finally, we evaluate  the approximation ratio of the QAOA  for the ML detection problem of  quadrature phase shift keying (QPSK) relying on   QAOA circuits of different depths.

\end{abstract}


\vspace{-0.7em}
\section{Introduction}
\label{sec:intro}
Signal detection determines the transmitted signals using mathematical tools,  which is an essential component of various communication systems. Maximum likelihood (ML)  detection is capable of finding the optimal solution  by a brute-force search over all possible transmitted signal vectors. Hence, its  complexity increases according to $M^N$ \cite{hanzo2005ofdm}, where $M$ is  the modulation order and $N$  is the number of the transmitted antennas/users.  Hence, the ML detector has to solve a  combinatorial optimization problem, which is proved to be NP-hard \cite{Verd89alg,Micciancio01IT}.  The promise of powerful parallel quantum computing  has attracted substantial    attention  in diverse fields.  
In the current  era of noisy intermediate-scale quantum 
(NISQ) computers, it is of salient significance to deal with their limited coherence time  \cite{preskill2018quantum}.   To circumvent these practical impediments  of NISQ devices, hybrid quantum-classical algorithms  that can be readily  implemented and tested in  NISQ devices have become prominent for demonstrating the much sought-after quantum advantage.  Inspired by this ambitious goal,   we solve  the ML detection problem with general constellations by the quantum approximate optimization algorithm (QAOA).

The QAOA is a variational method designed by Farhi \textit{et al.} \cite{farhi2014QAOA} in 2014   to find an approximation of the ground state of the problem Hamiltonian  associated with  combinatorial optimization problems.   More explicitly, the QAOA  creates a parameterized quantum circuit  by alternately applying the problem Hamiltonian representing the  original optimization problem as well as  the initial Hamiltonian  and it  then optimizes the parameters by a classical computer, while conducting the associated function evaluations on a quantum computer \cite{Wang18PhysRevA,streif2019comparison,Zhou20PhysRevX}.  The QAOA has attracted significant attention both from  industry and academia in the context of  different hardware platforms based on such as superconducting qubits \cite{Harrigan21NaturePhy,karamlou2021analyzing,willsch2020benchmarking,alam2019analysis}, trapped-ion qubits \cite{Pagano20PNAS} and photonics \cite{qiang18photonics}, where the problem size was reported to be  up to 23 qubits \cite{Harrigan21NaturePhy}.  
However, the performance of the QAOA critically depends both on the problem instances \cite{willsch2020benchmarking,Akshay2020PhyRewL} as well as  on the circuit depth \cite{farhi2014QAOA,Harrigan21NaturePhy,basso2021quantum}.    The lower bounds of the approximation ratio achieved by the QAOA are investigated in  \cite{farhi2014QAOA,Wang18PhysRevA,marwaha2021local,Wurtz21PhysRevA},  revealing that the  approximation ratio achieved by the QAOA for a  circuit of depth 3 is at least 0.7924 for all the three-regular graphs investigated \cite{Wurtz21PhysRevA}.   Moreover, some analytical and numerical results on the limitations of the QAOA in  some problem instances are investigated in 
 \cite{hastings2019classical,Bravyi20PhysRevLett,farhi2020quantumworst,farhi2020quantum}.    To overcome the limitations  both in terms of the associated  state preparation and parameter optimization,  the quantum symmetry properties  are exploited in \cite{Egger2021warm,Bravyi20PhysRevLett,shaydulin2021QIP} for improving the QAOA.
 As a further advance, the authors of \cite{basso2021quantum} evaluated  the QAOA relying on high-depth circuits with the objective of  outperforming some classical algorithms in finding the Max-Cut both on a large-girth graph and on the Sherrington-Kirkpatrick Model of \cite{Sherrington75PRL}. This  shed further light on the promising  potential of the QAOA.

Against this rich background, we demonstrate the benefits  of  the QAOA in solving the ML detection problem of  high-order modulation schemes, which is a generalization of  our previous work on configuring the QAOA for solving the ML detection of BPSK \cite{cui2021quantum}.   In contrast to the binary   symbols  which can be mapped to the qubits directly, the constellation of  higher order  quadrature phase shift keying (QPSK) and M-ary quadrature amplitude modulation (MQAM) has multiple constellation points with each representing a complex symbol that might be transmitted by the transmitter. 
Recall that  the basic requirement of using the QAOA is to map the classical optimization problem to the quantum Hamiltonian operator acting on qubits, whose ground state encodes the optimal solution of the classical optimization problem.  
Explicit Hamiltonian constructions for a bunch of  problem instances  on a graph  are provided  in  \cite{Lucas14Ising}.  Then the author of  \cite{hadfield2018representation}  conceived the general representation of the Hamiltonian for handling both pseudo Boolean functions as well as   constrained optimization problems.  As the optimization of the ML detection problem depends on  a set of complex variables,  a suitable Hamiltonian construction method was introduced in \cite{Kim19sigcom} by treating the real and imaginary components of the complex symbols separately, which requires  the design of  a bespoke transformation. Given this Hamiltonian representation method, the performance of the  uplink ML detection problem in  massive multiple-input and multiple-output (MIMO) systems was investigated by harnessing both  quantum annealing \cite{Kim19sigcom,kim2020towards}  and the coherent Ising machine of  \cite{singh2021ising}. 

In contrast to the  Hamiltonian representation used in \cite{Kim19sigcom,kim2020towards,singh2021ising}, in this paper we conceive the technique of constructing the Hamiltonian representation of the ML detection problem from the perspective of  the Boolean satisfiability (SAT) problem.   
More specifically, the SAT problem deals with  an assignment satisfying a particular ensemble of Boolean expressions, which is NP-complete \cite{cook1971complexity}.   The SAT problem can be modelled by pseudo Boolean functions in polynomial representation \cite{boros2002pseudo}, which helps us to construct the Hamiltonian acting on qubits.  Expecitly, it has been widely exploited \cite{Lucas14Ising,hadfield2018representation} that the quadratic unconstrained binary optimization (QUBO) relying on  a quadratic formalism can be readily transformed to the task of finding the ground state of a quadratic Hamiltonian with respect to a two-body quantum system, which can then be solved by the QAOA \cite{Glover2019QuantumBA,Egger2021warm}. Furthermore, many quantum algorithms have been  designed based on the  quadratic Hamiltonian such as the variational quantum eigensolver (VQE)  \cite{Peruzzo14nature} and the quantum annealing \cite{cruz2019qubo},  both of which also compute the eigenstates of the Hamiltonian.  Therefore,  the efficient representation of  the problem Hamiltonian plays a key role in the implementation of these quantum algorithms designed for computing the ground state. However, the Hamiltonian representation of a cubic or even higher-degree  pseudo Boolean function  contains either three-body or even many-body  interactions in the quantum system, which has to be reduced to a quadratic form for using the QAOA.

To expound a little further, the high-degree pseudo Boolean function constitutes   a high-degree multilinear polynomial. The corresponding process of  reducing the high-degree function  into a quadratic form is referred to in parlance  as `quadratization' \cite{boros2014quadratization,dattani2019quadratization}.
There are several techniques of reducing the corresponding high-degree  polynomials into  quadratic forms are summarized in \cite{dattani2019quadratization}.  More explicitly, 
the techniques of reducing the degree of the corresponding function  may rely on  auxiliary variables \cite{Ishikawa09conf,boros2014quadratization,chancellor2016direct,
dattani2019all,boros2020compact}, but it may also be arranged   without introducing auxiliary variables \cite{tanburn2015reducingp1,okada2015reducingp2,dridi2017prime,Ishikawa14conf}. 
A typical technique requiring auxiliary variables is  the method of reduction by substitution proposed in \cite{rosenberg1975reduction},  which was further improved by numerous researchers  \cite{Ishikawa09conf,Ishikawa14conf,boros2014quadratization, boros2018quadratizations, boros2020compact}.   
By contrast,  for  reduction without introducing auxiliary variables,   the authors  of \cite{okada2015reducingp2}  proposed a technique of  iteratively decomposing  the high-degree function into  a set of quadratic subfunctions 
corresponding to the leaves of a binary tree. However, this method  requires  post-processing after implementing the Hamiltonians of the subfunctions for attaining the ground state of the Hamiltonian of the original function.
Since there are several efficient  quadratization methods for both cubic and quaternary functions  \cite{Ishikawa09conf,dattani2019all,boros2020compact}, a hybrid quadratization  framework combining the split-reduction method of  \cite{tanburn2015reducingp1} and the techniques associated with auxiliary variables was also  introduced in  \cite{okada2015reducingp2}, which seeks to deal with the limited number of  qubits available in NISQ devices.   The rich variety of methods introduced for reducing the degree of functions provides  an abundance  of options   for  quadratization.  Hence, for  decomposing the high-degree function representing  the ML-detection problem,  we can opt for  one of these strategies to perform the reduction.

Based on the above discussion, we focus our attention on characterizing the Hamiltonian concerning the constellation diagram as well as the implementation of the QAOA for solving the ML detection of general  modulation constellations.  Our main contributions  are summarized as follows: 
\begin{enumerate}

\item We provide a new procedure for the Hamiltonian construction of the ML detection problem of general constellations by transforming it into a weighted minimum $N$-satisfiability  (WMIN-$N$-SAT) problem\footnote{  
A WMIN-$N$-SAT problem can be treated as an extension of the minimum satisfibility (MIN-SAT) problem \cite{Kohli94SIAM} by attaching  a non-negative weight to each clause,  where the goal  is to minimize the sum of weights of the satisfied clauses  by checking all possible assignments  \cite{Umair20SIP,LARROSA2008AI}. }. Specifically,   the objective  of our WMIN-$N$-SAT problem can be represented as  a pseudo Boolean function, which results in the unconstrained optimization problem of minimizing the pseudo Boolean function of interest.
 
 \item We formalize the connection between the problem Hamiltonian of the ML detection  and the Gray-labelled modulation constellations.   Explicitly, due to having only a  single bit-change between adjacent constellation points in  the Gray labelling, we exploit for QPSK  that  the sum of the squared  Euclidean distances from the received signal to the two pairs of  diagonal points are the same.  By harnessing this property, we demonstrate that  the degree of  the objective function  can be  beneficially reduced, hence simplifying the implementation of the Hamiltonian. 
 
 \item  We also exploit that there are no interactions between the qubits encoding  the in-phase  and    the quadrature bits in a Gray-labelled MQAM constellation, which indicates that the degree of the Hamiltonian associated with the MQAM is equal to the number of in-phase bits. 
We further extend  the  results  provided for  a single-input constellation to  MIMO channels  associated with a joint constellation. 

\item  We compare the architecture of communication systems relying on the QAOA assisted ML detector (QML) to that of  the classical ML detector (CML). Explicitly, the communication system associated with the  QML does not need the conventional signal demapping procedure. Finally, we  provide the simulation results for quantifying the approximation ratio of the QML for QPSK.
\end{enumerate}

The rest of the paper is organized as follows. In Section \ref{sec:sysmod} we commence with a brief  recap of the ML detection problem. In Section \ref{sec:siso} we focus on the  Hamiltonian construction of  single-input and single-output (SISO) systems. More explicitly,  we transform the ML detection problem of a general constellation  to the WMIN-$N$-SAT problem  in Section \ref{sec:probref}, followed by characterizing the Hamiltonian construction in  Section \ref{sec:ham_construct}, where the explicit Hamiltonians of  different constellations and our conclusions on the Gray-labelled MQAM constellations are given. In Section  \ref{sec:mimo} we introduce the Hamiltonian construction of the  MIMO system. Section \ref{sec:qaoa_ml} illustrates the QAOA assisted ML detection, followed by the simulation results of the QAOA implementation of  QPSK in Section \ref{sec:sim}.  Section \ref{sec:conclus}   concludes  the paper.

\vspace{-0.7em}
\section{System Model}
\label{sec:sysmod}

We consider a $N_r \times N_t$ MIMO communication system having $N_t$ transmit antennas (TAs) and $N_r$ receive antennas (RAs). The received vector $\mathbf{y}$ is given by
\begin{equation}
\mathbf{y} = \mathbf{H}\mathbf{s} + \boldsymbol{\eta}, \label{eq:r_mimo}
\end{equation}
where we have $\mathbf{y} \in \mathcal{C}^{N_r}$,   $\mathbf{H} \in \mathcal{C}^{N_r \times N_t}$  is the channel matrix,  $\mathbf{s}  \in \mathcal{C}^{N_t}$ is the vector  of  transmit  signals, and  $\boldsymbol{\eta}$ is the complex-valued  zero-mean noise vector with $\boldsymbol{\eta} \sim \mathcal{CN}(0, \mathbf{I}_{N_r})$.  Here, the receiver has perfect knowledge of the random channel $\mathbf{H}$.
We consider a  general constellation, where  $\mathbf{s} = [s_0, \cdots,s_{N_t-1}]$ is independently selected from a finite set of complex numbers $\mathcal{A}$  according to the data to be transmitted. Hence, $\mathbf{s} \in \mathcal{A}^{N_t}$. Finally, we define  $M  = |\mathcal{A}|$ as the number of points in $\mathcal{A}$, where $|(\cdot)|$ denotes the cardinality of a set $(\cdot)$. The ML detector   can  thus be interpreted as  finding the closest lattice point in an $N_t$-dimensional complex space, which can be expressed as 
\begin{eqnarray}
\begin{aligned}
\hat{\mathbf{s}} = \min_{\mathbf{s} \in \mathcal{A}^{N_t}} \| \mathbf{y} - \mathbf{H}\mathbf{s}\|^2 . \label{eq:ML_mimo}
\end{aligned}
\end{eqnarray}

\vspace{-0.7em}
\section{SISO Channel}
\label{sec:siso}

For a SISO channel,  i.e. $N_t = N_r = 1$,   the ML solution for the constellation $\mathcal{A}$ is given by
\begin{eqnarray}
\begin{aligned} \label{eq:ML_siso}
\hat{s} &= \mathrm{arg} \min_{s \in \mathcal{A}}   | y - h s |^2.
\end{aligned}
\end{eqnarray}
Due to the quantum parallelism,  quantum computers are capable of evaluating  all possible solutions at the same time.  Furthermore, quantum computers operate in Hilbert space relying on binary  variables. As a result, for solving the ML detection problem on the quantum computers, the solutions to the classical ML detection problem have to be encoded into bit strings.  Given a constellation $\mathcal{A}$ of $M$ points ($M$ possible solutions associated with the ML detection problem),  $N = \log_2(M)$ bits are required for representing each point as an  unique bit string. Let  $ z_{0},\cdots, z_{N-1} \in \{0,1\}^N$  represent the binary variables associated with  each possible solution of the ML detection problem. We  see that there are $2^N = M$ possible solutions in total with respect to  the binary variables $z_{0},\cdots, z_{N-1} $.
 
 \vspace{-0.7em}
\subsection{Problem reformulation}
\label{sec:probref}
Next, we introduce the mapping rule transforming the constellation points to the binary variables, which allows us to transform the ML detection problem into a WMIN-$N$-SAT problem defined in \cite{Umair20SIP,LARROSA2008AI}, where a  problem instance  of WMIN-$N$-SAT is a collection of clauses, with each clause being a Boolean function of $N$ binary variables $z_{0},\cdots, z_{N-1}$. 

Let $a_i$ denote the $i$-th point ($i$ is a decimal number) in the constellation of $\mathcal{A}$ with $i \in \mathcal{M} = \{ 0,\cdots,M-1\}$, where each $i$ has  a $N$-bit binary representation of $[i]_2 = b_0 \cdots b_{N-1}$.
The square of the Euclidean distance between the received signal $y$ and $a_i$ is given by
\begin{equation}
d_i = |y - ha_i|^2, \label{eq:weight}
\end{equation}
where $d_i \ge 0$. Correspondingly,  the goal of the ML detection problem is to find the point $a_i$ having the minimum $d_i$, for $i \in \mathcal{M}$.

\begin{figure} [t!]
\centering
\begin{subfigure}{0.45\textwidth}
\includegraphics[width = 0.95\linewidth]{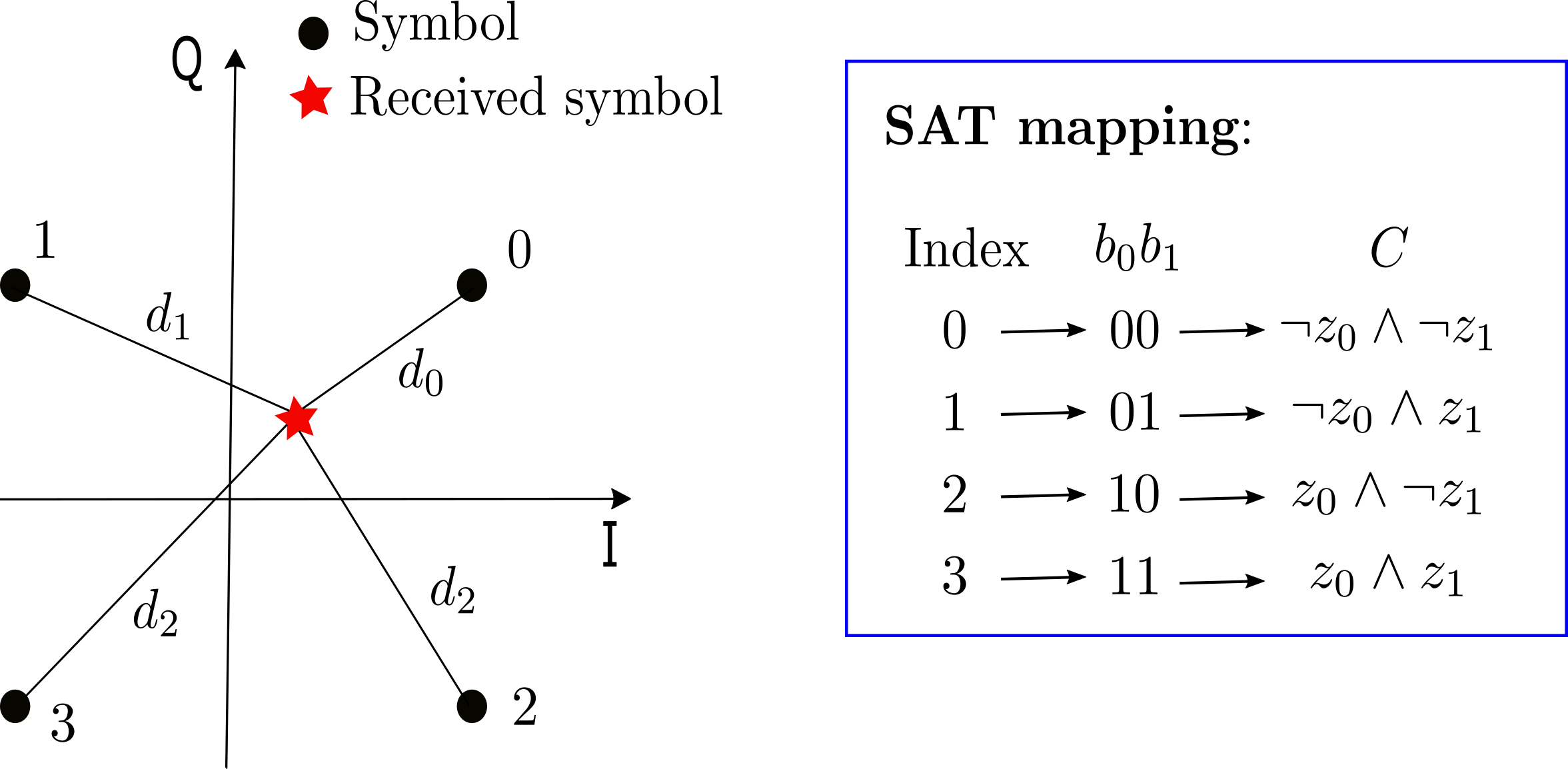}
 \vspace{-0.5em}  \caption{{\small QPSK:  A  weighted 2-SAT }}
 \label{fig:qpsk_mapping}
 \end{subfigure}
 \begin{subfigure}{0.45\textwidth}
\includegraphics[width = 0.95\linewidth]{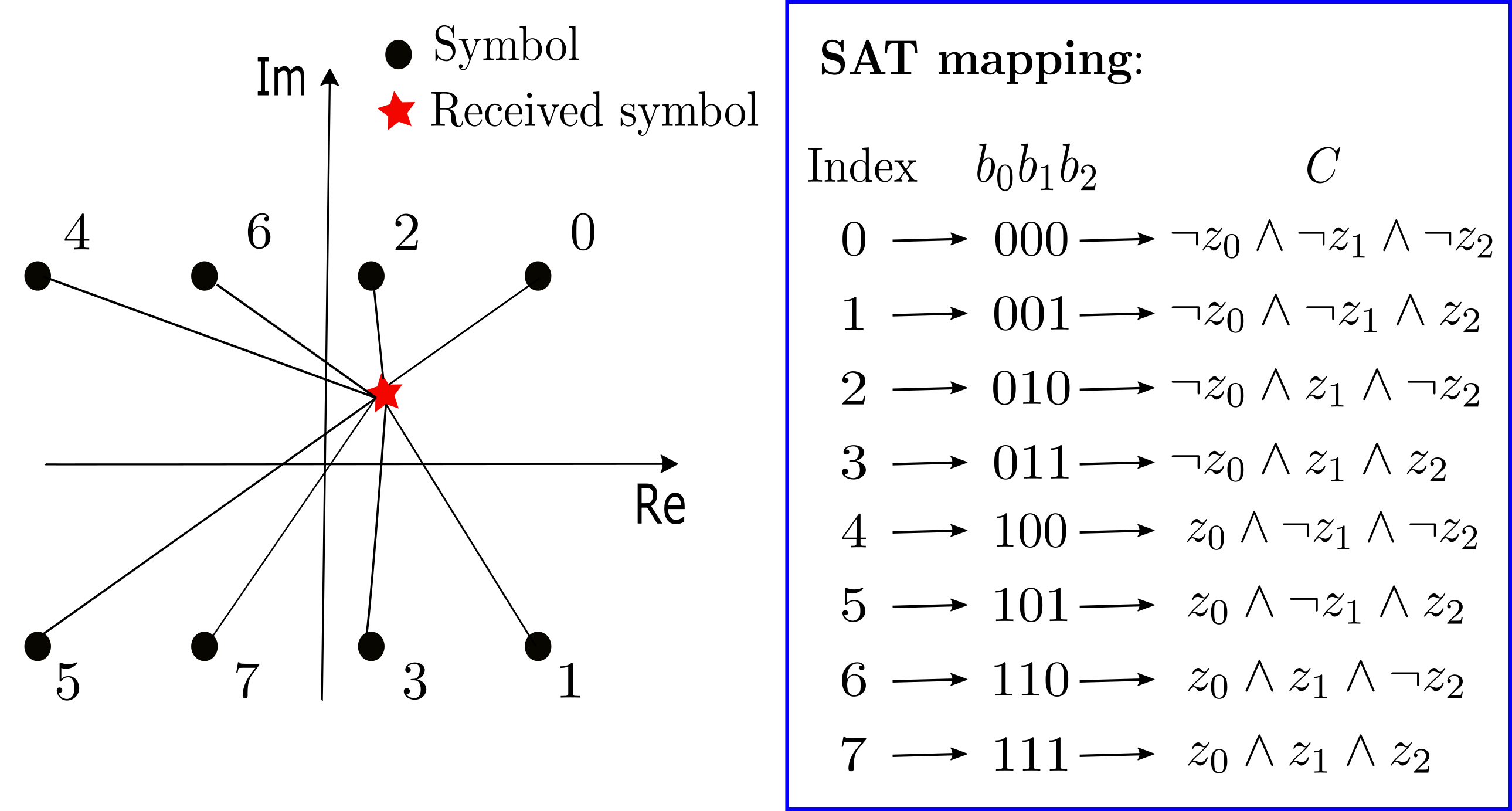}
 \vspace{-0.5em}  \caption{{\small  8QAM:   A weighted 3-SAT}  }
 \label{fig:8qam_mapping}
 \end{subfigure}
   \vspace{-1.0em} \caption{{\small Illustrations of  mapping the ML detection problem into SAT for QPSK and 8QAM.} }
 \label{fig:mappings}
   \vspace{-1.5em}
\end{figure}

Here, we define a weighted clause as $w(C_i)$  associated with  the  non-negative weight $d_i$ obtained from \eqref{eq:weight},  where $C_i$ is a clause associated with the $N$ binary variables  $z_{0},\cdots, z_{N-1} $. Explicitly, the clause $C_i$  is defined as follows:
\begin{eqnarray}
\begin{aligned}
C_i =  B(z_0)\wedge B(z_1) \wedge \cdots \wedge B(z_{N-1}), \label{eq:clause}
\end{aligned}
\end{eqnarray}
with $B(z_n)$ being
\begin{eqnarray}
\label{eq:binmap}
\begin{aligned}
B(z_n) = \begin{cases}
\neg z_n,& \mathrm{if}~b_n=0,\\
z_n, &  \mathrm{if}~b_n=1,
\end{cases}
\end{aligned}
\end{eqnarray}
where $b_n$ is the $n$-th bit of the binary number $b_0\cdots b_{N-1}$ associated with the decimal number $i$.   Here $\wedge$ and  $\neg$ denote the operators `And' and 'Not', respectively. We portray  the SAT mapping  of  both QPSK and  8QAM in Fig. \ref{fig:mappings} for illustrating the connection between the  clauses  and the constellation points. 
 Fig. \ref{fig:qpsk_mapping}  shows an  example of mapping the QPSK symbols into clauses of a WMIN-2-SAT  problem.   The symbols on the QPSK constellation diagram  are  depicted  at the left-hand side of  Fig. \ref{fig:qpsk_mapping}, where the symbol labels follow a Gray mapping  order. The square of the Euclidean distance $d_i$, $i = 0, \cdots,3$, can be computed by \eqref{eq:weight}. Furthermore, the SAT mapping process from the ordering of the points to the  clause is demonstrated at the right-hand side of Fig. \ref{fig:qpsk_mapping}.  
 According to   \eqref{eq:clause}, we  construct four clauses as shown in  Fig. \ref{fig:qpsk_mapping}, where each clause relies on both of the two binary variables.  Furthermore, we also demonstrate the mapping  process for a Gray-labelled 8QAM  constellation in Fig. \ref{fig:8qam_mapping}, which encodes the decimal index of the constellation points into clauses having three variables. Therefore, the ML detection of the 8QAM  constellation can be encoded into a WMIN-3-SAT problem containing 8 clauses in total.

As a result, given an assignment of the binary variables  $z_{0},\cdots, z_{N-1} $,  the weighted clause  $w(C_i)$ representing a clause $C_i$ with a weight $d_i$, $i \in \mathcal{M}$,   can be expressed as 
\begin{eqnarray}
\begin{aligned}
w(C_i) = d_i C_i = d_i [ B(z_0)\wedge B(z_1) \wedge \cdots \wedge B(z_{N-1}) ].\label{eq:weighted_clause}
\end{aligned}
\end{eqnarray}

Additionally, the Boolean formula  $C_i(z_{0},\cdots, z_{N-1} )$ can be rewritten as a function of  binary variables, since   $B(z_i) \wedge B(z_j)$ and $\neg z_i$ can be equivalently expressed as  $B(z_i) \cdot B(z_j)$ and $(1 - z_i)$, respectively. 
The objective function of the WMIN-$N$-SAT at hand can thus be formulated as a Boolean pseudo function given by 
\begin{eqnarray}
\begin{aligned}
f(z_{0},\cdots, z_{N-1} ) = \sum_{i=0}^{M-1} w(C_i) = \sum_{\mathcal{S} \subseteq \mathcal{N}} \bar{d}_{\mathcal{S}} \prod_{n \in \mathcal{S} } z_n, \label{eq:obj_sat}
\end{aligned}
\end{eqnarray}
where  $\mathcal{N} = \{0,\cdots,N-1\}$ is the set of all variable indices and $\mathcal{S}$ is a subset of $\mathcal{N} $ such that  $\mathcal{S} \subseteq \mathcal{N}$.  Furthermore, $\bar{d}_{\mathcal{S}} $ represents the coefficient of the term $\prod_{n \in \mathcal{S} } z_n$ in the expansion of $f(z_0, \cdots,z_{N-1})$.  Note that the objective function  $f(z_0, \cdots,z_{N-1})$ is  a multilinear polynomial  of degree $N$,  where the degree of $f(z_0, \cdots,z_{N-1})$ is  the maximum number of distinct variables occurring in any of its monomials  \cite{Fine2017book}.  Correspondingly, the degree of a monomial associated with $\mathcal{S}$ can be denoted as $|\mathcal{S}|$ representing the number of elements in $\mathcal{S}$. Note that we omit the constant term in \eqref{eq:obj_sat}, which does not affect the optimal solution of the ML detection. For notational simplicity, the constant terms in the function $f(z_{0},\cdots, z_{N-1} )$ and in the associated Hamiltonian function  will be omitted throughout the paper. 
Therefore, the ML detection problem of \eqref{eq:ML_siso} can now  be equivalently transformed into the  WMIN-$N$-SAT problem as follows:
\begin{eqnarray}
\begin{aligned}
\min_{z_{0},\cdots, z_{N-1}  \in \{0,1\}^N}  f(z_{0},\cdots, z_{N-1} )
\end{aligned}
\end{eqnarray}

Following the rules of the polynomial expansion applied to  \eqref{eq:obj_sat},  we formulate a pair of  remarks from the expansion of $f(z_0,\cdots, z_{N-1})$, which will help us to discover some interesting properties of   $f(z_0,\cdots, z_{N-1})$ concerning different constellations.

\vspace{-0.9em}
\begin{remark}\label{term:remark_expan1}
The expansion  of $C_i$ contains the monomial of the form $\prod_{n \in \mathcal{S} } z_n$, $\mathcal{S} \subseteq \mathcal{N}$, if and only if the bit string $b_0\cdots b_{N-1}$ such that $b_l=0$, for all $ l \in \mathcal{N} \setminus \mathcal{S} = \{l: l\in \mathcal{N} ~\mathrm{and} ~l \notin \mathcal{S}\}$.
\end{remark}
\vspace{-0.9em}

Note that {\bf Remark \ref{term:remark_expan1}} can  be equivalently interpreted in a more neat manner as follows.  The expansion  of $C_i$ contains the monomial of the form $\prod_{n \in \mathcal{S} } z_n$, $\mathcal{S} \subseteq \mathcal{N}$, if and only if  $i \in \mathcal{M}'$, where
\begin{eqnarray}
\mathcal{M}' = \{i:  [i]_2=b_0\cdots b_{N-1}, ~\mathrm{s.t.}~ b_l = 0,~  \forall l \in \mathcal{N} \setminus \mathcal{S}\}. \label{eq:M_prime}
\end{eqnarray}

\vspace{-0.9em}
\begin{remark}\label{term:remark_expan2}
{\color{black}There are $2^ {|\mathcal{S} |}$ clauses containing the monomial of the form $\prod_{n \in \mathcal{S}}z_n$, $\mathcal{S} \subseteq \mathcal{N}$.}
\end{remark}
\vspace{-0.9em}

\vspace{-0.9em}
\begin{remark}\label{term:remark_expan3}
The coefficient $\bar{d}_{\mathcal{S}} $ can be expressed as follows:
\begin{equation}
\bar{d}_{\mathcal{S}}  = \sum_{i \in \mathcal{S}'} d_{i} \prod_{n \in \mathcal{S}} (-1)^{1-b_n},
\end{equation}
with $\mathcal{S}' = \{i:  [i]_2=b_0\cdots b_{N-1}, ~\mathrm{s.t.}~ b_n = 0,~  \forall n \in \mathcal{N} \setminus \mathcal{S}\}$  and $|\mathcal{S}'| = 2^{|\mathcal{S}|}$.
\end{remark}
\vspace{-0.9em}

 \vspace{-0.7em}
\subsection{Constructing Hamiltonians}
\label{sec:ham_construct}
In the QAOA, the objective function is encoded into a problem Hamiltonian containing a sequence of Pauli Z operators \cite{farhi2014QAOA,hadfield2018representation}, diagonally acting on qubits corresponding to the computational basis vectors. 
From \eqref{eq:weighted_clause},  we  can construct the  Hamiltonian operator associated with clause $C_i$  as follows:
\begin{eqnarray}
H_{f,C_i}  |z_0\cdots z_{N-1}\rangle  = w(C_i) |z_0 \cdots z_{N-1}\rangle,
\end{eqnarray}
where we see that $w(C_i)$ corresponds to the eigenvalues of $H_{f,C_i}$ associated with the eigenstates $|z_0\cdots z_{N-1}\rangle$.
Given the expression of binary variables to the clause  $C_i(z_0,\cdots,z_{N-1})$,   the Hamiltonian $H_{f,C_i} $ can be readily obtained  by replacing $z_{i}$ with $\frac{1 - \sigma_z^{(i)}}{2}$.
Therefore, the general form for  the problem Hamiltonian of $f(z_0, \cdots, z_{N-1})$ in \eqref{eq:obj_sat} can be expressed as 
\begin{eqnarray}
\begin{aligned}
H_f  = \sum_{\mathcal{S} \in \mathcal{N}} g_{\mathcal{S}} \prod_{n \in \mathcal{S}} \sigma_z^{(n)}
 ,  \label{eq:Hf_general}
\end{aligned}
\end{eqnarray}
where $g_{\mathcal{S}}$ is the coefficient of the term $\prod_{n \in \mathcal{S}} \sigma_z^{(n)}$,  which describes the interactions between the qubits in the quantum system. 
Given an arbitrary system state $|\psi \rangle$,  the problem Hamiltonian $H_f$ obeys  $\langle \psi | H_f |\psi \rangle \ge 0$, $H_f |\psi \rangle = f_{\min}  |\psi \rangle$ if and only if $|\psi \rangle$ is  the ground state of $H_f$ \cite{Hadfield19Alg}, which corresponds to the assignment  that  only satisfies the associated constellation point of having the minimum Euclidean distance. The goal  of the QAOA is thereby to seek the ground state of  $H_f$.

Furthermore, a common choice of the initial Hamiltonian $H_B$ and the initial state $\psi(0)$ used in the QAOA is formulated as follows:
\begin{eqnarray}
\begin{aligned}
H_B = \sum_{i=0}^{N-1} \sigma_x^{(i)},
\end{aligned}
\end{eqnarray}
and  
\begin{eqnarray}
| \psi(0) \rangle= \frac{1}{2^N} \sum_{z_{N-1}} \cdots \sum_{z_0} |z_0\rangle \cdots |z_{N-1}\rangle,
\end{eqnarray}
where  $\sigma_x^{(i)}$ is  the Pauli X operator acting on the $i$-th qubit and $| \psi(0) \rangle$ is a uniform superposition over the computational basis.

\subsubsection{Problem Hamiltonian  for QPSK}

We  see  from Fig. \ref{fig:qpsk_mapping}  that  the QAOA harnessed the ML detection of QPSK requires $2$ bits for  encoding the 4 symbols of the QPSK constellation,
where  the Boolean formula for each clause is also given.
Based on  \eqref{eq:clause} and \eqref{eq:weighted_clause}, we  rewrite the weighted clause $w(C_i)$ as a  function of binary variables as follows:
\begin{eqnarray}
\begin{aligned} \label{eq:fun_weight_clause}
w(C_0):~ d_0 (1 - z_0) (1 - z_1), ~
 w(C_1):~d_1 (1 - z_0) z_1, ~
 w(C_2): ~d_2 z_0 (1 - z_1), ~
w(C_3):~ d_3 z_0 z_1.
\end{aligned}
\end{eqnarray}
Here each $w(C_i)$ is a quadratic function, which involves a product term $z_0z_1$. The objective function $f_{QPSK}(z_0,z_1)$   is thus given by 
\begin{equation}
f_{QPSK}(z_0,z_1) = \bar{d}_{0,1}z_0 z_1 + \bar{d}_0 z_0 +  \bar{d}_1 z_1, \label{eq:fqpsk_exp}
\end{equation}
where $\bar{d}_{0,1} = d_0 - d_1 - d_2+d_3$, $\bar{d}_0 = d_2-d_0$ and $\bar{d}_1 = d_1 - d_0$. 

\vspace*{-0.9em} 
\begin{proposition}\label{pro:property1}
Following the Gray mapping order, the coefficient $\bar{d}_{0,1}$ of the  quadratic term is $0$, and thus $f(z_0,z_1)$ can be simplified as
\begin{eqnarray}
f_{QPSK}(z_0,z_1) =  \bar{d}_0 z_0 +  \bar{d}_1 z_1 . \label{eq:f_qpsk_simp}
\end{eqnarray}
\begin{proof}
See Appendix~A.
\end{proof}
\end{proposition}
\vspace*{-0.9em}

\begin{figure} [t!]
\centering
\includegraphics[clip,width = 0.7\textwidth]{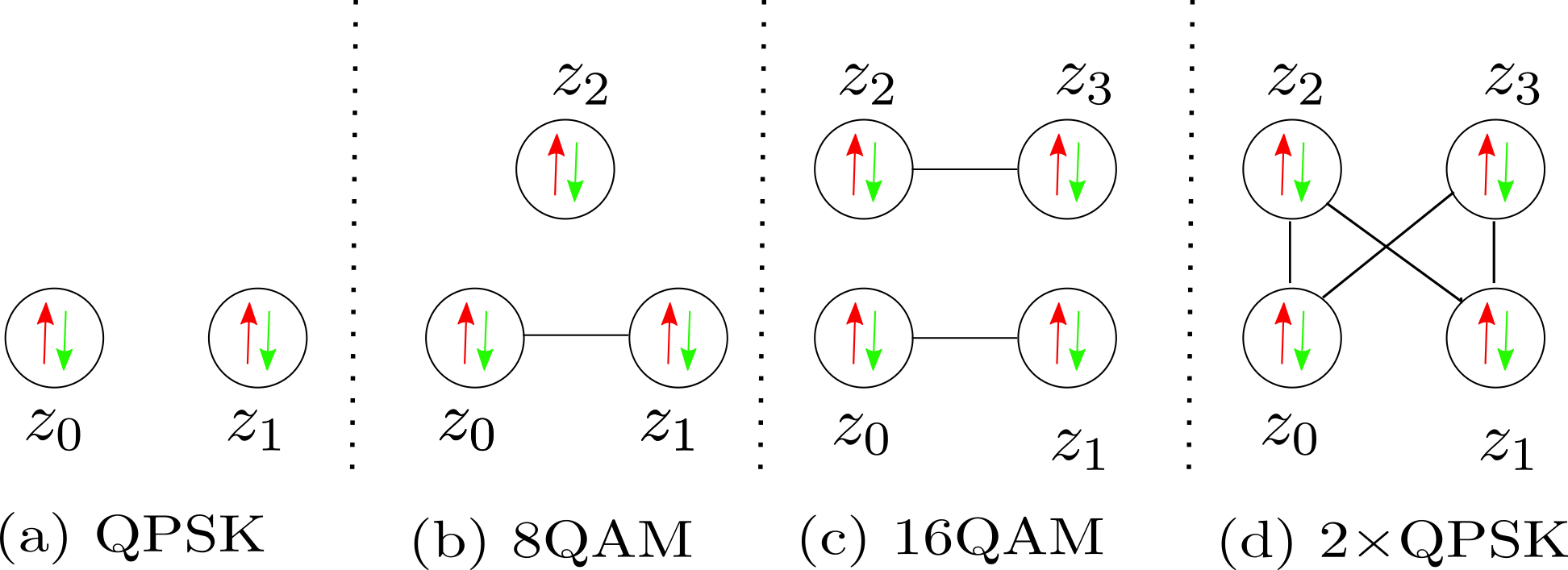}
    \vspace{-1.0em} \caption{ {\small An exemplary illustration of the interactions between the   qubits for different modulation schemes, where ${\color{red}\uparrow}$ and ${\color{green}\downarrow}$ represent the two possibilities of a bit value in classical systems.  In contrast, a single qubit  can be represented as a superposition of classical states: $|z_k\rangle = \alpha_k |{\color{red}\uparrow} \rangle + \beta_k| {\color{green}\downarrow} \rangle, \alpha_k,\beta_k \in \mathcal{C}$.}}
 \label{fig:qubits_illu}
   \vspace{-1.5em}
\end{figure}

Correspondingly,  the problem Hamiltonian $H_{f_QPSK}$ can be constructed immediately upon replacing $z_i$ by $\frac{1}{2}(1 - \sigma_z^{(i)})$, $i = 0,1$, which can be expressed as
\begin{eqnarray}
\begin{aligned} \label{eq:Hf_qpsk}
H_{f_{QPSK}}=  -\bar{d}_0 \sigma_z^{(0)} - \bar{d}_1\sigma_z^{(1)}.
\end{aligned}
\end{eqnarray}
Here we have omitted the common constant coefficient $\frac{1}{2}$ and the constant terms, since they do not affect the  ground state of $H_{f_{QPSK}}$. Therefore, we will also omit the common constant coefficient  in the Hamiltonian function throughout the paper.    We find that there are no quadratic terms containing $\sigma_z^{(0)}\sigma_z^{(1)}$ in $H_{f_{QPSK}}$, which indicates that there are no interactions between the two qubits, as shown in Fig. \ref{fig:qubits_illu}a. Therefore,  for implementing the ML detection of QPSK we can use a  single-qubit quantum device  twice or a pair of single-qubit quantum devices in a parallel manner.  As a further advance, we provide another proposition for further extending the property to a rectangle.

\vspace*{-0.9em} 
\begin{proposition} \label{term:remark_rectangle}
If four points of  the constellation diagram form a rectangle, then
the sum of the squared distance of  the received signal from the two  pairs of  diagonal points are the same.
\end{proposition}
\vspace*{-0.9em} 

The proof of  {\bf Proposition \ref{pro:property1}} was formulated  for a square, which is shown in {\bf Remark \ref{term:remark_square}}.
We can immediately extend this property to a rectangle in {\bf Proposition \ref{term:remark_rectangle} } following the   methods used for proving {\bf Proposition \ref{pro:property1}}.

\vspace*{-0.9em} 
\begin{remark}
If using the binary mapping order, the  points $0$ and $3$ are neighbours instead of being  diagonal points.  Therefore, the quadratic term $z_0 z_1$ in $f_{QPSK}(z_0,z_1)$ cannot be cancelled.
\end{remark}
\vspace*{-0.9em} 

\vspace*{-0.9em} 
\begin{theorem} \label{term:theorem}
For a Gray-labelled constellation such as a rectangular QAM  or  a M-ary phase-shift keying
(MPSK),  with $M=2^N$and $N\in \mathcal{Z}^{+}$, the coefficient of  the monomial having the highest degree $N$ i.e. the monomial of  the form $\prod_{n=0}^{N-1}z_n$,  in $f(z_0,\cdots,z_{N-1})$  is always $0$.
\begin{proof}
See Appendix~B.
\end{proof}
\end{theorem}
\vspace*{-0.9em} 

\begin{figure} [t!]
\centering
\begin{subfigure}{0.24\textwidth}
\begin{center}
\includegraphics[clip,width = 0.97\linewidth]{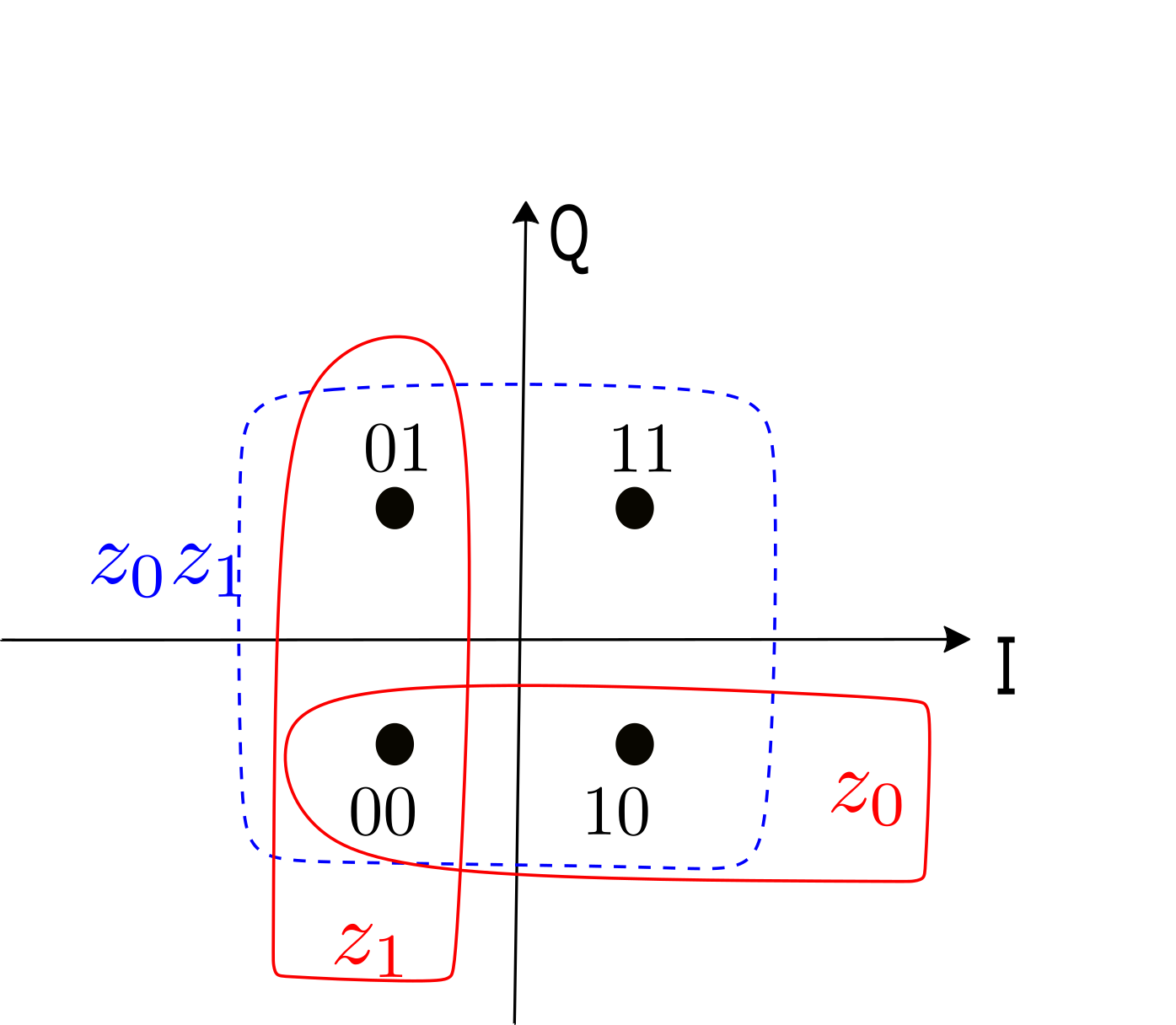}
 \caption{{\small QPSK: $ \mathcal{S}=1,2$}}
  \label{fig:qpsk_cg}
  \end{center}
\end{subfigure}
\begin{subfigure}{0.24\textwidth}
\begin{center}
\includegraphics[clip, width = 0.97\linewidth]{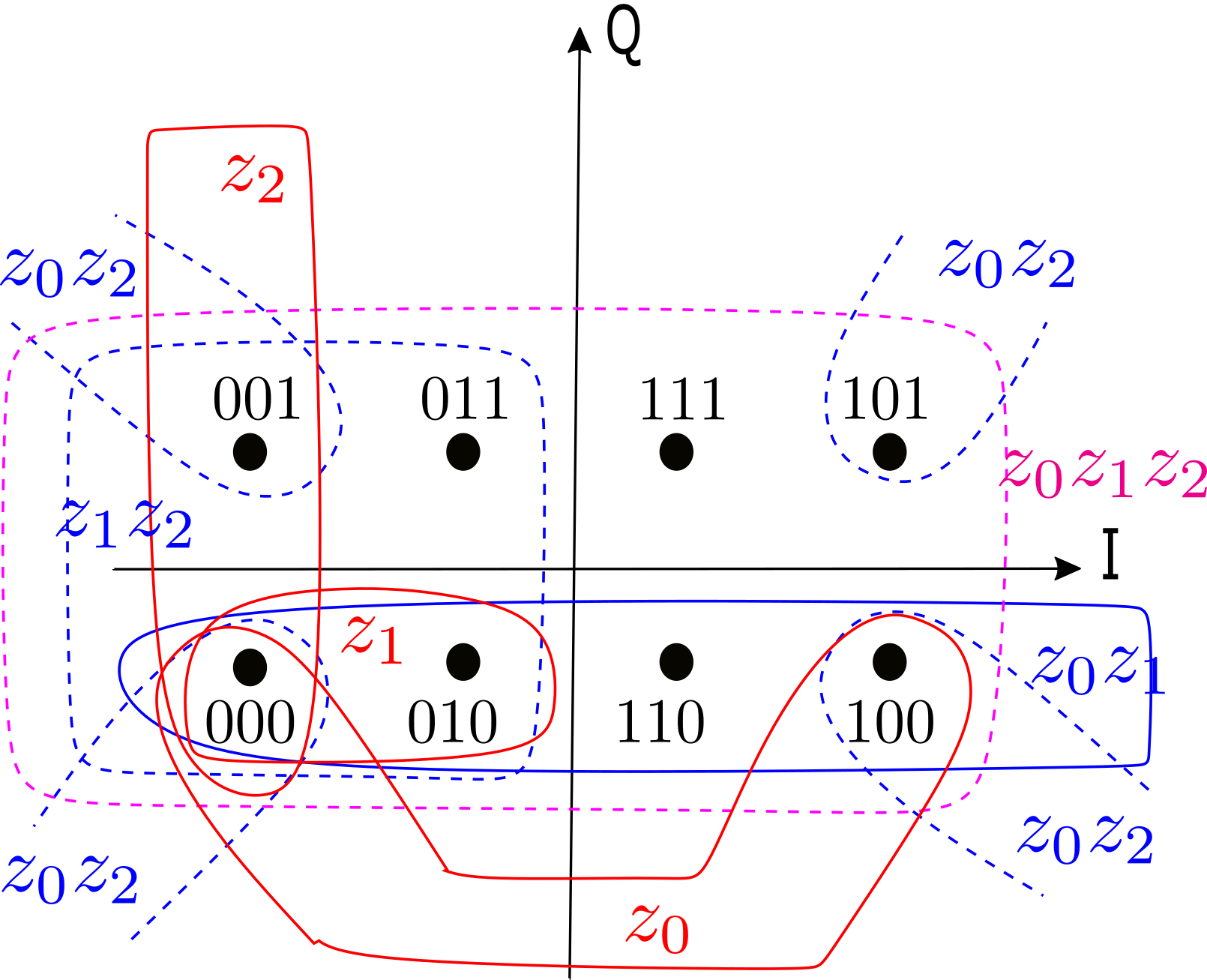}
\caption{{\small 8QAM: $ \mathcal{S}=1,2,3$} }
 \label{fig:8qam_cg}
 \end{center}
\end{subfigure}
\begin{subfigure}{0.24\textwidth}
\begin{center}
\includegraphics[clip,width = 0.97\linewidth]{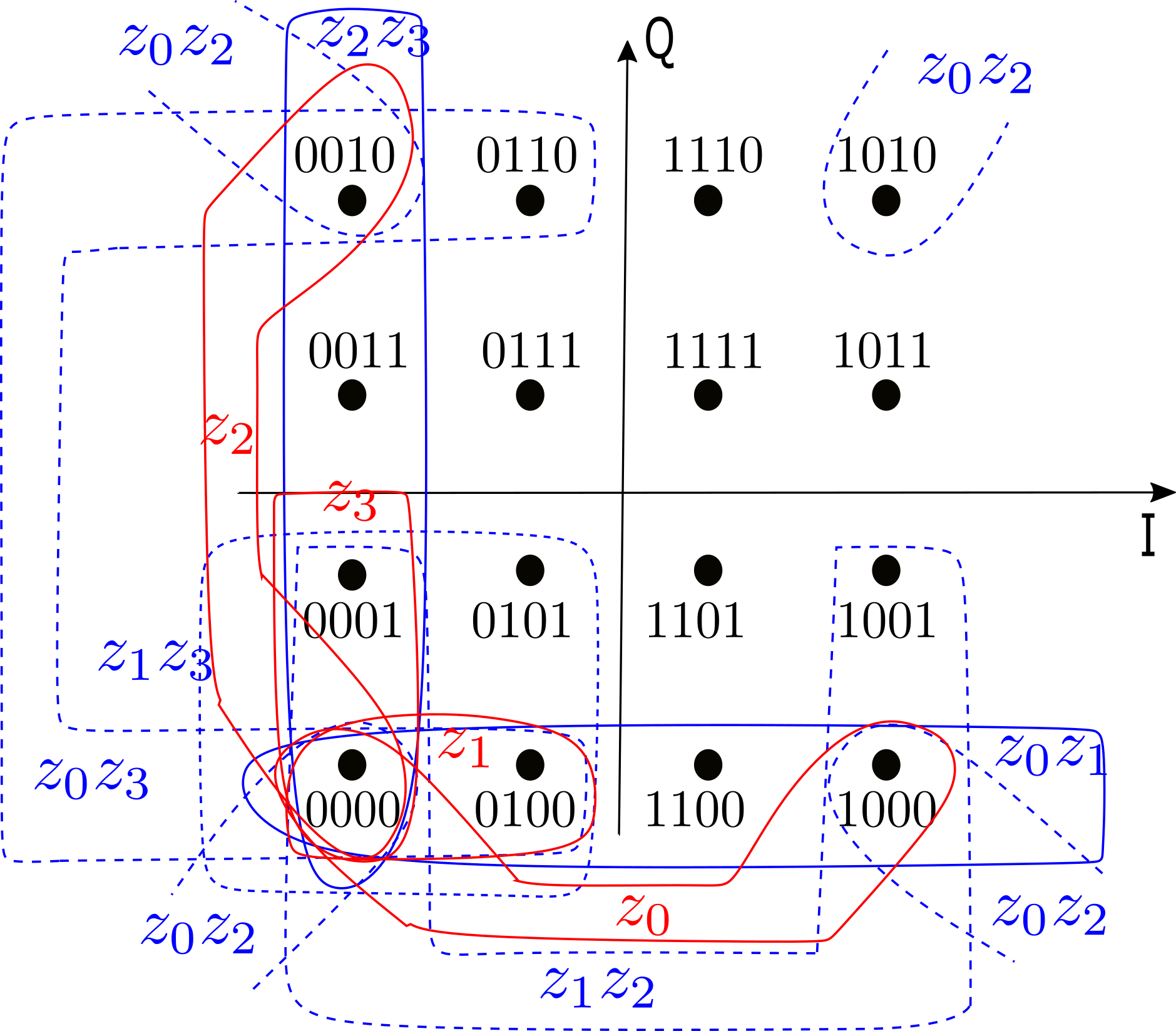}
 \caption{{\small 16QAM: $ \mathcal{S}=1,2$} }
  \label{fig:16qam_cg}
  \end{center}
\end{subfigure}
\begin{subfigure}{0.24\textwidth}
\begin{center}
\includegraphics[clip, width = 0.97\linewidth]{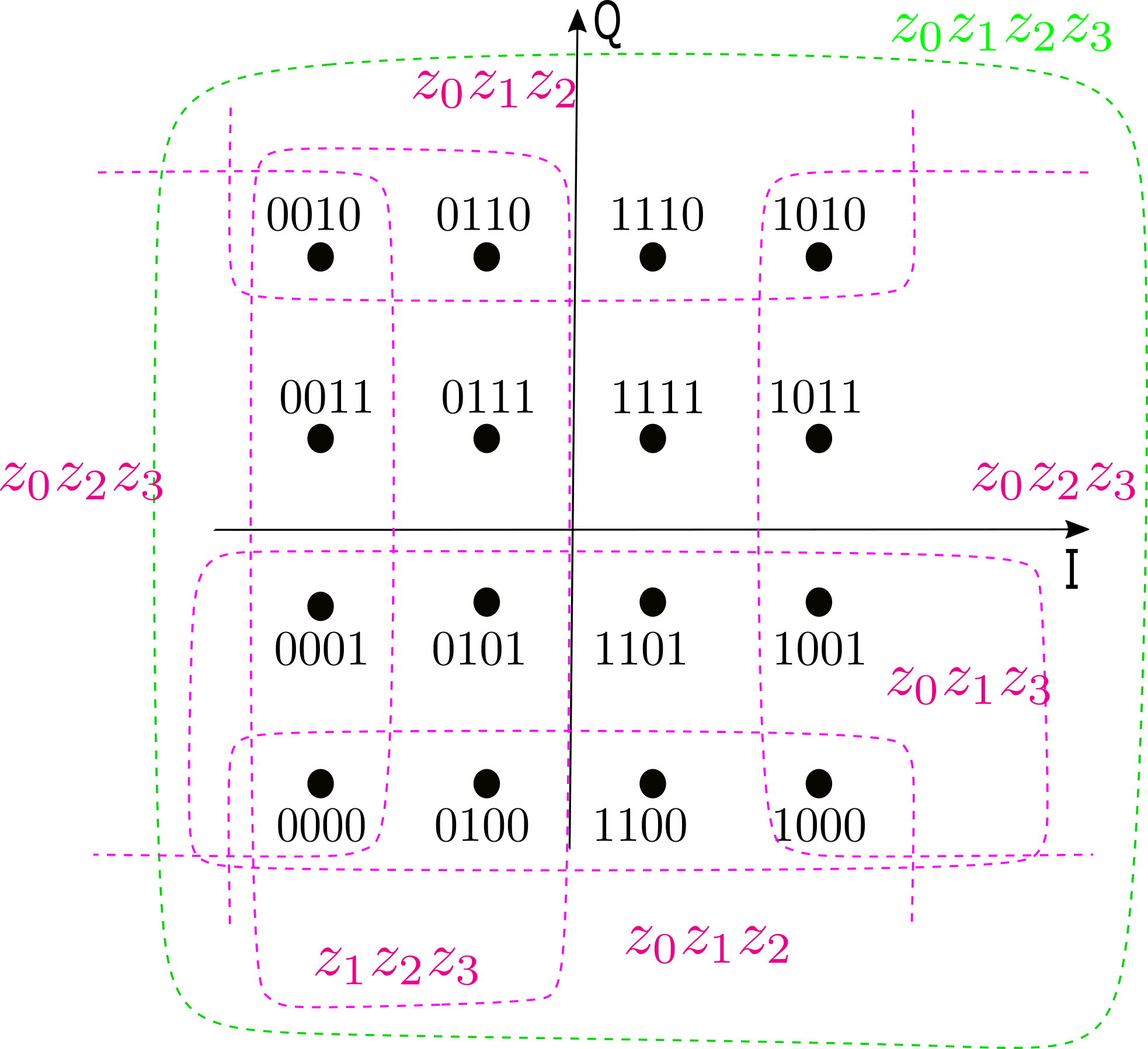}
\caption{{\small 16QAM:  $\mathcal{S}=3,4$} }
 \label{fig:16qam_cgp2}
 \end{center}
\end{subfigure}
\vspace{-1.5em}  \caption{Patterns for the possible monomials in $f(z_0,\cdots,z_{N-1})$ in terms of QPSK, 8QAM and 16QAM using Gray mapping, where the constellation points associated with a monomial of same variables are grouped together. For clarity,  Fig. \ref{fig:16qam_cg} shows the possible monomials of degree 1 and 2, while Fig. \ref{fig:16qam_cgp2} shows the  possible monomials of degree 3 and 4.  Note that the coefficient of a monomial  is zero if the set of its associated constellation points is surrounded by  dashed lines, and  the  set of the constellation points associated with the  monomials having non-zero coefficients  are surrounded by solid lines.}
 \label{fig:ex_cg}
   \vspace{-1.0em}
\end{figure}

\begin{figure} [t!]
\centering
\begin{subfigure}{0.3\textwidth}
\begin{center}
\includegraphics[clip,width = 0.9\linewidth]{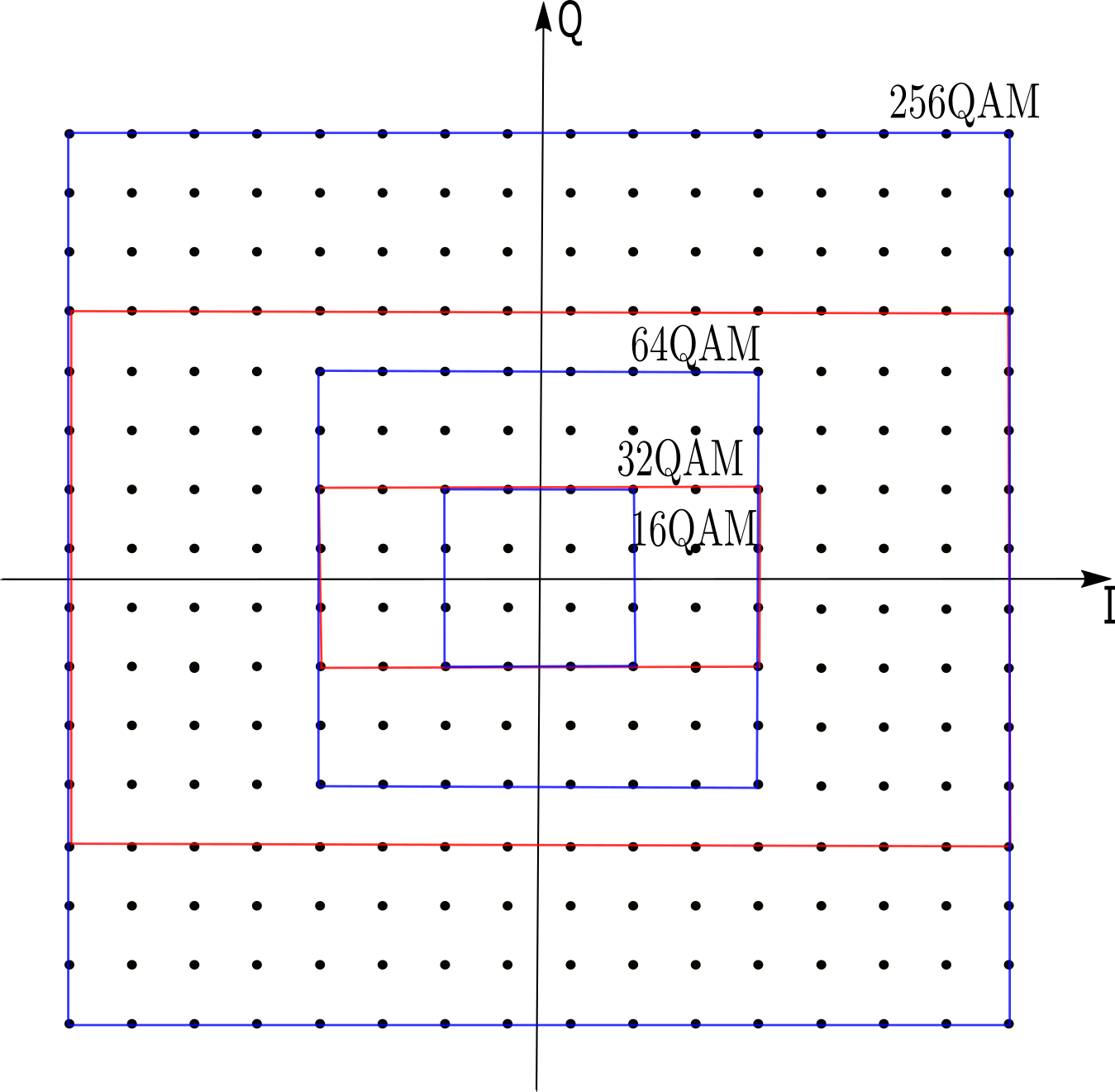}
\caption{{\small Rectangular constellation}}
 \label{fig:mqam_rec}
  \end{center}
 \end{subfigure}
 \begin{subfigure}{0.3\textwidth}
  \begin{center}
\includegraphics[clip,width = 0.9\linewidth]{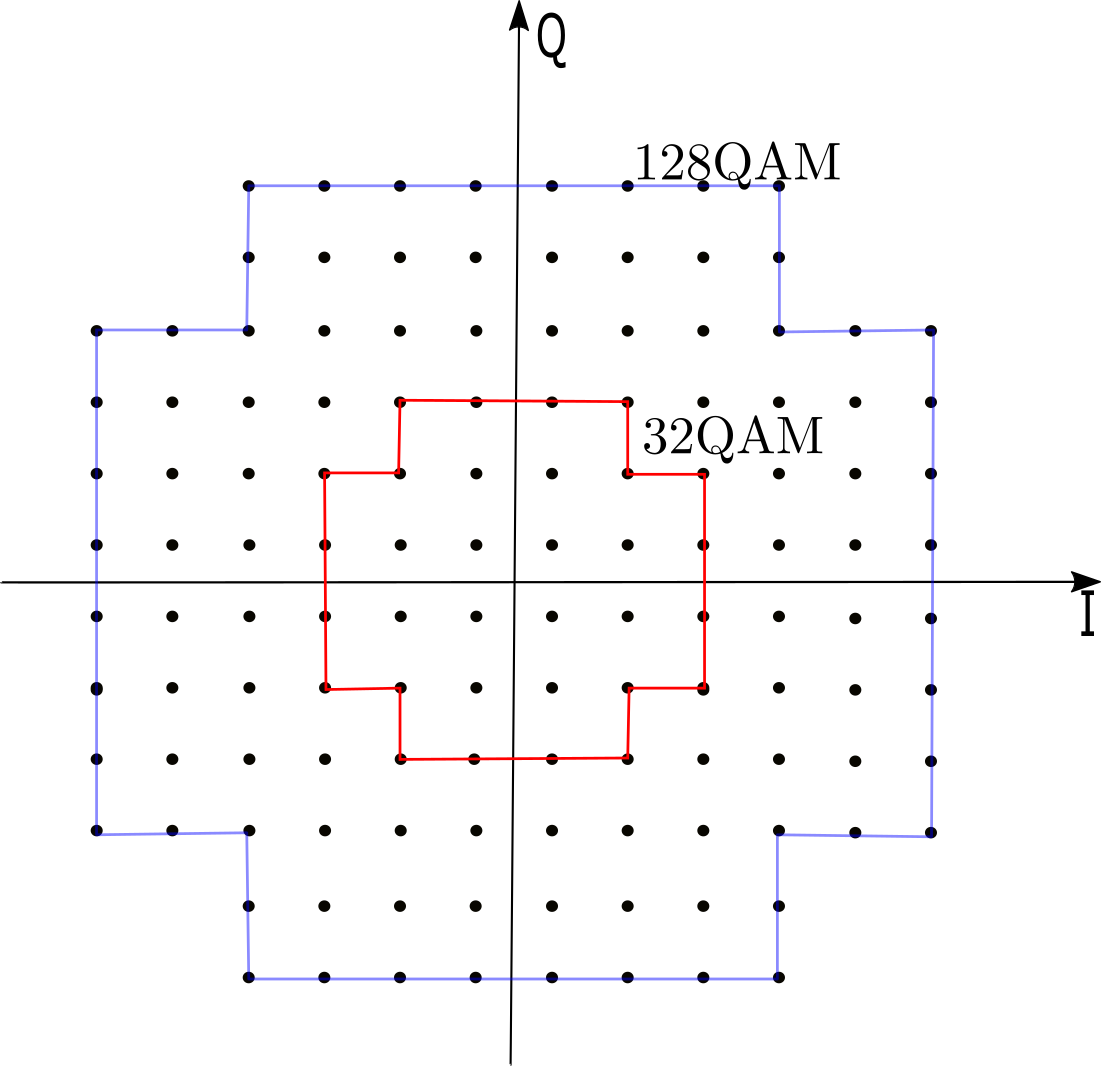}
\caption{{\small Cross constellation}}
 \label{fig:mqam_cross}
  \end{center}
 \end{subfigure}
   \vspace{-1.5em} \caption{{\small  Various constellation diagrams for MQAM. }}
 \label{fig:Mqam_mapping}
   \vspace{-1.5em}
\end{figure}

We can immediately infer from  {\bf Theorem \ref{term:theorem}}  that  the coefficient of the term $z_0z_1$ in $f_{QPSK}$ of  \eqref{eq:fqpsk_exp} is zero.  Fig. \ref{fig:qpsk_cg} illustrates the constellation points associated with the monomials of the form $z_0,~z_1$ and $z_0z_1$, respectively.  It can be readily seen that  the constellation points associated with the monomial of the form $z_0z_1$ create a rectangle and the pair of the diagonal points $(0,3)$ has  even numbers of  $0$ in their binary representations, while the binary orders for  the other pair of  the diagonal points $(1,2)$ have  odd numbers of $0$.  
In communication systems, QAM is  widely adopted for data modulation \cite{recommendation2012detailed},  where we will investigate the connection between the degree of the objective function formulated  and the associated constellation diagram.   Fig. \ref{fig:mqam_rec} portrays   various MQAM constellation diagrams  for $M=16, 32, 64,128$ and $256$,  which can be labelled using Gray mapping order. 
Furthermore,  Fig. \ref{fig:mqam_cross}  shows the cross constellations for $32$QAM and $128$QAM, which cannot be labelled following the Gray mapping rule \cite{Smith75TCOM}. 
The methods of Gray mapping  for rectangular QAM constellations and   quasi-Gray mapping for cross QAM constellations can be found in  \cite{Wesel01IT,Vitthaladevuni05TWC}.

\vspace*{-0.9em} 
\begin{corollary} \label{term:coro_nongray}
If a constellation is  not based on  Gray mapping,  the monomial of the form $\prod_{n\in \mathcal{N}}z_n$ in $f(z_0,\cdots, z_{N-1})$ has a coefficient of zero, if there exists a bunch of  rectangles with the constellation points as  their vertices, such that the following properties:  
\begin{enumerate}
\item   The vertices of these rectangles  cover all points in the constellation; \label{cor:item1}
\item  There are no shared vertices between any two rectangles;  and \label{cor:item2}
\item  For each rectangle the orders  associated with the two pairs of the diagonal points  contain odd(even) and even(odd) numbers of $0$s, respectively. \label{cor:item3}

\end{enumerate}

\end{corollary}
\vspace*{-0.9em}

\vspace*{-0.9em} 
\begin{corollary} \label{term:coro_nongray2}
Given a constellation with any reasonable mapping order,  the monomial of the form $\prod_{n \in \mathcal{S}}z_n$ in  $f(z_0,\cdots,z_{N-1})$ has  a coefficient of  zero, if  the constellation points of order $i$, for all $i\in \mathcal{M}'$,  form a bunch of rectangles such that  the properties in  \ref{cor:item2}) and \ref{cor:item3}) of {\bf Corollary \ref{term:coro_nongray}}.

\end{corollary}
\vspace*{-0.9em} 

The proofs of  {\bf Corollary \ref{term:coro_nongray}} and {\bf Corollary \ref{term:coro_nongray2}}  are provided in  Appendix~C. Note that {\bf Corollary \ref{term:coro_nongray2}} shows the zero-coefficient conditions for a monomial of a  constellation with  general labelling orders.

\subsubsection{Problem Hamiltonian  for 8QAM Constellation}

As shown in Fig. \ref{fig:8qam_mapping},  there are eight clauses in terms of the eight constellation points of the 8QAM constellation diagram.    By combining these clauses  into a function of  binary variables $z_0$, $z_1$ and $z_2$,  we have the objective function  of 
\begin{eqnarray}
\begin{aligned} \label{eq:f_8qam}
& f_{8QAM}(z_0,z_1,z_2) =  d_0(1-z_0)(1-z_1)(1-z_2)+d_1(1-z_0)(1-z_1)z_2 + d_2(1-z_0)z_1(1-z_2) \\&  \quad +  d_3(1-z_0)z_1z_2 + d_4 z_0(1-z_1)(1-z_2) + d_5 z_0(1-z_1)z_2 + d_6 z_0z_1(1-z_2) + d_7 z_0 z_1 z_2\\
& = \bar{d}_{0,1,2}  z_0 z_1 z_2 +  \bar{d}_{0,1}  z_0z_1 + \bar{d}_{0,2}  z_0 z_2+ \bar{d}_{1,2}  z_1z_2 + \bar{d}_0 z_0  +  \bar{d}_1  z_1 + \bar{d}_{2}  z_2 + d_0,
\end{aligned}
\end{eqnarray}
where $\bar{d}_{0,1,2} = -d_0 +d_1 +d_2 -d_3 +d_4 -d_5 -d_6 +d_7$,  $\bar{d}_{0,1} = d_0 -d_2 -d_4 +d_6$,  $\bar{d}_{0,2} =  d_0 -d_1 -d_4 +d_5$,  $\bar{d}_{1,2} = d_0 -d_1 -d_2 +d_3 $, $\bar{d}_{2} =  -d_0+d_1$, $\bar{d}_{1} = -d_0 +d_2$ and $\bar{d}_0 = -d_0 +d_4 $.  In Fig. \ref{fig:8qam_cg}, we demonstrate the sets of constellation points associated with all possible  monomials, i.e.  for all $\mathcal{S} \subseteq \mathcal{N}$.  Following the property in {\bf Theorem  \ref{term:theorem}} and {\bf Corollary \ref{term:coro_nongray2}}, we have $\bar{d}_{0,1,2} = 0$, $\bar{d}_{0,2} = 0$, and $\bar{d}_{1,2} = 0$. Hence  $f_{8QAM}(z_0,z_1,z_2)$  is  simplified as: 
\begin{eqnarray}
f_{8QAM}(z_0,z_1,z_2)  =   \bar{d}_{0,1}  z_0z_1 + \bar{d}_0 z_0  +  \bar{d}_1  z_1 + \bar{d}_{2}  z_2,
\end{eqnarray}
which is a quadratic function involving a product item $z_0 z_1$.

Correspondingly,  the quantum Hamiltonian $H_f$  can be expressed as
\begin{eqnarray}
\begin{aligned} \label{eq:Hf_8qam}
H_{f_{8QAM}} =  \bar{d}_{0,1} \sigma_z^{(0)} \sigma_z^{(1)} - (2\bar{d}_0+\bar{d}_{0,1}) \sigma_z^{(0)} - (2\bar{d}_1 + \bar{d}_{0,1}) \sigma_z^{(1)}  - 2 \bar{d}_2  \sigma_z^{(2)}, 
\end{aligned}
\end{eqnarray}
which  indicates that  implementing the ML detection  of 8QAM requires 3 qubits, but there is only a single  interaction between qubit $0$ and qubit $1$, as illustrated in Fig. \ref{fig:qubits_illu}(b).
As a result,   at least two-qubit quantum devices are needed for performing the ML detection of 8QAM.

\subsubsection{Problem Hamiltonian  for $16$QAM}
For the 16QAM  constellation having 16 points, we need $N = \log_2(16) = 4$ qubits for encoding the solutions of the ML detection problem. By transforming the clause  into a function of  binary variables $z_0$, $z_1$, $z_2$ and $z_3$,  we have the objective function
\begin{eqnarray}
\begin{aligned} \label{eq:f_16qam}
f_{16QAM}(z_0,z_1,z_2,z_3) =&  d_0(1-z_0)(1-z_1)(1-z_2)(1-z_3)+d_1(1-z_0)(1-z_1)(1-z_2)z_3  \\&  +  \cdots+d_{15} z_0 z_1 z_2z_3\\
=& \underbrace{\bar{d}_{0,1}z_0z_1+ \bar{d}_0 z_0 + \bar{d}_1 z_1}_{f(z_0,z_1)} +\underbrace{\bar{d}_{2,3}z_2z_3 +\bar{d}_2z_2+\bar{d}_3z_3}_{f(z_1,z_2)},
\end{aligned}
\end{eqnarray}
where $\bar{d}_{0,1} = d_0-d_4-d_8+d_{12},~\bar{d}_{2,3} = d_0-d_1-d_2+d_3, ~\bar{d}_0 = d_8-d_0, ~\bar{d}_1 = d_4-d_0,~\bar{d}_2=d_2 - d_0$ and $\bar{d}_{3} = d_1-d_0$.  Given  $f_{16QAM}(z_0,z_1,z_2,z_3) $, we can now immediately write out  $H_{f_{16QAM}}$ explicitly. Here we focus on studying the connection between the patterns in the constellation diagram and the monomials in $f_{16QAM}(z_0,z_1,z_2,z_3)$. 
 The patterns of the constellation points associated with the monomials from degree 1 to degree 4  in $f_{16QAM}(z_0,z_1,z_2,z_3)$ are illustrated in Fig. \ref{fig:16qam_cg} and Fig. \ref{fig:16qam_cgp2}.   Observe  from Fig. \ref{fig:16qam_cg}  that  all of the monomials of degree 1 exist in the expansion of  $f_{16QAM}(z_0,z_1,z_2,z_3)$, while there are only two monomials of degree 2.  Explicitly, in Fig. \ref{fig:16qam_cg} the two monomials of  the forms  $z_0z_1$ and $z_2z_3$  correspond to the constellation points of the bottom row and the far left column in the constellation diagram, respectively. 
Furthermore, we  see from Fig. \ref{fig:16qam_cgp2} that all of the  coefficients of the monomials with degree 3 and degree 4 are zero, since the set of their associated constellation points form rectangles.  Correspondingly,  the simplified expansion of $f_{16QAM}(z_0,z_1,z_2,z_3)$ is obtained in \eqref{eq:f_16qam}.
 Moreover, upon comparing \eqref{eq:f_16qam} to \eqref{eq:f_8qam}, we find that  the expansion of $f_{16QAM}(z_0,\cdots,z_3)$ can be viewed as the sum of  $f_{8QAM}(z_0,z_1)$ and $f_{8QAM}(z_2,z_3)$.   This indicates that in the four-qubit system associated with  the  ML detection problem of 16QAM,  there are two pairs of spins in parallel,  as illustrated in Fig. \ref{fig:qubits_illu}c, where interactions only happen within each pair. Therefore,  it is  possible to implement the ML detection of 16QAM in a two-qubit quantum device.

As described in \cite{Smith75TCOM,Vitthaladevuni05TWC}, for a rectangular QAM constellation, the  bits associated with each point can be  split  into two groups,
denoted as $b_0 \cdots b_{\lceil \frac{N}{2}\rceil - 1}$ referred to as the in-phase bits and $b_{\lceil \frac{N}{2}\rceil} \cdots b_{N-1}$ referred to as the quadrature phase bits, respectively.   For the simplicity of  notations, we  define a pair of  sets  $\mathcal{N}_I=\{0,\cdots,\lceil \frac{N}{2}\rceil - 1\}$ and $\mathcal{N}_Q = \{\lceil \frac{N}{2}\rceil ,\cdots,N-1\}$ corresponding to the indices of  the in-phase and quadrature phase bits, respectively.  Hence,  we have $N_I = |\mathcal{N}_I| = \lceil \frac{N}{2}\rceil $ and $N_Q=|\mathcal{N}_Q| = N-\lceil \frac{N}{2}\rceil = \lfloor \frac{N}{2}\rfloor$. The two groups of bits can then be  arranged following  Gray mapping  along each axis and  the  final  order is thus obtained by the combination of  the in-phase and quadrature phase bits.

\vspace{-0.9em}
\begin{theorem} \label{term:theory_2}
Consider a rectangular MQAM constellation $\mathcal{A}$ following a Gray mapping order, where  $N = \log_2 M$ denotes the number of bits per symbol.
\begin{enumerate}
\item The monomials of the form $\prod_{n \in \mathcal{S}}z_n$ in $f(z_0,\cdots, z_N)$ are zero if  there exists at least one pair of  bits denoted by $b_l=0$ and $b_m=0$ with $l,m \in  \mathcal{S}$, such that $b_l$ and $b_m$ belong to different groups (including the group of in-phase bits $\mathcal{N}_I$ and the group of quadrature bits $\mathcal{N}_Q$). \label{theorem2_item1}

\item The degree of $f(z_0,\cdots, z_N)$ is $N_I = \lceil \frac{N}{2}\rceil$. \label{theorem2_item2}

\end{enumerate}

\begin{proof}
See Appendix~D
\end{proof}
\end{theorem}
\vspace{-0.9em}

\vspace{-0.9em}
\begin{remark}\label{term:rem_IQp1}
{\bf Theorem \ref{term:theory_2}} shows that  the monomials of the form $\prod_{n \in \mathcal{S}} z_n$ in the expanded $f(z_{0},\cdots,z_{N-1})$ will be cancelled if  $\exists (l,m) \in  \mathcal{S}$ such that $b_l=b_m = 0$ with $l \in \mathcal{N}_I$ and $m \in \mathcal{N}_{Q}$.
\end{remark}
\vspace{-0.9em}

\vspace*{-0.9em} 
\begin{remark}\label{term:remark_7}
{\bf Theorem \ref{term:theory_2}} indicates that no interactions happen between  the qubits encoding  the in-phase  and  quadrature bits in the quantum system associated with the ML detection problem of interest.  
\end{remark}
\vspace*{-0.9em}

As shown in Fig. \ref{fig:qubits_illu}a-c,   we can use a  graph to illustrate the interactions among qubits  based on the problem Hamiltonian constructed.  Note that the  graph is disconnected for  a Gray-labelled MQAM constellation, since the qubits corresponding to $\mathcal{N}_I$ and $\mathcal{N}_Q$ are independent.  In particular, the graph  will contain isolated  nodes if $N_I= 1$ or $N_Q =1 $,  seen in the graphs generated from  QPSK and 8QAM in Fig. \ref{fig:qubits_illu}.

\vspace*{-0.9em} 
\begin{remark}
Observe from {\bf Theorem \ref{term:theory_2}}  that  the degrees of  64QAM, 256QAM and 1024QAM are 3, 4 and 5, respectively.
\end{remark}
\vspace*{-0.9em}

Let us now turn to  the cross constellation of odd-bit QAM  following pseudo-Gray mapping.  {\bf Corollary \ref{term:coro_nongray}} demonstrates that the degree of $f(z_0,\cdots, z_N)$ is no higher than  $N-1$, 
if the cross constellation obeys  the properties in \ref{cor:item1})-\ref {cor:item3}) of {\bf Corollary \ref{term:coro_nongray}}.  Furthermore,  the pseudo-Gray mapping breaks the pattern of four points forming a rectangle, where a pair of points in each edge share the same in-phase bits or the quadrature phase bits. Therefore, the constellation points associated with a monomial do not always form rectangles.  We have to  check the coefficient of each monomial  individually. For instance, for the cross 32QAM constellation following the quasi-Gray mapping orders of \cite{Wesel01IT,Vitthaladevuni05TWC}, there are always some monomials of degree $4$ which  cannot be cancelled,  this results in  the objective function $f_{32QAM}(z_0,\cdots, z_4)$ of degree $4$.

\subsubsection{Problem Hamiltonian of 64QAM}
We finally consider the objective function $f(z_0,\cdots,z_5)$ for  64QAM,  the degree of which is $3$  based on   {\bf Theorem \ref{term:theory_2}}.  Following {\bf Remark \ref{term:remark_7}}, 
we  simplify $f(z_0,\cdots,z_5)$  as follows:
\begin{eqnarray}
\begin{aligned}
f_{64QAM}(z_0,\cdots, z_{5}) = f(z_0,z_1,z_2) + f(z_3,z_4,z_5),
\end{aligned}
\end{eqnarray}
where 
\begin{eqnarray}
\label{eq:f_deg3}
\begin{aligned}
 f(z_l,z_m ,z_n) = \bar{d}_{l,m,n}z_lz_mz_n+\bar{d}_{l,m}z_lz_m+\bar{d}_{l,n}z_lz_n+\bar{d}_{m,n}z_mz_n+\bar{d}_lz_l +\bar{d}_mz_m\bar{d}_nz_n, 
\end{aligned}
\end{eqnarray}
with $(l,m,n)=(0,1,2)$ and $(3,4,5)$, respectively. We can obtain the coefficients $\bar{d}_{\mathcal{S}}$ from {\bf Remark \ref{term:remark_expan3}}, which are  given by 
\begin{eqnarray}
\begin{aligned}
&\bar{d}_{0,1,2} = -d_0+d_{8}+d_{16}-d_{24}+d_{32}-d_{40}-d_{48}+d_{56}, 
\bar{d}_{0,1} = d_0-d_{16}-d_{32}+d_{48}, \\
&\bar{d}_{0,2} = d_0 -d_{8}- d_{32}+d_{40},
\bar{d}_{1,2} = d_0-d_{8} - d_{16}+d_{24},
\bar{d}_{0} = -d_0 + d_{32}, 
\bar{d}_{1} = -d_{0}+d_{16},  \\
& \bar{d}_{2} = -d_{0}+d_8, 
 \bar{d}_{3,4,5} = -d_0+d_{1}+d_{2}-d_{3}+d_{4}-d_{5}-d_{6}+d_{7}, 
\bar{d}_{3,4} = d_0-d_2-d_4+d_6, \\
& \bar{d}_{3,5} = d_0 -d_1-d_4+d_5,
\bar{d}_{4,5} = d_0-d_1-d_2+d_3,
\bar{d}_{3} = -d_0 + d_{4}, 
\bar{d}_{4} = -d_{0}+d_{2}, 
\bar{d}_{5} = -d_{0}+d_1.
\end{aligned}
\end{eqnarray}
Then the problem Hamiltonian $H_{f_{64QAM}}$ can  be obtained upon replacing  $z_i$ by $\frac{1}{2}(1-\sigma_z^{(i)})$ for all $i$. Correspondingly, the Hamiltonians of $f(z_0,z_1,z_2)$ and  $f(z_3,z_4,z_5)$ are independent, which can thus be implemented using two independent quantum systems. Note that the degrees of  $f(z_0,z_1,z_2)$ and  $f(z_3,z_4,z_5)$ are three,  
which are associated with a $3$-local Hamiltonian, respectively. 
We have to reduce the $3$-local Hamiltonian into a  two-local Hamiltonian, since the QAOA is typically  limited to allowing no more than two-local interactions.  This task can be carried out by following the   reduction method introduced in Section \ref{sec:intro}, such as the method of reduction by substitution \cite{boros2020compact}  requiring auxiliary variables or the split-reduction method  of 
\cite{okada2015reducingp2} operating without adding auxiliary variables.

\vspace{-0.9em}
\section{Problem Hamiltonian Constructions for MIMO Channels}
\label{sec:mimo}
Following the investigations in Section \ref{sec:siso}, in this section we consider the problem Hamiltonian of  the $N_r \times N_t$ MIMO channel. Firstly, we rewrite \eqref{eq:r_mimo} as follows:
\begin{eqnarray}
{\small
\begin{aligned}
 \begin{bmatrix}
y_0 \\
\vdots \\
y_{N_r-1}
\end{bmatrix} &= \begin{bmatrix}
h_{0,0} & \cdots &h_{0,N_t-1} \\
\vdots & \ddots & \vdots \\
h_{N_r-1,0}&\cdots& h_{N_r-1,N_t-1}
\end{bmatrix} \begin{bmatrix}
s_0 \\ \vdots \\s_{N_t-1}
\end{bmatrix} + \begin{bmatrix}
\eta_0 \\ \vdots \\\eta_{N_t - 1}
\end{bmatrix} 
= \begin{bmatrix}
\sum_{k=0}^{N_t-1} h_{1,k}s_{k} +\eta_0 \\
 \vdots \\
 \sum_{k=0}^{N_t-1} h_{N_r-1,k}s_{k} +\eta_{N_r-1}
\end{bmatrix},
\end{aligned}}
\end{eqnarray}
which can be viewed as $N_r$ parallel SISO channels.  The objective function of the ML detection problem in \eqref{eq:ML_mimo} can be rewritten as
\begin{eqnarray}
{\small
\begin{aligned}
 f_{ML}(s_0,\cdots,s_{N_t-1})=& \begin{Vmatrix}
 \begin{bmatrix}
y_0 \\
\vdots \\
y_{N_r-1}
\end{bmatrix} - \begin{bmatrix}
\sum_{k=0}^{N_t-1} h_{1,k}s_{k}  \\
 \vdots \\
 \sum_{k=0}^{N_t-1} h_{N_r-1,k}s_{k} 
\end{bmatrix}
\end{Vmatrix}^2 
 = \begin{Vmatrix}
 \begin{bmatrix}
y_1 -\sum_{k=0}^{N_t-1} h_{1,k}s_{k}  \\
\vdots \\
y_{N_r-1}-\sum_{k=0}^{N_t-1} h_{N_r-1,k}s_{k} 
\end{bmatrix}
\end{Vmatrix}^2  \\
=& \sum_{l=0}^{N_r-1}  \begin{vmatrix}y_l -\sum_{k=0}^{N_t-1} h_{l,k}s_{k} \end{vmatrix}^2.
\end{aligned}}
\end{eqnarray}
Let $\mathcal{A} = \mathcal{A}_0 \times \cdots\times\mathcal{A}_{N_t-1}$ denote the joint constellation, with `$\times$' being the Cartesian product, where a constellation $\mathcal{A}_k$, $k \in \mathcal{N}_t=\{0,\cdots,N_t-1\}$, contains $\mathcal{M}_k$ constellation points.  Hence, $\mathcal{A}$ contains $M=\sum_{k\in \mathcal{N}_t} M_k$ constellation points in  total, with the indices set $\mathcal{M}= \{0,\cdots,M-1\}$.
Therefore, the ML detection problem is to find a point from the joint constellation $\mathcal{A}$  which has the minimum sum of the distance from the received signal over each antenna.  
Numbering the points in  $\mathcal{A}$ requires $N = \sum_{k \in \mathcal{N}_t}N_k$ bits, i.e  the qubits required  for representing the solutions, where $N_k = \log_2(M_k)$ is the number of bits required for  $\mathcal{A}_k$, $k \in \mathcal{N}_t$.
Furthermore, the orders of a point in $\mathcal{A}$ can be constructed by the Cartesian product of $N_t$ sets, in which the element is a $N_t$-tuple denoted as $(i_0, \cdots, i_{N_t - 1})$ with $i_k$ being the decimal order for  $\mathcal{A}_k$ and $i_k \in \mathcal{M}_k=\{0,\cdots,M_k-1\}$. 
As a result,  an element $i \in \mathcal{M}$ is defined as 
\begin{eqnarray}
\begin{aligned}
 [i]_2 = [i_0]_2 + \cdots + [i_{N_t-1}]_2, \label{eq:dec_bin}
\end{aligned}
\end{eqnarray}
where `$+$'  represents the concatenation operator which glues the binary orders together.  Explicitly, the mapping rule between the $i$-th point in $\mathcal{A}$ and the binary variables $z_0,\cdots,z_{N-1}$ is given by
\begin{eqnarray}
\label{eq:bin_var_map}
\begin{aligned}
\begin{matrix}
[i]_2 = \overbrace{b_0 \cdots  b_{N_0-1}}^{[i_0]_2} \cdots  \cdots b_{N-1} \leftrightarrow \overbrace{z_0,\cdots, z_{N_0-1}}^{[i_0]_2},\cdots, \cdots,z_{N-1}.
\end{matrix}
\end{aligned}
\end{eqnarray}
In addition, for a constellation $\mathcal{A}_k, k\in \mathcal{N}_t$,  we define $\mathcal{N}_{k,I}=\{\sum_{l=0}^{k-1}N_l,\cdots,\sum_{l=0}^{k-1}N_l+ \lceil \frac{N_k}{2} \rceil-1\}$ and $\mathcal{N}_{k,Q} = \{\sum_{l=0}^{k-1}N_l+\lceil \frac{N_k}{2} \rceil,\cdots, \sum_{l=0}^{k}N_l-1\}$ to be  the sets of the in-phase and quadrature phase bit indices, respectively.  Hence we have $N_k = N_{k,I}+N_{k,Q} $ with $N_{k,I} = |\mathcal{N}_{k,I}|$ and  $N_{k,Q} = |\mathcal{N}_{k,Q}|$. 
 
For  the $l$-th antenna, the received signal $y_l= \sum_{k \in \mathcal{N}_t} h_{l,k}s_{k} +\eta_l$ can be viewed as the output of a multiple access channel (MAC)  having $N_t$ users.  Similar to \eqref{eq:weight}, the squared distance between  $y_l$ and the $i$-th constellation point in  $\mathcal{A}$, $i \in \mathcal{M}$,  can  be expressed as
\begin{eqnarray}
d_{l,i} = \begin{vmatrix}y_l -\sum_{k \in \mathcal{N}_t} h_{l,k}s_{k}  \end{vmatrix}^2, \label{eq:weight_mimo}
\end{eqnarray}
where the connection between $i$ and $k, k \in \mathcal{N}_t$ satisfies \eqref{eq:dec_bin}. Following the transformations of \eqref{eq:clause}-\eqref{eq:binmap},  the objective function of the ML detection of MIMO systems can be expressed as 
\begin{eqnarray}
\label{eq:f_mimo}
\begin{aligned}
f(z_0,\cdots,z_{N-1}) = \sum_{l=0}^{N_r - 1} \sum_{i \in \mathcal{M}} w_l(C_{i}) 
= \sum_{l=0}^{N_r - 1}  \sum_{\mathcal{S} \in \mathcal{N}} d_{\mathcal{S}} \prod_{n \in \mathcal{S} } z_n,
=  \sum_{l=0}^{N_r - 1}  f_l(z_0,\cdots,z_{N-1}),
\end{aligned}
\end{eqnarray}
where $w_l(C_i)$ and $f_l(z_0,\cdots,z_{N-1})$ represent the weighted function and the corresponding component of the objective function for  the $l$-th received antenna, respectively.   
Similar to \eqref{eq:weighted_clause},  $w_l(C_i)$ can be expressed as
\begin{eqnarray}
\begin{aligned}
w_l(C_i) =d_{l,i} C_i = d_{l,i} [B(z_0) \wedge \cdots \wedge B(z_{N-1})].  \label{eq:weighted_clause_mimo}
\end{aligned}
\end{eqnarray}

Next we investigate the connection of the zero-coefficient characteristics of a monomial associated with a single constellation studied  in Section \ref{sec:siso} and the joint constellation of  the MIMO system, which would allow us  to simplify  the problem Hamiltonian representation.  
Since it is straightforward to express the problem Hamiltonian from $f(z_0,\cdots,z_{N-1})$, we focus on simplifying $f(z_0,\cdots,z_{N-1})$.

\vspace*{-0.9em} 
\begin{proposition}\label{term:remark_mimo_prop1}
  The monomial of the form $\prod_{n\in \mathcal{S}}z_n$ in $f(z_0,\cdots,z_{N-1})$  of \eqref{eq:f_mimo} has a coefficient  of zero,   if the constellation points in $\mathcal{A}_k$ contributing the component of  the variables in $\mathcal{S}\cap \mathcal{N}_k$  form a bunch of rectangles obeying the conditions in 2) and 3) of {\bf Corollary \ref{term:coro_nongray2}}. 
 \begin{proof}

 Following \eqref{eq:f_mimo} stating that $f(z_0,\cdots,z_{N-1})$  is the sum of $f_l(z_0,\cdots,z_{N-1})$, we can  infer that the zero-coefficient monomial in $f_l(z_0,\cdots,z_{N-1})$ has a coefficient of zero in $f(z_0,\cdots,z_{N-1})$. Therefore, we focus on the monomials in  $f_l(z_0,\cdots,z_{N-1})$.
The variables $z_n$,  $n \in \mathcal{S}_{k}=\mathcal{S} \cap \mathcal{N}_k$ come from the bits representing $\mathcal{A}_k$ and there are $2^{|\mathcal{S}|}=\prod_{k \in \mathcal{N}_t} 2^{|\mathcal{S}_k|}$ constellation points contributing to the monomial of $\mathcal{S}$. Fixing the orders of the constellation point in $\mathcal{A}_{k'}, k'\ne k$, there are $2^{|\mathcal{S}_k|}$ constellation points in $\mathcal{A}_k$ containing the monomial of  $\mathcal{S}$. 
 We then have that $d_{l,i}$ associated with the clause $C_{l,i}$ for the $l$-th receive antenna is $ d_{l,i} = |\bar{y}_i - h_{l,k} s_k  |^2$,
where $\bar{y}_{i} = y_l - \sum_{k' \ne k} h_{l,k'}s_{k'}$.  This indicates that if the constellation points in $\mathcal{A}_k$ satisfy the zero-coefficient conditions, the sum of the coefficients for $i \in \mathcal{M}'$ with fixing $i_{k'}, k' \ne k$, is zero, where $M'$ is given in \eqref{eq:M_prime}.   Furthermore, by traversing $ 2^{|\mathcal{S}|-|\mathcal{S}_{k}|}$ constellation points in $\mathcal{A}_{k'},k'\ne k$, the  final coefficient of  $\prod_{n \in \mathcal{S}} z_{n}$ can be obtained, which is the sum of $2^{|\mathcal{S}|-|\mathcal{S}_{k}|}$ $0$s. Hence, the zero-coefficient characteristics of the monomials on a single constellation are suitable for the monomials of  a MIMO channel.
 
  \end{proof}
\end{proposition}
\vspace*{-0.9em}

\vspace*{-0.9em} 
\begin{remark}\label{term:remark_8}
For   a MIMO system of Gray-labelled constellations,  the  coefficients of the monomials containing the product of the variables both from the in-phase  and from the quadrature phase bits are zero.  Therefore, there are still no interactions between the qubits encoding the in-phase  and  quadrature bits. 
\end{remark}
\vspace*{-0.9em}

\vspace*{-0.9em} 
\begin{corollary}\label{term:corollary_4}
For a $N_r \times N_t$ MIMO system with Gray-labelled QPSK,  the degree of the objective function $f(z_0,\cdots,z_N)$  is  $N_t$. 
\end{corollary}
\vspace*{-0.9em} 

\vspace*{-0.9em} 
\begin{corollary}\label{term:corollary_5}
For a $N_r \times N_t$ MIMO system  using  Gray-labelled MQAM, the degree of the objective function $f(z_0,\cdots,z_N)$ is  $N_{I} {N_t}$.
\end{corollary}
\vspace*{-0.9em} 

The proofs of {\bf Corollary \ref{term:corollary_4}} and {\bf Corollary \ref{term:corollary_5}} are given in Appendix~E.  Based on these two Corollaries,  we provide the following theorem for characterizing the degree of the objective function concerning a MIMO system. 

\vspace*{-0.9em} 
\begin{theorem}\label{term:theorem_3}
For a $N_r \times N_t$ MIMO channel associated with QAM,  the joint constellation is given by $\mathcal{A}$, where  each constellation follows  Gray mapping.   The degree of the objective function $f(z_0,\cdots,z_N)$  is $\sum_{k=0}^{N_t - 1} N_{k,I} $. 
\begin{proof}
See Appendix~F
\end{proof}
\end{theorem}
\vspace*{-0.9em}

Finally, we present an example of an $1\times 2$ MIMO channel using QPSK for illustrating its problem Hamiltonian, since the number of receiver antennas does not affect the degree of the objective function.  
Hence,  the ML detection problem  is defined over  the joint constellation $\mathcal{A}_{QPSK}^2$, denoted as $2\times$QPSK, which  requires $4$ bits for representing each  constellation point.    From \eqref{eq:f_mimo} the objective function of our ML detection problem for  the $1\times 2$ MIMO  scenario can be expressed as 
\begin{eqnarray}
\label{eq:f_2QPSK}
\begin{aligned}
f_{2\times QPSK}(z_0,\cdots,z_3) = &  d_0(1-z_0)(1-z_1)(1-z_2)(1-z_3)+d_1(1-z_0)(1-z_1)(1-z_2)z_3  \\&  +  \cdots+d_{15} z_0 z_1 z_2z_3
= \sum_{(m,n) \in \mathcal{E}}\bar{d}_{m,n}z_mz_n + \sum_{n \in \mathcal{N}}\bar{d}_{n}z_n,
\end{aligned}
\end{eqnarray}
 where $\mathcal{E} = \{(0,2),(0,3),(1,2),(1,3)\}$ and $\mathcal{N} = \{0,1,2,3\}$. Following  {\bf Remark \ref{term:remark_expan3}},  the coefficients can be expressed as $
 \bar{d}_{0,2} = d_0-d_2 - d_8+d_10,   \bar{d}_{0,3} = d_0-d_1-d_8+d_9, 
 \bar{d}_{1,2} = d_0-d_2-d_4+d_6, \bar{d}_{1,3} = d_0-d_1-d_4+d_5,
 \bar{d}_{0} = -d_0+d_8, \bar{d}_1 = -d_0+d_4, \bar{d}_2 = -d_0+d_2, \bar{d}_3 = -d_0+d_1.$
Note  that \eqref{eq:f_2QPSK} is a quadratic function and the problem Hamiltonian thus contains two-local interactions  at most. The interactions between the qubits are illustrated  in Fig. \ref{fig:qubits_illu}(d). We  can see that there are  no interactions within the two pairs of qubits $(z_0,z_1)$  and  $(z_2,z_3)$.

 \vspace{-0.7em}
\section{QAOA Assisted ML detection}
\label{sec:qaoa_ml}

 \vspace{-0.7em}
\subsection{Implementation of the QAOA}

\begin{figure} [t!]
\centering
\includegraphics[width = 0.69\linewidth]{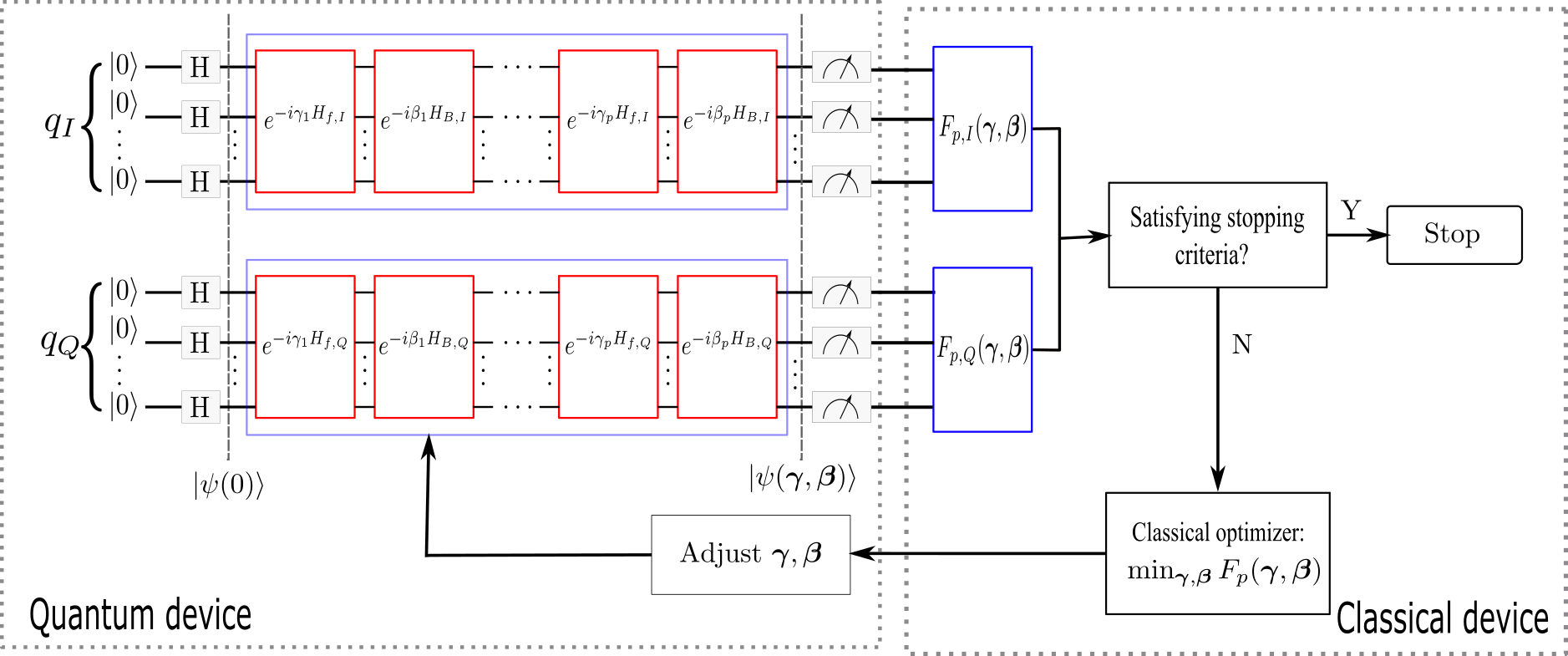}
  \vspace{-1.1em}  \caption{Schematic diagram of the QAOA for Gray-labelled constellations of a SISO channel. }
 \label{fig:qaoa_diag}
   \vspace{-1.0em}
\end{figure}

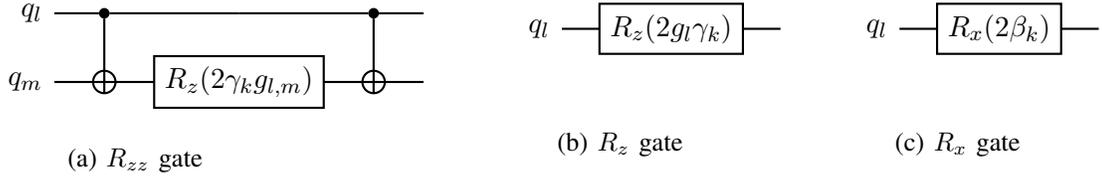
\begin{figure}[ht!]
\centering
\begin{subfigure}[t]{.22\textwidth}
\begin{center}
\begin{quantikz}
&\lstick{$q_l$} & \ctrl{1} & \qw & \ctrl{1} & \qw  \\
&\lstick{$ q_m$}  & \targ{} & \gate{R_z(2\gamma_k g_{l,m})} & \targ{} & \qw
\end{quantikz}
\caption{$R_{zz}$ gate}
\label{fig:cir_rzz}
\end{center}
\end{subfigure}
\qquad \qquad \qquad \qquad
\begin{subfigure}[t]{.17\textwidth}
\begin{center}
\begin{quantikz}
&\lstick{$q_l$}  &  \gate{R_z(2 g_l \gamma_k )}  & \qw \\
\end{quantikz}
\caption{$R_z$ gate}
\label{fig:cir_rz}
\end{center}
\end{subfigure}
\qquad \qquad
\begin{subfigure}[t]{.17\textwidth}
\begin{center}
\begin{quantikz}
&\lstick{$q_l$}  &  \gate{R_x(2\beta_k )}  & \qw \\
\end{quantikz}
\caption{$R_x$ gate}
\label{fig:cir_rx}
\end{center}
\end{subfigure}
 \vspace{-1.5em}\caption{{\small Schematic of basic quantum gates for implementing QAOA circuits}}
\label{fig:cir_gates}
 \vspace{-1.5em}
\end{figure}

In this section, we  revisit the basic principles of the QAOA. The QAOA is a approximate hybrid quantum-classical  optimization algorithm proposed  in \cite{farhi2014QAOA}, which alternately  applies the problem Hamiltonian $H_f$ and the mixing Hamiltonian $H_B$ to the initial state prepared by the quantum system.  For a SISO channel associated with a Gray-labelled QAM constellation, following {\bf Remark \ref{term:remark_7}} that there are no interactions between the qubits encoding the in-phase  and  quadrature bits, we can  employ a pair of   QAOA  circuits for each, respectively, as shown in Fig. \ref{fig:qaoa_diag}. 
In Fig. \ref{fig:qaoa_diag},    $\boldsymbol{\gamma} = [\gamma_1,\cdots,\gamma_p]$, $\boldsymbol{\beta} = [\beta_1,\cdots,\beta_p]$ and $p$ is a parameter controlling the depth of the circuits.  Moreover, $q_I$ and $q_Q$ represent the qubits  encoding the  in-phase and quadrature phase bits, respectively.   
For each level $k$, $k=1,\cdots,p$,  the  parameterized  unitary operators $U(H_f,\gamma_k)$ and $U(H_B, \beta_k) $  in terms of $H_f$ and $H_B$ can be represented as follows:
\begin{align}
U(H_f,\gamma_k) &=  \underbrace{e^{-i \gamma_k H_{f,I}}}_{U(H_{f,I},\gamma_k)} + \underbrace{ e^{-i \gamma_k H_{f,Q}}}_{U(H_{f,Q},\gamma_k)}, \\
U(H_B, \beta_k) &=  \underbrace{ e^{-i \beta_k H_{B,I}} }_{U(H_{B,I}, \beta_k) } + \underbrace{ e^{-i \beta_k H_{B,Q}}}_{U(H_{B,Q}, \beta_k) },
\end{align}
where $H_{\cdot,I}$ and $H_{\cdot,Q}$ represent the Hamiltonian operators associated with $q_I$ and $q_Q$, respectively. 
Here we  discuss the implementation of  the QAOA circuits of $q_I$ as an example, since the QAOA circuits in terms of $q_I$ and $q_Q$ have the same  structure.
For each level $k$, the unitary evolution of $U(H_{f,I},\gamma_k)$ involves a sequence  of two-qubit unitary operators  $R_{zz}(2\gamma_k g_{l,m}) = e^{-i\gamma_k g_{l,m} \sigma_z^{(l)} \sigma_z^{(m)}}$ and a set of  single-qubit unitary operators $R_z(2 g_{l } \gamma_k) =  e^{-i g_{l } \gamma_k \sigma_z^{(l)} }$,  $l,m \in \mathcal{N}_I,l\ne m$, which  can be  realized by the  circuits shown in Fig. \ref{fig:cir_rzz} and Fig. \ref{fig:cir_rz}, respectively.  Here, $g_{l,m}$ and $g_{l}$ are the coefficients obtained in \eqref{eq:Hf_general}, which are related to the channel realizations.  We use $q_l$ and $q_m$ to represent the  quantum states of the $l$-th and $m$-th qubit, respectively.   Furthermore,  $R_z(\theta) $ refers to  the rotation-Z operator having parameter $\theta$ with $R_z(\theta) = e^{-\frac{i \theta}{2}}$.   More specifically,  each  unitary operator $R_{zz}$ as seen in Fig. \ref{fig:cir_rzz},  consists of two CNOT gates and a $R_z$ gate.  We then have the matrix representation  of $R_{zz}(2\gamma_k d_{l,m})$, which is  given by 
\begin{eqnarray}
\label{eq:r_zz}
\begin{aligned}
R_{zz}(2\gamma_k d_{l,m}) &=   CX_{l,m} R_z(2\gamma_k d_{l,m})  CX_{l,m} = e^{-i \gamma_k d_{l,m} CX_{l,m} \sigma_z^{(m)} CX_{l,m}} \\
&=  e^{-i \gamma _k d_{l,m} \sigma_z^{(l)} \otimes  \sigma_z^{(m)}}
= \mathrm{diag}[e^{-i \gamma_k d_{l,m}},e^{-i \gamma_k d_{l,m}} ,e^{-i \gamma_k d_{l,m}} ,e^{-i \gamma_k d_{l,m}} ],
\end{aligned}
\end{eqnarray}
where $CX_{l,m}$ represents the CNOT gate, with $q_l$ and $q_m$ being the control  and the target qubits, respectively.
In addition, the  unitary evolution of $U(H_{B,I},\beta_k)$ for level $k$ is realized by a series of rotation-X operators with parameter $\beta_k$, where the associated circuit unit is illustrated in Fig. \ref{fig:cir_rx} and $R_x(\theta) = e^{\frac{-i\theta}{2}}$. 
\begin{figure} [t!]
\centering
\begin{subfigure}{0.9\textwidth}
\begin{center}
\includegraphics[clip,width = 0.9\textwidth]{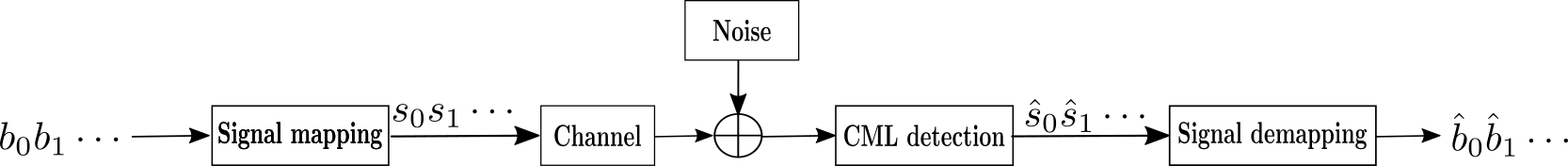}
 \caption{CML detector assisted uncoded communication system}
  \label{fig:CML}
  \end{center}
\end{subfigure}
\begin{subfigure}{0.9\textwidth}
\begin{center}
\includegraphics[clip, width = 0.9\textwidth]{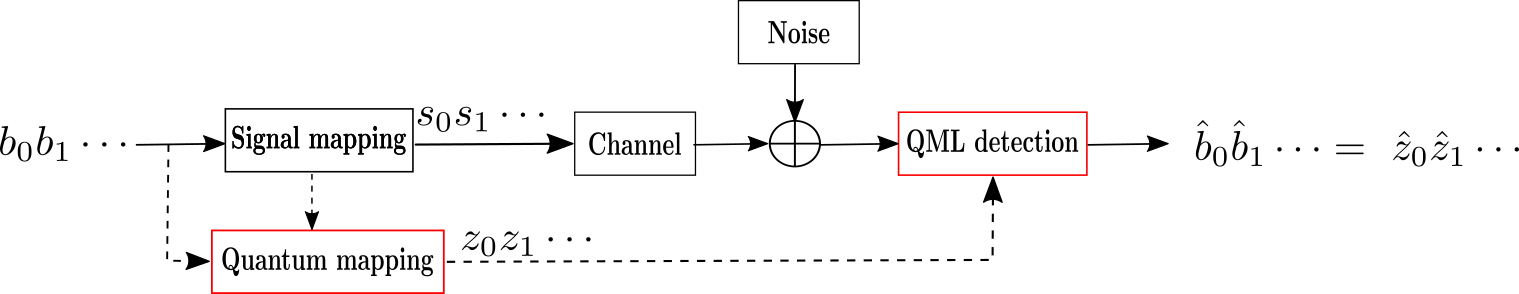}
\caption{QML detector assisted uncoded communication system }
 \label{fig:QML}
 \end{center}
\end{subfigure}
    \vspace{-1.5em} \caption{Comparison of uncoded  CML and QML receivers.}
 \label{fig:QML_CML}
   \vspace{-1.5em}
\end{figure} 

Following the unitary evolution associated with the problem Hamiltonian $H_f$ and the initial Hamiltonian $H_B$,   the parameterized quantum sate $|\psi(\boldsymbol{\gamma},\boldsymbol{\beta} ) \rangle $ in Fig. \ref{fig:qaoa_diag} can be expressed as
\begin{eqnarray}
\begin{aligned}
|\psi(\boldsymbol{\gamma},\boldsymbol{\beta} ) \rangle & = U(H_B, \beta_p) U(H_f,\gamma_p) \cdots U(H_B, \beta_1) U(H_f,\gamma_1) |\psi(0) \rangle, \\
 &= U(H_{B,I}, \beta_p) U(H_{f,I},\gamma_p) \cdots U(H_{B,I}, \beta_1) U(H_{f,I},\gamma_1) |\psi_{I}(0) \rangle + \\& \quad U(H_{B,Q}, \beta_p) U(H_{f,Q},\gamma_p) \cdots U(H_{B,Q}, \beta_1) U(H_{f,Q},\gamma_1) |\psi_{Q}(0) \rangle.
\end{aligned}
\end{eqnarray} 
 By measuring the state $|\psi(\boldsymbol{\gamma},\boldsymbol{\beta} ) \rangle $ in the computational basis, we glean an output $|\langle z_0 \cdots z_{N-1} | \psi(\boldsymbol{\gamma},\boldsymbol{\beta} ) \rangle |^2$ associated with a single candidate solution $|z_0 \cdots z_{N-1} \rangle$.  As illustrated in Fig. \ref{fig:qaoa_diag},  repeating the preparation of the parameterized state $|\psi(\boldsymbol{\gamma},\boldsymbol{\beta})\rangle$ and the measurements,    the expectation value of the objective function to our ML detection can be expressed as
\begin{eqnarray}
\begin{aligned}
F_p(\boldsymbol{\gamma},\boldsymbol{\beta}) &= \langle \psi(\boldsymbol{\gamma},\boldsymbol{\beta} ) | H_f | \psi(\boldsymbol{\gamma},\boldsymbol{\beta} ) \rangle \\
&= \langle \psi_I(\boldsymbol{\gamma},\boldsymbol{\beta} ) | H_{f,I} | \psi_I(\boldsymbol{\gamma},\boldsymbol{\beta} ) \rangle + \langle \psi_Q(\boldsymbol{\gamma},\boldsymbol{\beta} ) | H_{f,Q} | \psi_Q(\boldsymbol{\gamma},\boldsymbol{\beta} ) \rangle.
\end{aligned}
\end{eqnarray}
Therefore, we can express $F_1$ for QPSK as 
\begin{eqnarray}
\label{eq:F_1qpsk} 
\begin{aligned}
F_1(\gamma_1,\beta_1) =& \langle
\psi(\gamma_1,\beta_1) | H_f|\psi(\gamma_1,\beta_1)\rangle 
= \bar{F}_{0} + \bar{F}_{1}, ~~\text{where}~ \\
\bar{F}_{0} =& \bar{d}_{0}\sin(2\beta_1)  \sin(2\bar{d}_{0} \gamma_1)   \cos(2\bar{d}_{01} \gamma_1) ~\text{and}~
\bar{F}_{1} =  \bar{d}_{1}\sin(2\beta_1) \sin(2\bar{d}_{1} \gamma_1)  \cos(2\bar{d}_{01} \gamma_1)
\end{aligned}
\end{eqnarray}
correspond to the terms of $\sigma_z^{(0)}$ and $\sigma_z^{(1)}$  in the $H_f$, respectively.

 \vspace{-0.7em}
\subsection{ Quantum Maximum Likelihood (QML) detection}
Recall that in the QAOA, the solutions of the ML detection problem are encoded into the eigenstates and the optimal solution corresponds to the ground state of the problem Hamiltonian of the ML detection problem. For the problem Hamiltonian constructed in Section \ref{sec:siso} and Section \ref{sec:mimo}, we encode the candidate solutions (complex symbols represented in the constellation diagram) of the  original ML detection problem into binary variables,  in which the optimal binary solution to the ground state of the problem Hamiltonian is  the specific transmit bit string corresponding to the optimal constellation point. As a result,  the output of the QAOA is the estimated binary bit string sent by the transmitter, which is different from the output of  the CML detector. Fig. \ref{fig:QML_CML} compares  the CML  and  QML  receivers. Explicitly,   Fig. \ref{fig:CML} shows the CML detector assisted uncoded communication system, where the estimated bit string is  attained  by performing signal demapping after  CML detection. We see from Fig. \ref{fig:QML} that the estimated bit string is attained directly from the QML detector.  Furthermore,  quantum mapping  in Fig. \ref{fig:QML} corresponds to  the objective function constructed in \eqref{eq:obj_sat} for SISO systems or in \eqref{eq:f_mimo} for MIMO systems.

Finally, we introduce the  approximation ratio metric \cite{farhi2014QAOA,Zhou20PhysRevX} for  evaluating the quality of the solution  provided  by the QAOA. Since the  ML detection problem is a minimization problem, we define the approximation ratio  as follows: 
\begin{eqnarray}
\begin{aligned}
\rho = \frac{f_{CML}}{F_p(\boldsymbol{\gamma}^*,\boldsymbol{\beta}^*)},
\end{aligned}
\end{eqnarray}
where $f_{CML}$ is the objective function value obtained by the classical ML detection method.

 \vspace{-0.7em}
\section{Simulation Results}
\label{sec:sim}

In this section,  we first  visualize the  expectation values $F_1$ for QPSK concerning a set of concrete noise and channel coefficients. Then,  we quantify
the performance of the  QAOA  assisted QPSK ML detection,  where the QAOA is implemented using Qiskit Aer \cite{Qiskit}.

\begin{figure} [t!]
\centering
\begin{subfigure}{0.24\textwidth}
\includegraphics[width = 0.95\linewidth]{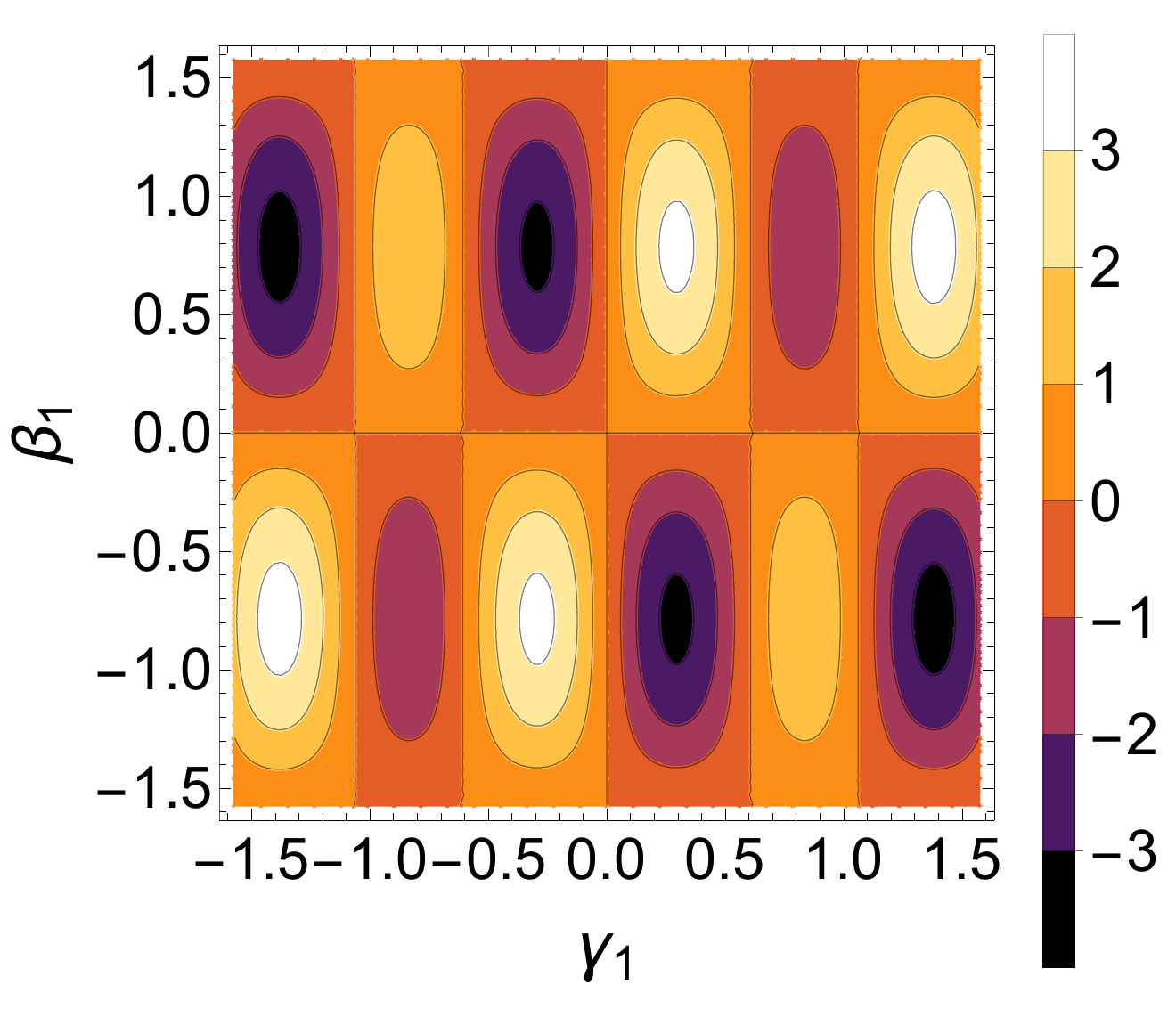}
  \vspace{-0.5em}
  \caption{{\small SNR = 0dB (AWGN)}}
 \label{fig:F1_snr_0_awgn}
 \end{subfigure}
  \begin{subfigure}{0.24\textwidth}
\includegraphics[width = 0.95\linewidth]{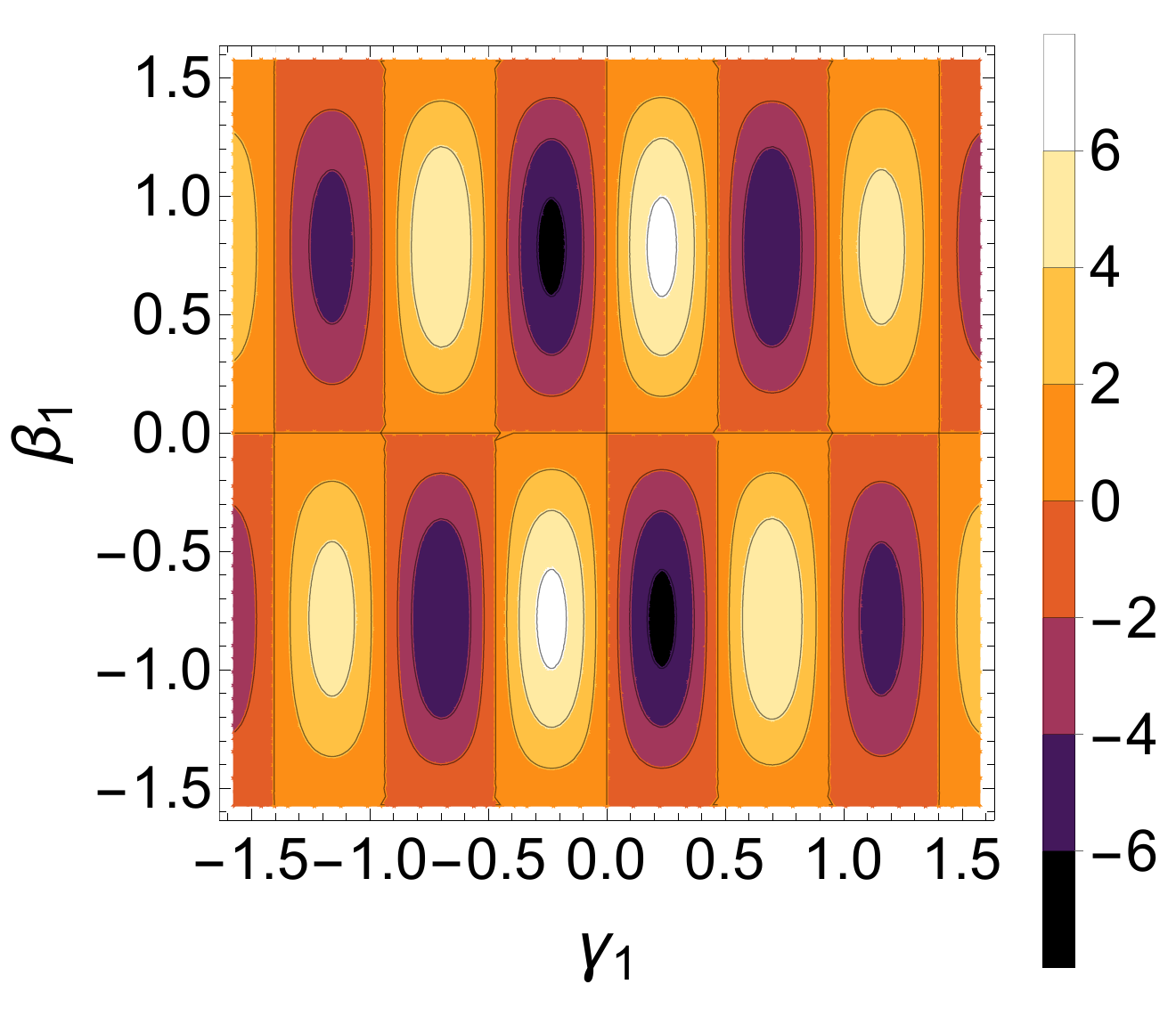}
  \vspace{-0.5em}
  \caption{{\small SNR = 10dB (AWGN)} }
 \label{fig:F1_snr_10_awgn}
 \end{subfigure}
\begin{subfigure}{0.24\textwidth}
\includegraphics[width = 0.95\linewidth]{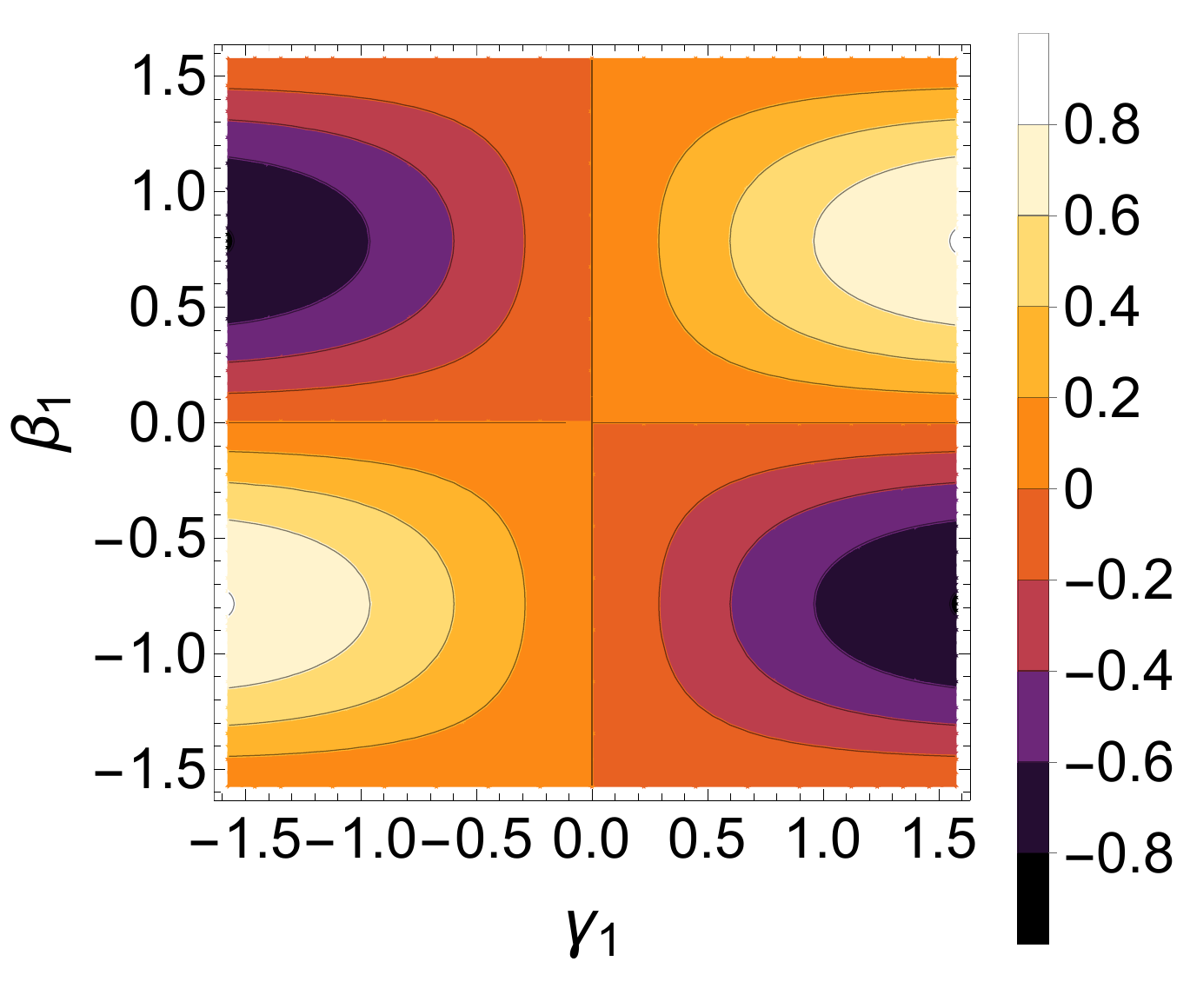}
  \vspace{-0.5em}
  \caption{{\small SNR = 0dB (Rayleigh)} }
 \label{fig:F1_snr_0_ray}
 \end{subfigure}
  \begin{subfigure}{0.24\textwidth}
\includegraphics[width = 0.95\linewidth]{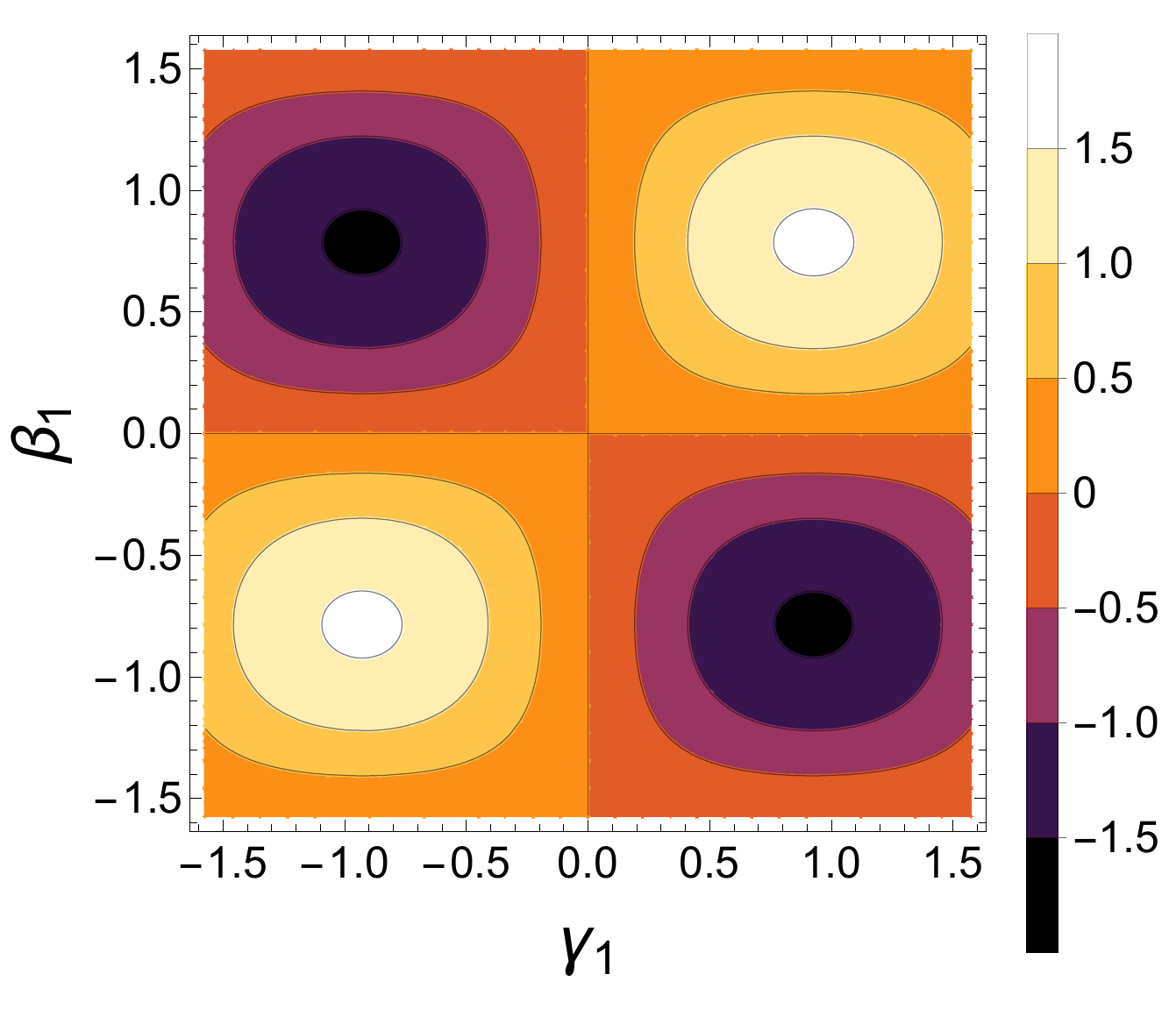}
  \vspace{-0.5em}
  \caption{{\small SNR = 10dB (Rayleigh)} }
 \label{fig:F1_snr_10_ray}
 \end{subfigure}
 
 \begin{subfigure}{0.25\textwidth}
\includegraphics[width = 0.95\linewidth]{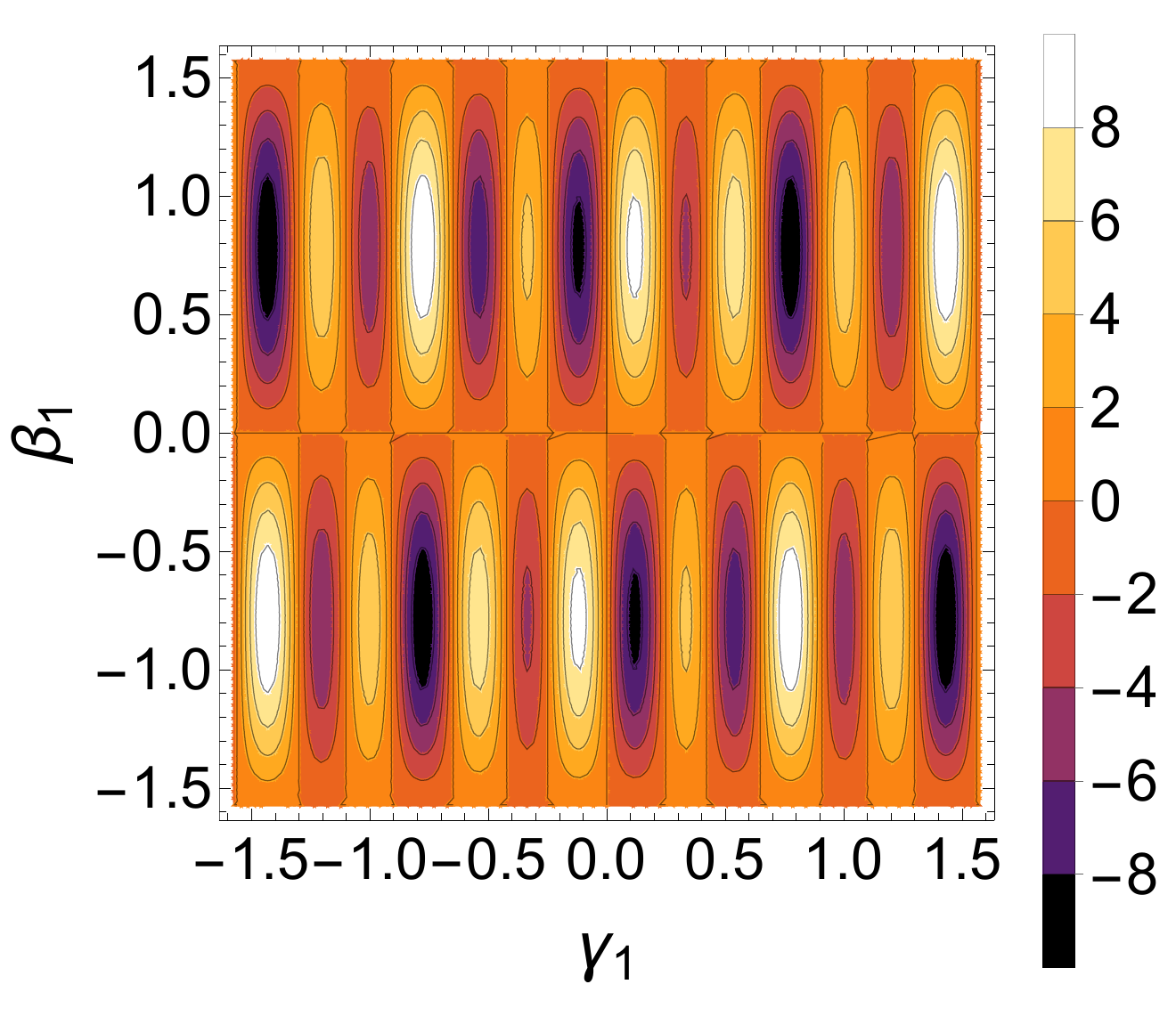}
  \vspace{-0.5em}
  \caption{$s= -1+i$}
 \label{fig:F1_snr_0_awgnsym2}
 \end{subfigure}
 \begin{subfigure}{0.25\textwidth}
\includegraphics[width = 0.95\linewidth]{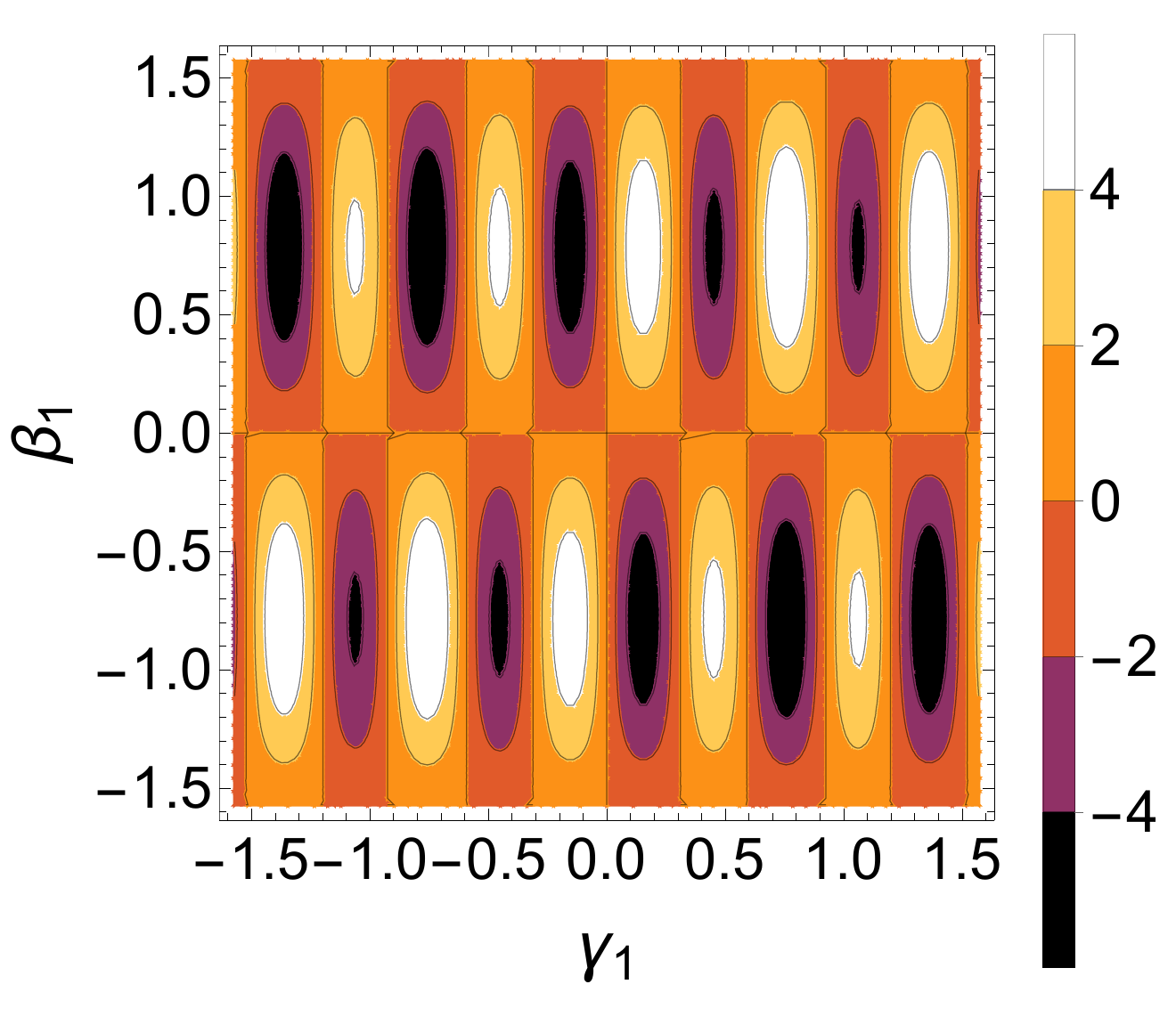}
  \vspace{-0.5em}
  \caption{$s= 1-i$}
 \label{fig:F1_snr_0_awgnsym3}
 \end{subfigure} 
  \begin{subfigure}{0.25\textwidth}
\includegraphics[width = 0.95\linewidth]{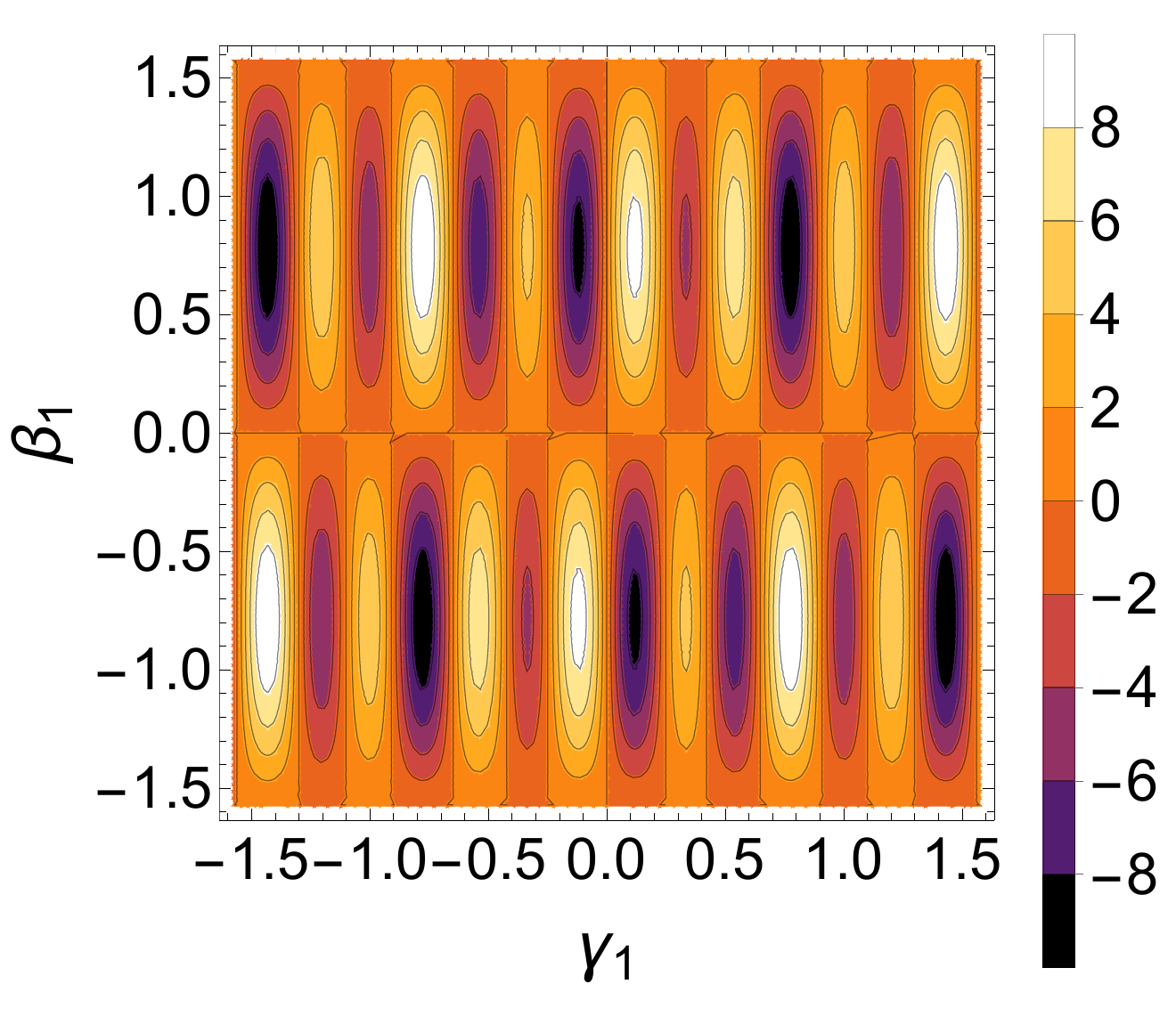}
  \vspace{-0.5em}
  \caption{$s= -1-i$}
 \label{fig:F1_snr_0_awgnsym4}
 \end{subfigure} 
  \vspace{-1.5em} \caption{{\small Landscapes of $F_1(\gamma_1,\beta_1)$ for QPSK with  different ML detection problem instances. Fig. \ref{fig:F1_snr_0_awgn}-Fig. \ref{fig:F1_snr_10_ray} are for the same transmit symbol $s=1+i$, while Fig. \ref{fig:F1_snr_0_awgnsym2}-Fig. \ref{fig:F1_snr_0_awgnsym4} are for  the same configurations as Fig. \ref{fig:F1_snr_0_awgn}, except for the transmit symbols. }}
 \label{fig:F1s}
   \vspace{-1.0em}
\end{figure}
Fig. \ref{fig:F1s} illustrates the landscape of the expectation values $F_1(\gamma_1,\beta_1)$ for QPSK, where multiple problem instances  are illustrated in terms of the AWGN channel and the Rayleigh channel.  Following the analytical expression derived in \eqref{eq:F_1qpsk}, the landscape of $F_1(\gamma_1,\beta_1)$ can be plotted using Mathematica, where the  values of the noise and the channel coefficients are generated using a seed value 100 for  reproducing the results. In the landscape plots,  the smaller (darker) is  better.  Specifically, the darkest spots indicate the global minima of $F_1(\gamma_1,\beta_1)$ that we are seeking to find using the classical optimizer.  
Fig. \ref{fig:F1_snr_0_awgn} and Fig. \ref{fig:F1_snr_10_awgn} plots the landscape of $F_1(\gamma_1,\beta_1)$ for  different SNRs under an AWGN channel, while Fig. \ref{fig:F1_snr_0_ray} and Fig. \ref{fig:F1_snr_10_ray} plot the landscape under an uncorrelated Rayleigh channel with different SNRs. Moreover, in Fig. \ref{fig:F1_snr_0_awgnsym2}-Fig. \ref{fig:F1_snr_0_awgnsym4} we plot the landscape with  different transmit symbols. 
By comparing the landscapes plotted in Fig. \ref{fig:F1s},  we see  that the factors affecting the positions of the global minima includes  both the SNR, the noise and  the channel coefficients as well as the transmit symbols, which complicates the evaluation of  the QAOA when solving the ML detection problem.

\begin{figure} [t!]
\centering
\begin{subfigure}{0.4\textwidth}
\includegraphics[width = 0.9\linewidth]{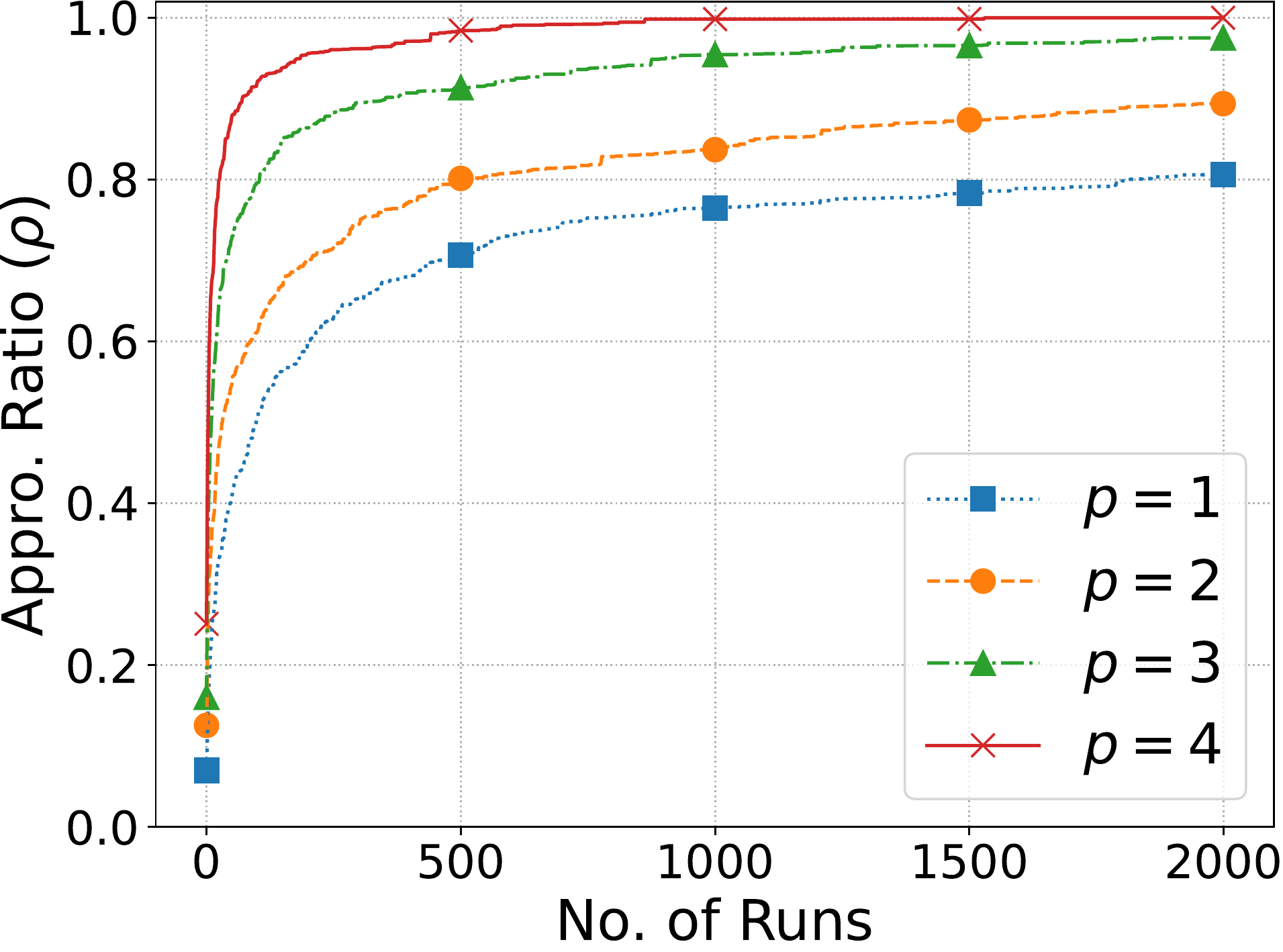}
  \vspace{-0.5em}
  \caption{AWGN channel}
 \label{fig:rho_fp_agwn}
 \end{subfigure}
 \begin{subfigure}{0.4\textwidth}
\includegraphics[width = 0.9\linewidth]{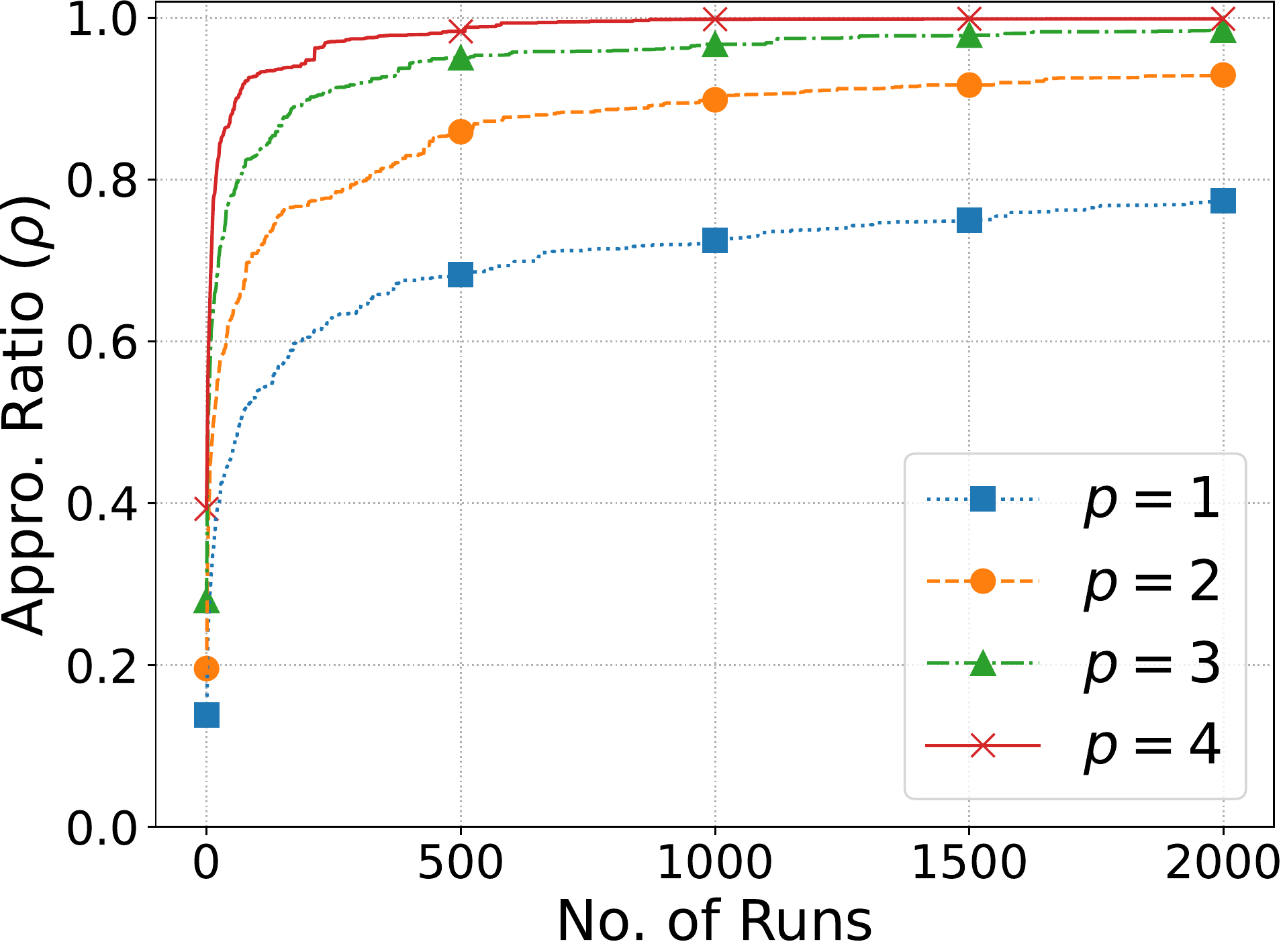}
  \vspace{-0.5em}
  \caption{Rayleigh channel}
  \label{fig:rho_fp_ray}
 \end{subfigure}
    \vspace{-1.5em}\caption{Comparison of  the approximation ratio versus the number of runs parameterized by $p$.}
   \label{fig:rho_fp}
   \vspace{-1.5em}
\end{figure}

\begin{figure} [htp!]
\centering
\begin{subfigure}{0.4\textwidth}
\includegraphics[width = 0.90\linewidth]{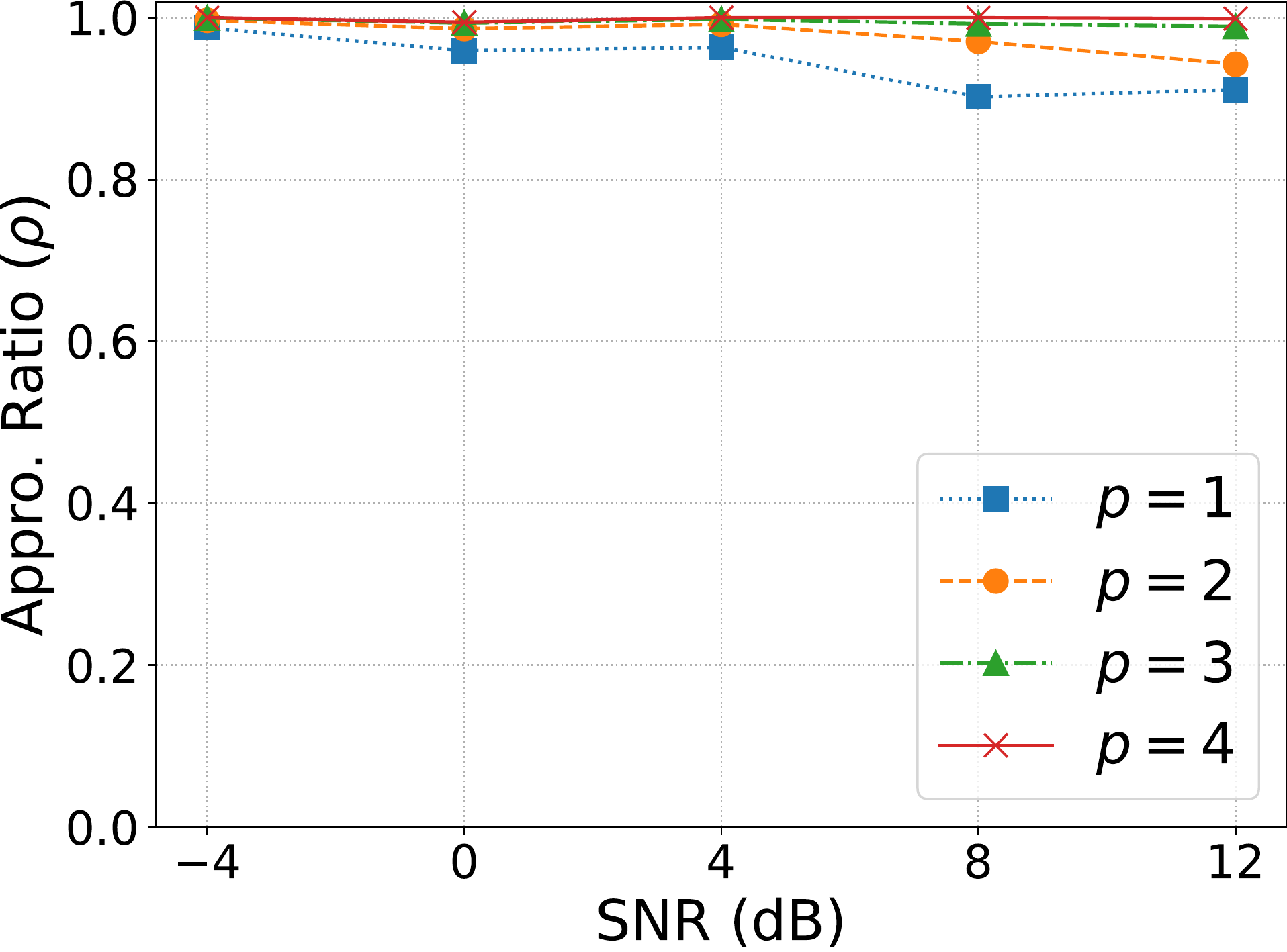}
  \vspace{-0.5em}  \caption{AWGN channel}
 \label{fig:sub1_rho_snr_awgn}
 \end{subfigure}
 \begin{subfigure}{0.4\textwidth}
\includegraphics[width = 0.90\linewidth]{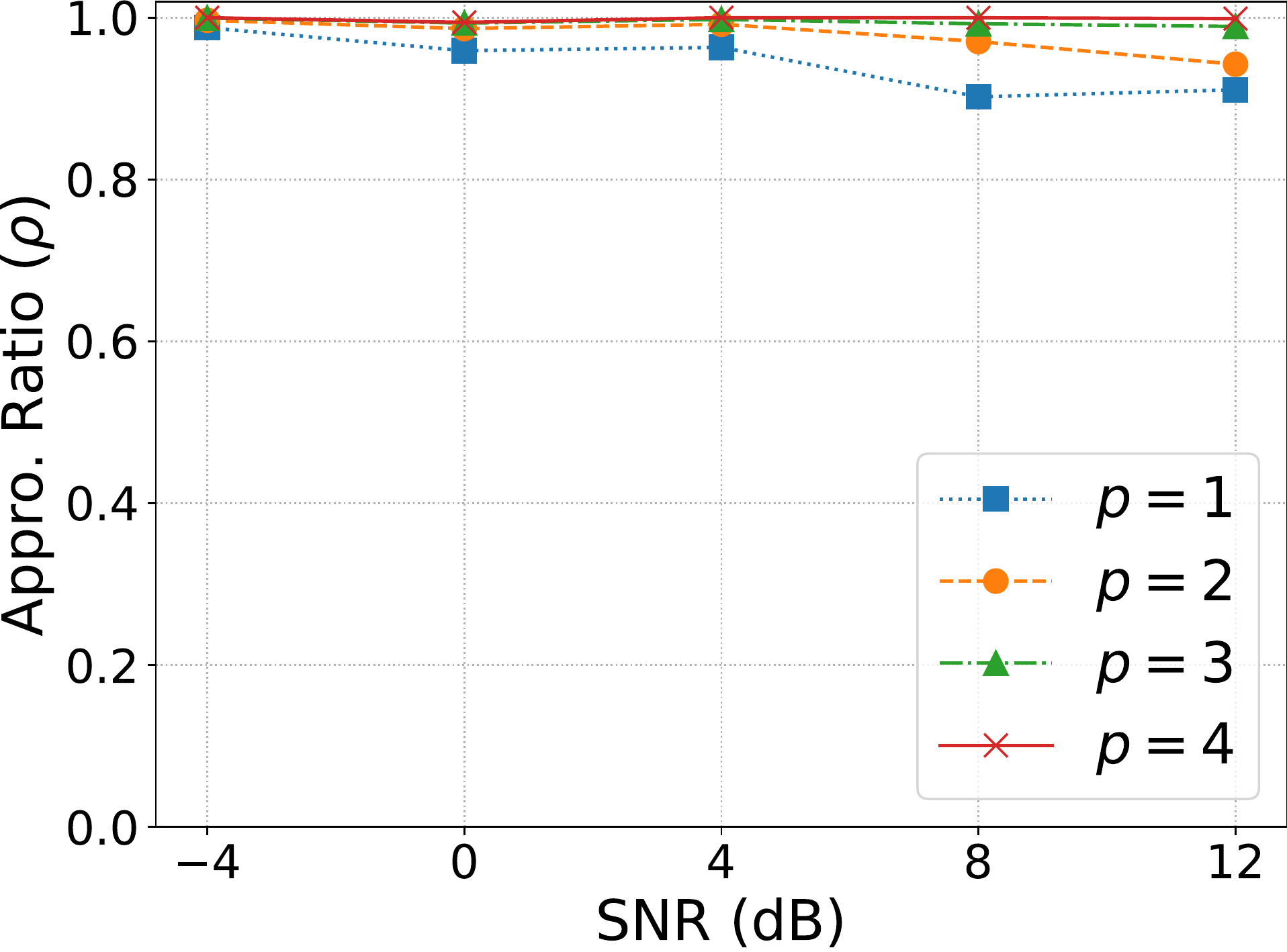}
  \vspace{-0.5em}  \caption{Rayleigh channel}
 \label{fig:sub2_rho_snr_awgn}
 \end{subfigure}
   \vspace{-1.5em}\caption{Comparison of the  approximation ratio versus SNR parameterized by $p$.}
 \label{fig:rho_snrs}
   \vspace{-1.5em}
\end{figure}
Fig. \ref{fig:rho_fp}  and Fig. \ref{fig:rho_snrs}  show the approximation ratio $\rho$  achieved by the QAOA when  solving the ML detection problem formulated for QPSK, where each $\rho$ is attained over 100 different channel realizations, given the noise $\eta  \in \mathcal{CN}(0,1)$ and the channel $h\in \mathcal{CN}(0,1)$.   
 In the simulations,  the expectation values of $F_p(\boldsymbol{\gamma},\boldsymbol{\beta})$ are attained by measuring the output of the QAOA circuits  generated by  Qiskit Aer \cite{Qiskit} and the parameters $\boldsymbol{\gamma},\boldsymbol{\beta}$ are updated using the classical optimizer--COYBLA \cite{powell2007view} which is a derivative-free optimization algorithm starting from a given initial point.  For obtaining a high-quality solution of the QAOA,  we set a complexity budget of 2000 runs for each problem instance  with respect to a single channel realization in the simulations, where the COYBLA starts  from different random initial points in each run.  
In Fig. \ref{fig:rho_fp}, we plots the approximation ratio $\rho$ for  different $p$ versus the number of runs at SNR=15dB under the AWGN channel and the Rayleigh channel, respectively.   We observe  that  $\rho$ becomes better and converges to a stable value as the number of runs increases. Moreover,  the maximum of $ \rho$ achieved tends to one as $p$ grows.  Observe in Fig. \ref{fig:rho_fp} that   $\rho$ achieves $ 0.7892$ for $p=1$, which increases to $ 0.9973$ for $p=4$ in the AWGN channel, while  $\rho$ reaches $0.7316$ for $p=1$ and tends to $0.9923$ for $p=4$ in the Rayleigh channel.  
Here  high approximation ratio is achieved because the  Hamiltonian formulated for  QPSK  in \eqref{eq:Hf_qpsk} only includes two spins without interactions, as shown in Fig. \ref{fig:qubits_illu}(a).  
Additionally,  Fig.\ref{fig:rho_snrs} shows the variation of $\rho$  with different $p$ versus  SNR in the AWGN channel and the Rayleigh channel, respectively. Specifically,  each value of $\rho$ is also attained over 100  channel realizations with respect to 100 different problem instances and  2000 runs are performed. We observe that the impact of the SNR on $\rho$ is more obvious for $p=1$ than  for $p\ge2$, which indicates that $\rho$  is sensitive to the specific problem instances of different SNRs for $p=1$, while it becomes less sensitive when $p\ge 2$.

 \vspace{-0.7em}
\section{Conclusions}
\label{sec:conclus}
A new   Hamiltonian construction procedure was conceived for the ML detection problem by using the SAT formulation, which allows us to simplify the problem Hamiltonian's  representation.  In our formulation, encoding the ML detection problem of an  $M$-dimensional constellation,  each constellation point  corresponding to a  $N= \log_2(M)$ bit string  needs  at most $N$ qubits.   In particular, for an MQAM constellation following  a Gray mapping order, the number of qubits required for encoding the problem can be reduced to $\lceil N/2 \rceil$.
Furthermore, we  demonstrated  the connection between the degree of the objective function with respect to the problem Hamiltonian and the Gray-labelled constellation diagram. In contrast to the classical ML detector  designed for outputting the estimated complex symbol, the QAOA assisted ML detector directly delivers the estimated  input bit string, without  any signal demapping procedure required in the classical communication system.   Our future work will extensively evaluate the results derived on the available QAOA platforms in NISQ devices and further quantify the performance of the QML detector in comparison to  classical detectors. 

 \vspace{-0.7em}
\numberwithin{equation}{section}
\section*{Appendix~A: Proof of Proposition \ref{pro:property1}} \label{Appdx:A}
\renewcommand{\theequation}{A.\arabic{equation}}
\setcounter{equation}{0}
\renewcommand{\thefigure}{A.\arabic{figure}}
\setcounter{figure}{0}
\renewcommand{\theremark}{A.\arabic{remark}}
\setcounter{remark}{0}

\begin{figure} [t!]
\centering
\begin{subfigure}{0.28\textwidth}
\includegraphics[width = 0.95\linewidth]{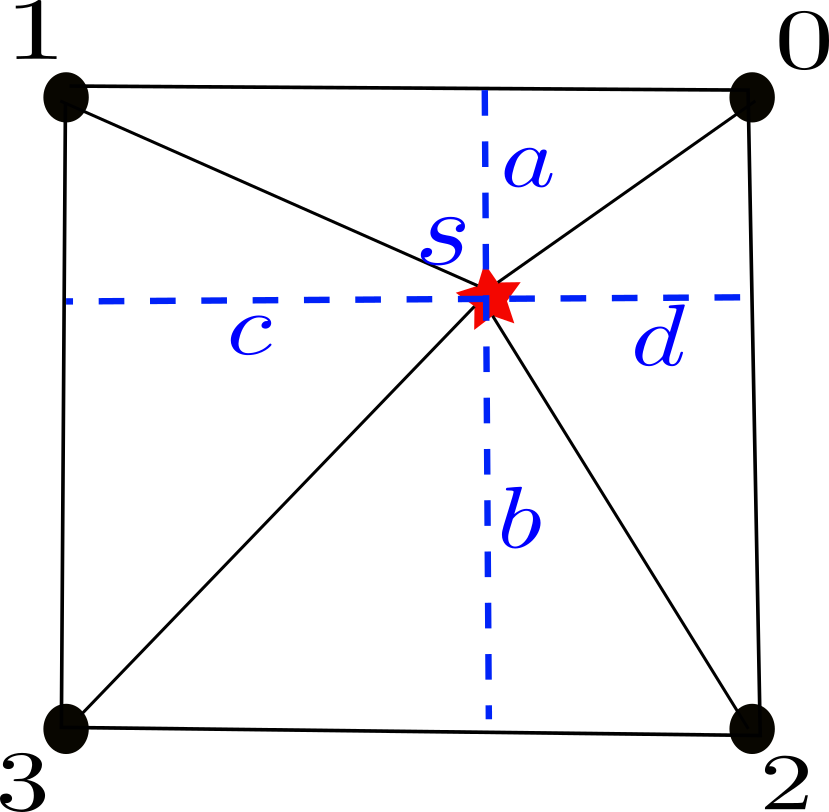}
 \vspace{-0.5em}  \caption{{\small $s$  inside the square }}
 \label{fig:qpsk_proofp1}
 \end{subfigure}
 \begin{subfigure}{0.25\textwidth}
\includegraphics[width = 0.95\linewidth]{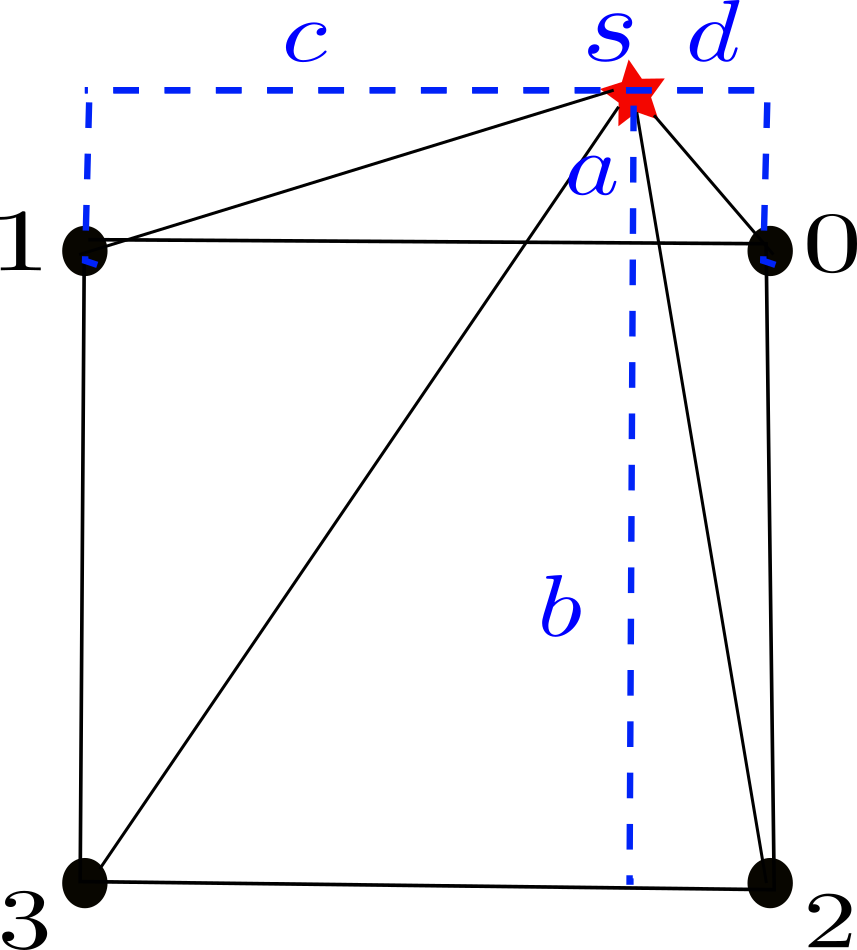}
 \vspace{-0.5em}  \caption{{\small $s$  outside the square}  }
 \label{fig:qpsk_proofp2}
 \end{subfigure}
   \vspace{-1.5em} \caption{{\small Illustrations of a point inside and outside the square constructed by the  constellation points.} }
 \label{fig:qpsk_proof}
   \vspace{-1.5em}
\end{figure} 

Here, we provide two methods of  proving that  $d_{0,1} = 0$.  We first  prove it using  geometric properties.  Observe from Fig. \ref{fig:qpsk_mapping} that the constellation points of  QPSK are arranged on  a squared grid.  For a received signal, its position will always fall inside, outside or on the square formed by the constellation points of  Fig. \ref{fig:qpsk_proof}, where we  denote the position of the received signal by $s$.
If $s$ is on the square, it is straightforward to see that $d_0+d_3 = d_1+d_2$, since they are equal to the squared length of the diagonal line.  Now we consider more general cases, where $s$ is  either inside or outside the square.
 We add a pair of  straight lines passing through $s$, while being parallel to the two edges of a right angle, respectively, represented by  the dotted lines in Fig. \ref{fig:qpsk_proof},  which leads to  four line segments from $s$.   The  length of the line segments is denoted by $a,b,c$ and $d$, respectively, as shown in Fig. \ref{fig:qpsk_proof}.  For $s$ inside  the square in Fig. \ref{fig:qpsk_proofp1},    we can see that $d_0 = a^2+d^2$, $d_1 = a^2+c^2$, $d_2 = b^2+d^2$ and $d_3 = b^2+c^2$  following the  geometric properties.   We see that  $d_0+d_3 = d_1+d_2 = a^2+b^2+c^2+d^2$. Thus, $\bar{d}_{0,1} = 0$ is proved.  For $s$ outside  the square in Fig. \ref{fig:qpsk_proofp2}, we observe that  $d_0 = a^2+d^2$, $d_1 = a^2+c^2$, $d_2 = (a+b)^2+d^2$ and $d_3 = (a+b)^2+c^2$.  Here we have  $d_0+d_3 = d_1+d_2 = a^2+ (a+b)^2+c^2+d^2$. Hence $\bar{d}_{0,1}$ is  $0$. Therefore, we conclude the above proved   property of $d_0+d_3 = d_1+d_2 $ in the following remark.

\vspace*{-0.9em} 
\begin{remark}\label{term:remark_square}
Given a point and a square, the sum of the squared distances of  the received signal from the two pairs of diagonal points are the same.
\end{remark}
\vspace*{-0.9em} 

 In the second method, we prove that $\bar{d}_{0,1} = 0$ from the perspective of  the communication system.  We assume that   the complex channel coefficient $h$ and the complex received signal $y$ are represented  as  $h = a+bj$ and $y = c+dj$, respectively, where $a,b,c$ and $d$ are real numbers.  For the constellation of Figure \ref{fig:qpsk_mapping}, we have $\mathcal{A} = \{1+j, -1+j,1-j,-1-j\}$. We thus have
 \begin{eqnarray}
 \begin{aligned}
d_0+d_3 &= |y - h(1+j)|^2 +  |y - h(-1-j)|^2 \\
&= |c+dj - (a+bj)(1+j)|^2+ |c+dj - (a+bj)(-1-j)|^2 \\
&= 2(2a^2+2b^2+c^2+d^2),
 \end{aligned}
 \end{eqnarray}
 and
 \begin{eqnarray}
 \begin{aligned}
 d_1+d_2 &= |y - h(-1+j)|^2 +  |y - h(1-j)|^2 \\
&= |c+dj - (a+bj)(-1+j)|^2+ |c+dj - (a+bj)(1-j)|^2 \\
&= 2(2a^2+2b^2+c^2+d^2) = d_0+d_3.
 \end{aligned}
 \end{eqnarray}
Hence $\bar{d}_{0,1} = 0$ is thus proved.

Therefore, we have $f(z_0,z_1) = \bar{d}_{0,1}z_0z_1 + \bar{d}_0 z_0 +  \bar{d}_1 z_1 + d_0$.

 \vspace{-0.7em}
\numberwithin{equation}{section}
\section*{Appendix~B: Proof of {\bf Theorem \ref{term:theorem}}} \label{Appdx:B}
\renewcommand{\theequation}{B.\arabic{equation}}
\setcounter{equation}{0}
\renewcommand{\thefigure}{B.\arabic{figure}}
\setcounter{figure}{0}

In our SAT mapping process, given a point $a_i$ of order $i$ with a binary representation $b_0\cdots b_{N-1}$,  there is a clause $C_i$ associated with point $i$. From {\bf Remark \ref{term:remark_expan1}} and {\bf Remark \ref{term:remark_expan2}},
 we observe that  there are only two clauses,  whose  expansions  contain the monomial of the form $z_l$, $l \in \mathcal{N}$, i.e. $C_0$ associated with the bit string $0\cdots 0$ and $C_{2^{N-1-l}}$ with the bit string  $b_0 = 0, \cdots,b_{l-1}=0,b_{l}=1,b_{l+1}=0,\cdots,b_{N-1} = 0$.
Following  {\bf Remark \ref{term:remark_expan3}}, we can obtain the sign of the coefficient for the monomial of  the form  $z_l$ in the expansion of $w(C_i)$. Explicitly, when expanding a weighted clause $w(C_i), i \in \mathcal{N}$, a bit $b_{l'}, l' \in \mathcal{N}$  will contribute $+1$ to the final coefficient of the monomial of $z_l$  if $b_{l'}=1$ and will contribute  $-1$  if $b_{l'} = 0$.
 For  the monomial of the form $z_l$ the sign of the coefficient $d_i$,  which is the squared distance spanning from the received signal points to  the $i$-th  constellation points,  can then be known by multiplying these returned $-1$s or $+1$s together.
 Applying the above process to  the pair clauses $w(C_0)$ and $w(C_{2^{N-1-l}})$ that contain the monomial of the form $z_l$, the signs in  front of  $d_i, ~i\in \{0,2^{N-1-l}\}$,   can be determined. Correspondingly, we can obtain the  coefficient associated with term $z_l$ in  the expansion of $f(z_0,\cdots, z_{N-1})$ by summing the signed  squared distances $d_i,~i\in \{0,2^{N-1-l}\}$.

For  the  expansions  of the clauses  containing the monomial of the form $z_lz_m$,  there are $2^2 =4$ clauses associated with  $\mathcal{S} = \{l,m\}$.
For the coefficient of the monomial of the form $z_lz_m$, $l,m \in \mathcal{N}, l\neq m$, in the expansion of  the clause $w(C_i), i\in \mathcal{S}$,   the pair of bits $b_lb_m$  will contribute $+1$ to the sign of the coefficient to the monomial of the form $z_lz_m$ if $b_l \times b_m=1$ and  will contribute  $-1$ if $b_l \times b_m=0$. Here, we use $\times$ to denote the product of $b_l$ and $b_m$ for avoiding any confusion.  Repeating the above process for all $i \in \mathcal{S}$, the signs in  front of the squared distances $d_i$,  are thus obtained in terms of the monomial of the form $z_lz_m$ in the expansion of $w(C_i)$ for any $i \in \mathcal{S}$.  We therefore obtain the coefficient associated with the term $z_lz_m$ in the expanded$f(z_0,\cdots, z_{N-1})$ by summing the coefficients for the monomial of the form $z_lz_m$.

Following the above procedures, we can obtain the sign of the coefficient of the monomial having any form of $\prod_{n\in \mathcal{S}}z_n$, $\mathcal{S} \in \mathcal{N}$, with respect to each clause.  
  As a result, the coefficient of the monomial having the form  $\prod_{n\in \mathcal{N}}z_n$ i.e. $\mathcal{S}= \mathcal{N}$ in the expanded $f(z_0,\cdots,z_{N-1})$ contains all of the squared distances $d_i$, $i \in \mathcal{M}$.
For a rectangular QAM constellation obeying Gray mapping,   the coefficients associated with the diagonal points on the smallest lattice grid constructed by four constellation points have the same sign, while the signs of the coefficients associated with the two pairs of the diagonal points are opposite.  This comes from the property of Gray mapping where the orders in binary bits of a pair of neighbouring points differ in only one bit, and the order in  bits of the diagonal points thereby differ in two bits.    As a result, the coefficient of  $\prod_{n\in \mathcal{N}}z_n$ in $f(z_0,\cdots, z_{N-1})$ becomes $0$. In terms of a  PSK constellation following a Gray mapping order,  the PSK constellation would contain $M=2^N$, $N \in \mathcal{Z}^{+}$, constellation points \cite{Lee85tcom}, which are positioned at a  uniform angular spacing around a circular. 
Therefore, the smallest grid formed for the PSK constellation can be constructed  by choosing  a pair of  points separated by $\pi$  along with their neighbours having angular spacing $\pi$. Analogous to the smallest grid for the QAM constellation, the signs of the coefficients associated with the two pairs of the diagonal points are opposite corresponding to this rectangle.   Therefore, the coefficient of  the monomial of  the form  $\prod_{n\in \mathcal{N}}z_n$ is zero.

\vspace{-0.7em}
\numberwithin{equation}{section}
\section*{Appendix~C: Proofs of {\bf Corollary \ref{term:coro_nongray}} and {\bf Corollary \ref{term:coro_nongray2}}} \label{Appdx:C}
\renewcommand{\theequation}{C.\arabic{equation}}
\setcounter{equation}{0}
\renewcommand{\thefigure}{C.\arabic{figure}}
\setcounter{figure}{0}

\subsubsection*{Proof of {\bf Corollary \ref{term:coro_nongray}}}
Following the proof of {\bf Theorem \ref{term:theorem}}, the coefficient of the monomial having the form  $ \prod_{n\in \mathcal{N}}z_n$ in the expanded $f(z_0,\cdots,z_{N-1})$ contains all of the squared distances $d_i$, $i\in \mathcal{M}$.
In {\bf Corollary \ref{term:coro_nongray}},  the property in $\ref{cor:item3})$ indicates that  the coefficients  associated with a pair of diagonal  points of a rectangle share the same sign, while  the coefficients  associated with the  other  pair of diagonal  points have the opposite sign.  This makes the sum of the coefficients of  $\prod_{n\in \mathcal{N}}z_n$  over a rectangle  is  zero.  Furthermore, the properties in $\ref{cor:item1})$ and $\ref{cor:item2})$ indicate that all of the  points in the constellation can form a set of rectangles without shared vertices between any two of them. Therefore, the final coefficient of $\prod_{n\in \mathcal{N}}z_n$ in the expanded $f(z_0,\cdots,z_{N-1})$ is  zero,  since the coefficient  of $\prod_{n\in \mathcal{N}}z_n$ over a rectangle  is zero,  which accrues from the property in $\ref{cor:item3})$.

\subsubsection*{Proof of {\bf Corollary \ref{term:coro_nongray2}}}

Following the proof of {\bf Corollary \ref{term:coro_nongray}}, the property in  \ref{cor:item3})  of {\bf Corollary \ref{term:coro_nongray}} can be extended to demonstrate the coefficient of any monomial in $f(z_0,\cdots,z_{N-1})$.
\ref{cor:item3}) of {\bf Corollary \ref{term:coro_nongray}}   indicates that for each  rectangle constructed by the constellation points in $\mathcal{S}$,  the resultant coefficient for the monomial of the form $\prod_{n\in \mathcal{S}} z_n$   zero.  
For the monomial of the form $\prod_{n\in \mathcal{S}} z_n$, $\mathcal{S}\subseteq \mathcal{N}$,  we know that it is associated with   $2^{|\mathcal{S}|}$ constellation points, which can form $2^{{|\mathcal{S}|}-1}$  rectangles without sharing vertices. 
Recall that  \ref{cor:item2}) of {\bf Corollary \ref{term:coro_nongray}} indicates that the rectangles formed by the constellation points which do not share vertices.    Hence,  the coefficient of  the monomial of the form $\prod_{n\in \mathcal{S}} z_n$, $\mathcal{S} \subseteq \mathcal{N}$, is zero.

 \vspace{-0.7em}
\numberwithin{equation}{section}
\section*{Appendix~D: Proof of {\bf Theorem \ref{term:theory_2}}} \label{Appdx:D}
\renewcommand{\theequation}{D.\arabic{equation}}
\setcounter{equation}{0}
\renewcommand{\thefigure}{D.\arabic{figure}}
\setcounter{figure}{0}
\renewcommand{\theremark}{D.\arabic{remark}}
\setcounter{remark}{0}

Recall that for a rectangular QAM constellation, the  bits associated with each point can be represented as $b_l, l \in \mathcal{N}_I$ for the in-phase bits and as $b_m, m \in \mathcal{N}_Q$ for the quadrature phase bits, which are arranged in Gray mapping order  along each axis.  
Therefore, for a Gray-labelled QAM constellation,  the points on one line parallel to the in-phase axis or the quadrature axis contains odd and even numbers of '0's, alternatively. Correspondingly, we have the following properties for a  rectangular QAM constellation following a Gray mapping:

\vspace*{-0.9em} 
\begin{remark} \label{term:observ}
 Given two points in the same line,  one of the points has an odd numbers of '0's , while the other point  has an odd numbers of '0's if there are even numbers of points between them, and has an even numbers of '0's if they are separated by  odd numbers of points. 
\end{remark}
\vspace*{-0.9em} 

Now we turn to the proof of {\bf Theorem \ref{term:theory_2}}.
Given a set $\mathcal{S} \subseteq \mathcal{N}$ with $\mathcal{S}\cap \mathcal{N}_I \neq \emptyset$ and $\mathcal{S}\cap \mathcal{N}_Q \neq \emptyset$,  the variables in $\prod_{n \in \mathcal{S}} z_n$ come partially   from the in-phase bits  and  partially  from the quadrature phase bits.  Let us now define $L_I$ and $L_Q$ as the numbers of bits choosing from the in-phase bits and the quadrature phase bits, respectively, where we have  $l$ and $m$ such that $1 \le L_I \le N_I$, $1\le L_Q \le N_Q $.  Following {\bf Remark \ref{term:remark_expan1}},  we see that the constellation points  generating  the term $\prod_{n \in \mathcal{S}} z_n$ are the points lying at the crossings of the $2^{L_I}$ lines  parallel to the quadrature axis and $2^{L_Q}$ lines parallel to the in-phase axis.
Correspondingly, there are $2^{L_I} \times 2^{L_Q}$ constellation points associated with  the term $\prod_{n \in \mathcal{S}} z_n$. Here $2^{L_I}$ and $2^{L_Q}$ are the numbers of points  in a line along the in-phase axis sharing the same quadrature phase bits and the quadrature phase axis sharing the same in-phase bits, respectively.

For proving the statement in \ref{theorem2_item1}) of {\bf Theorem \ref{term:theory_2}},   we start by separating the points in a line  parallel to the in-phase axis  into two groups according to the number of '0' in each binary orders.  
For the $k$-th line parallel to the in-phase axis, $k \in 
\mathcal{N}_Q$, the group in which  the binary orders  contain  odd numbers of '0's  is denoted as $\mathcal{G}_{odd}^{k}$, while the group in which  the binary orders have even numbers of '0's is denoted as $\mathcal{G}_{even}^{k}$. 
 Thus, for each line there are $2^{L_{I}-1}$ points in each group. 
Given a $k$, we  randomly choose two constellation points $a_0$ and $a_1$ such that 
$ a_0 \in \mathcal{G}_{odd}$
 and $a_1 \in \mathcal{G}_{even}^{k}$. 
Because of the symmetric pattern of the Gray mapping, along the quadrature phase axis,  there exists a pair of points $a_2 \in G_{even}^{k'}$ and $a_3 \in \mathcal{G}_{odd}^{k'}$ in a line $k'$, $k'\ne k$,  parallel to the in-phase axis as well, such that  the binary orders of $a_2$ and $a_3$ contain the same quadrature phase bits with $a_0$  and $a_3$, respectively.  Since the constellation follows the Gray mapping order,  the four points $a_0, \cdots,a_3$ form a  rectangle such that the binary orders corresponding to one  pair of the diagonal points contains  odd numbers of '0's and those labelling the other pair of diagonal points contain  even numbers of '0's. 
 In the  proof of {\bf Theorem \ref{term:theorem}} we discussed the correlation of the sign in  front of  the squared distance,  where   in the expansion of a clause $w(C_i), i \in \{0, 2^{N-1-l}\}$, for the coefficient of a monomial containing the variable $z_l'$, $ l' \in \mathcal{N}$, the  bit $b_{l'}$  will contribute $+1$ to the final coefficient of the monomial in the expanded $w(C_i)$  if $b_{l'}=1$ and  it will contribute  $-1$  if $b_{l'} = 0$.  This leads to the fact that  the signs in  front of the  monomial at hand  associated with a pair of diagonal points are the same, while the signs of the monomial corresponding to the other pair of  diagonal points are  the opposite. Accordingly, the sum of the coefficients of the monomial associated with the four points is zero, where {\bf Proposition \ref{term:remark_rectangle}} is used.   
Based on the pattern given in  {\bf Remark \ref{term:observ}}, we can repeat the above process to construct $2^{L_I+L_Q-2}$ rectangles such that the sum of the coefficient of the monomials associated with each rectangle is zero. As a result,   the coefficient  of the monomial of the form $\prod_{n \in \mathcal{S}} z_n$ in $f(z_0,\cdots, z_{N-1})$ is zero, if  there exists at least one pair of  bits denoted by $b_l=0$ and $b_m=0$ with $l,m \in  \mathcal{S}$, such that $b_l$ and $b_m$ belong to different groups $\mathcal{N}_I$ and $\mathcal{N}_Q$.

Next we prove the statement in \ref{theorem2_item2}).  The statement  in \ref{theorem2_item1}) indicates that the  coefficient  of the monomial of the form $\prod_{n \in \mathcal{S}} z_n$ in $f(z_0, \cdots,z_{N-1})$, for any $|\mathcal{S}| > N_I$ is zero. 
Now we turn to the calculation of  the coefficient for the monomial of the form $\prod_{n \in \mathcal{S}} z_n$, for any $|\mathcal{S}| = N_I$.    Consider the monomial of the form $\prod_{n \in \mathcal{N}_I} z_n$  associated with the in-phase bits $b_0 \cdots b_{N_I -1}$.   From {\bf Remark \ref{term:remark_expan1}}, we know that a constellation point contributes a coefficient of the term $\prod_{n\in \mathcal{N}_I} z_n$ if and only if the quadrature bits are zeros i.e., $b_{N_I}=0, \cdots, b_{N-1} = 0$. Note that the constellation points corresponding to $b_0 \cdots b_{N-1}$ such that $b_0 \cdots b_{N_I -1} \in \{0,1\}^{N_I}$  and $b_{N_I}=0, \cdots, b_{N-1} = 0$, are lying on the line parallel to the in-phase axis. The coefficient of  the monomial having the form $\prod_{n \in \mathcal{N}_I} z_n$  thus cannot be cancelled. Therefore, there exits at least one  monomial of degree $N_I$ in $f(z_0,\cdots,z_{N-1})$, which indicates that the monomial of degree $N_I$ exits in the expanded $f(z_0,\cdots,z_{N-1})$. Accordingly, the degree of $f(z_0,\cdots,z_{N-1})$ is $N_I$.

\vspace{-0.7em}
\numberwithin{equation}{section}
\section*{Appendix~E: Proofs of {\bf Corollary \ref{term:corollary_4}} and {\bf Corollary \ref{term:corollary_5}}} \label{Appdx:E}
\renewcommand{\theequation}{E.\arabic{equation}}
\setcounter{equation}{0}
\renewcommand{\thefigure}{E.\arabic{figure}}
\setcounter{figure}{0}

\subsubsection*{Proof of {\bf Corollary \ref{term:corollary_4}}}

 For a $N_r \times N_t$ MIMO channel  associated with QPSK, we have $N = 2N_t$ and  $N_{0,I}=\cdots=N_{N_t-1,I} = N_I$. 
 We further denote the objective function as $f_{N_t\times QPSK}(z_0,\cdots,z_{N-1})$.  
Recall that the coefficient of the quadratic term in the objective function $f_{QPSK}(z_0,z_1)$ is zero and  the degree of  $f_{QPSK}(z_0,z_1)$ of \eqref{eq:f_qpsk_simp}  is $1$.  Following {\bf Proposition \ref{term:remark_mimo_prop1}} and  {\bf Remark \ref{term:rem_IQp1}}, we can see that the coefficients of the monomials with a degree higher than $N_t$ in the expanded $f_{N_t\times QPSK}(z_0,\cdots,z_{N-1})$ are  zero,   since they  at least contain a pair of bits $b_l=0$ and $b_m=0$ such that $l\in \mathcal{N}_I$, $m\in \mathcal{N}_Q$ and $\mathcal{N}_I, \mathcal{N}_Q$ associated with a single QPSK constellation.
 In  the expanded $f_{N_t\times QPSK}(z_0,\cdots,z_{N-1})$,  there is a class of the monomials of degree $N_t$, in which  the variables come from the $N_t$ QPSK constellations. In this case of monomials of degree $N_t$, the coefficient cannot be cancelled  and thus the degree of $f_{N_t\times QPSK}(z_0,\cdots,z_{N-1})$ is $N_t$.
 
\subsubsection*{Proof of  {\bf Corollary \ref{term:corollary_5}}}
For a $N_r \times N_t$ MIMO channel with Gray-labelled MQAM, $M\ge 4$,  we have  $N_{0,I}=\cdots=N_{N_t-1,I} = N_I$.
Recall  from {\bf Theorem \ref{term:theory_2}} that the degree of the objective function over  a single constellation is $N_I$. 
 Following the proof of {\bf Corollary \ref{term:corollary_4}}, the coefficients of the monomials with degree higher than $N_I {N_t}$ in the expanded $f_{z_0, \cdots,z_{N-1}}$ is zero due to {\bf Remark \ref{term:rem_IQp1}}. Furthermore,  the coefficients of the monomials containing the variables associated with the $N_I$ in-phase bits of each  MQAM  constellation,  cannot be cancelled from the proof of {\bf Theorem \ref{term:theory_2}}.

 \vspace{-0.7em}
\numberwithin{equation}{section}
\section*{Appendix~F: Proof of {\bf Theorem \ref{term:theorem_3}}} \label{Appdx:F}
\renewcommand{\theequation}{F.\arabic{equation}}
\setcounter{equation}{0}
\renewcommand{\thefigure}{F.\arabic{figure}}
\setcounter{figure}{0}
\renewcommand{\theremark}{F.\arabic{remark}}
\setcounter{remark}{0}
 
 It is plausible that the degree of $f(z_0,\cdots,z_N)$ is $N_{k,I} = \lceil \frac{N}{2}\rceil=N_I$ if $N_t = 1$, since this results in the ML detection problem on a single constellation. 
If $N_t \ge 2$, {{\bf Corollary \ref{term:corollary_4}} and  {\bf Corollary \ref{term:corollary_5}}} show that  for a $N_r \times N_t$ MIMO channel associated either with QPSK or with MQAM, the  degree of the objective function $f(z_0,\cdots,z_N)$ is  $N_{I} {N_t}$, where $N_I=1$ for QPSK.

Now we consider the $N_r \times N_t$ MIMO channel of hybrid MQAM, where the number of constellation points can be different from each other.  There are $N_t$   QAM constellations with each following a Gray mapping order.  
Following  {\bf Remark \ref{term:remark_8}}, the in-phase bits and the quadrature phase bits associated with a single constellation are still  independent. The highest degree of the monomials  produced from a single constellation is $N_{k,I}$, $k\in \mathcal{N}_t$, the product of which constitutes the  highest degree of the monomials in the expanded $f(z_0,\cdots,z_{N-1})$. Hence, the  degree of $f(z_0,\cdots,z_{N-1})$ is $\sum_{k\in \mathcal{N}_t} N_{k,I}$.
 
\vspace{-0.7em}
\bibliographystyle{IEEEtran}
 {\small \linespread{1.1}\selectfont
\bibliography{/users/jingjing.cui/dropbox/EndnoteLib/myrefv1}

\begin{thebibliography}{10}
\providecommand{\url}[1]{#1}
\csname url@samestyle\endcsname
\providecommand{\newblock}{\relax}
\providecommand{\bibinfo}[2]{#2}
\providecommand{\BIBentrySTDinterwordspacing}{\spaceskip=0pt\relax}
\providecommand{\BIBentryALTinterwordstretchfactor}{4}
\providecommand{\BIBentryALTinterwordspacing}{\spaceskip=\fontdimen2\font plus
\BIBentryALTinterwordstretchfactor\fontdimen3\font minus
  \fontdimen4\font\relax}
\providecommand{\BIBforeignlanguage}[2]{{%
\expandafter\ifx\csname l@#1\endcsname\relax
\typeout{** WARNING: IEEEtran.bst: No hyphenation pattern has been}%
\typeout{** loaded for the language `#1'. Using the pattern for}%
\typeout{** the default language instead.}%
\else
\language=\csname l@#1\endcsname
\fi
#2}}
\providecommand{\BIBdecl}{\relax}
\BIBdecl

\bibitem{hanzo2005ofdm}
L.~Hanzo, B.~Choi, T.~Keller \emph{et~al.}, \emph{OFDM and MC-CDMA for
  broadband multi-user communications, WLANs and broadcasting}.\hskip 1em plus
  0.5em minus 0.4em\relax John Wiley \& Sons, 2005.

\bibitem{Verd89alg}
S.~Verd{\'u}, ``Computational complexity of optimum multiuser detection,''
  \emph{Algorithmica}, vol.~4, no.~1, pp. 303--312, 1989.

\bibitem{Micciancio01IT}
D.~Micciancio, ``The hardness of the closest vector problem with
  preprocessing,'' \emph{{IEEE} Trans. Inf. Theory}, vol.~47, no.~3, pp.
  1212--1215, 2001.

\bibitem{preskill2018quantum}
J.~Preskill, ``Quantum computing in the {NISQ} era and beyond,''
  \emph{Quantum}, vol.~2, p.~79, 2018.

\bibitem{farhi2014QAOA}
E.~Farhi, J.~Goldstone, and S.~Gutmann, ``A quantum approximate optimization
  algorithm,'' \emph{arXiv preprint arXiv: 1411.4028}, 2014.

\bibitem{Wang18PhysRevA}
Z.~Wang, S.~Hadfield, Z.~Jiang, and E.~G. Rieffel, ``Quantum approximate
  optimization algorithm for maxcut: A {Fermionic} view,'' \emph{Phys. Rev. A},
  vol.~97, p. 022304, Feb 2018.

\bibitem{streif2019comparison}
M.~Streif and M.~Leib, ``Comparison of {QAOA} with quantum and simulated
  annealing,'' \emph{arXiv preprint arXiv:1901.01903}, 2019.

\bibitem{Zhou20PhysRevX}
L.~Zhou, S.-T. Wang, S.~Choi, H.~Pichler, and M.~D. Lukin, ``Quantum
  approximate optimization algorithm: Performance, mechanism, and
  implementation on near-term devices,'' \emph{Phys. Rev. X}, vol.~10, p.
  021067, Jun. 2020.

\bibitem{Harrigan21NaturePhy}
M.~P. Harrigan, K.~J. Sung \emph{et~al.}, ``Quantum approximate optimization of
  non-planar graph problems on a planar superconducting processor,''
  \emph{Nature Physics}, vol.~17, p. 332–336, 2021.

\bibitem{karamlou2021analyzing}
A.~H. Karamlou, W.~A. Simon \emph{et~al.}, ``Analyzing the performance of
  variational quantum factoring on a superconducting quantum processor,''
  \emph{npj Quantum Information}, vol.~7, no.~1, pp. 1--6, 2021.

\bibitem{willsch2020benchmarking}
M.~Willsch, D.~Willsch, F.~Jin, H.~De~Raedt, and K.~Michielsen, ``Benchmarking
  the quantum approximate optimization algorithm,'' \emph{Quantum Information
  Processing}, vol.~19, pp. 1--24, 2020.

\bibitem{alam2019analysis}
M.~Alam, A.~Ash-Saki, and S.~Ghosh, ``Analysis of quantum approximate
  optimization algorithm under realistic noise in superconducting qubits,''
  \emph{arXiv preprint arXiv:1907.09631}, 2019.

\bibitem{Pagano20PNAS}
G.~Pagano, A.~Bapat \emph{et~al.}, ``Quantum approximate optimization of the
  long-range {Ising} model with a trapped-ion quantum simulator,''
  \emph{Proceedings of the National Academy of Sciences}, vol. 117, no.~41, pp.
  25\,396--25\,401, 2020.

\bibitem{qiang18photonics}
X.~Qiang, X.~Zhou \emph{et~al.}, ``\BIBforeignlanguage{en}{Large-scale silicon
  quantum photonics implementing arbitrary two-qubit processing},''
  \emph{\BIBforeignlanguage{en}{Nature Photonics}}, vol.~12, no.~9, pp.
  534--539, Sep. 2018.

\bibitem{Akshay2020PhyRewL}
V.~Akshay, H.~Philathong, M.~Morales, and J.~Biamonte, ``Reachability deficits
  in quantum approximate optimization,'' \emph{Physical Review Letters}, vol.
  124, no.~9, Mar 2020.

\bibitem{basso2021quantum}
J.~Basso, E.~Farhi, K.~Marwaha, B.~Villalonga, and L.~Zhou, ``The quantum
  approximate optimization algorithm at high depth for {MaxCut} on large-girth
  regular graphs and the {Sherrington-Kirkpatrick} model,'' \emph{arXiv
  preprint arXiv:2110.14206}, 2021.

\bibitem{marwaha2021local}
K.~Marwaha, ``Local classical {MAX-CUT} algorithm outperforms $ p= 2$ {QAOA} on
  high-girth regular graphs,'' \emph{Quantum}, vol.~5, p. 437, 2021.

\bibitem{Wurtz21PhysRevA}
J.~Wurtz and P.~Love, ``{MaxCut} quantum approximate optimization algorithm
  performance guarantees for $p >1$,'' \emph{Phys. Rev. A}, vol. 103, p.
  042612, Apr 2021.

\bibitem{hastings2019classical}
M.~B. Hastings, ``Classical and quantum bounded depth approximation
  algorithms,'' \emph{arXiv preprint arXiv:1905.07047}, 2019.

\bibitem{Bravyi20PhysRevLett}
S.~Bravyi, A.~Kliesch, R.~Koenig, and E.~Tang, ``Obstacles to variational
  quantum optimization from symmetry protection,'' \emph{Phys. Rev. Lett.},
  vol. 125, p. 260505, Dec 2020.

\bibitem{farhi2020quantumworst}
E.~{Farhi}, D.~{Gamarnik}, and S.~{Gutmann}, ``The quantum approximate
  optimization algorithm needs to see the whole graph: Worst case examples,''
  \emph{arXiv preprint arXiv:2005.08747}, 2020.

\bibitem{farhi2020quantum}
E.~Farhi, D.~Gamarnik, and S.~Gutmann, ``The quantum approximate optimization
  algorithm needs to see the whole graph: {A} typical case,'' \emph{arXiv
  preprint arXiv: 2004.09002}, 2020.

\bibitem{Egger2021warm}
D.~J. Egger, J.~Mareček, and S.~Woerner, ``Warm-starting quantum
  optimization,'' \emph{Quantum}, vol.~5, p. 479, Jun 2021.

\bibitem{shaydulin2021QIP}
R.~Shaydulin, S.~Hadfield, T.~Hogg, and I.~Safro, ``Classical symmetries and
  the quantum approximate optimization algorithm,'' \emph{Quantum Information
  Processing}, vol.~20, no.~11, pp. 1--28, 2021.

\bibitem{Sherrington75PRL}
D.~Sherrington and S.~Kirkpatrick, ``Solvable model of a spin-glass,''
  \emph{Phys. Rev. Lett.}, vol.~35, pp. 1792--1796, Dec 1975.

\bibitem{cui2021quantum}
J.~Cui, Y.~Xiong, S.~X. Ng, and L.~Hanzo, ``Quantum approximate optimization
  algorithm based maximum likelihood detection,'' \emph{arXiv preprint
  arXiv:2107.05020}, 2021.

\bibitem{Lucas14Ising}
A.~Lucas, ``{Ising} formulations of many {NP} problems,'' \emph{Frontiers in
  Physics}, vol.~2, p.~5, 2014.

\bibitem{hadfield2018representation}
S.~Hadfield, ``On the representation of {Boolean} and real functions as
  {Hamiltonians} for quantum computing,'' \emph{ACM Trans. Quantum Computing},
  vol.~2, no.~4, dec 2021.

\bibitem{Kim19sigcom}
M.~Kim, D.~Venturelli, and K.~Jamieson, ``Leveraging quantum annealing for
  large mimo processing in centralized radio access networks,'' in
  \emph{Proceedings of the ACM Special Interest Group on Data Communication},
  2019, pp. 241--255.

\bibitem{kim2020towards}
------, ``Towards hybrid classical-quantum computation structures in
  wirelessly-networked systems,'' \emph{Proc. of the ACM Wkshp on Hot Topics in
  Networks}, pp. 110--116, 2020.

\bibitem{singh2021ising}
A.~K. Singh, K.~Jamieson, D.~Venturelli, and P.~McMahon, ``{Ising} machines'
  dynamics and regularization for near-optimal large and massive {MIMO}
  detection,'' \emph{arXiv preprint arXiv:2105.10535}, 2021.

\bibitem{cook1971complexity}
S.~A. Cook, ``The complexity of theorem-proving procedures,'' in \emph{Proc. of
  the third annual ACM symposium on Theory of computing}, 1971, pp. 151--158.

\bibitem{boros2002pseudo}
E.~Boros and P.~L. Hammer, ``{Pseudo-Boolean} optimization,'' \emph{Discrete
  applied mathematics}, vol. 123, no. 1-3, pp. 155--225, 2002.

\bibitem{Glover2019QuantumBA}
F.~W. Glover, G.~A. Kochenberger, and Y.-T. Du, ``Quantum bridge analytics i: a
  tutorial on formulating and using {QUBO} models,'' \emph{4OR}, vol.~17, pp.
  335--371, 2019.

\bibitem{Peruzzo14nature}
A.~Peruzzo, J.~McClean, P.~Shadbolt, M.-H. Yung, X.-Q. Zhou, P.~J. Love,
  A.~Aspuru-Guzik, and J.~L. O'Brien, ``A variational eigenvalue solver on a
  photonic quantum processor,'' \emph{Nature Communications}, vol.~5, no.~1, p.
  4213, 2014.

\bibitem{cruz2019qubo}
W.~Cruz-Santos, S.~E. Venegas-Andraca, and M.~Lanzagorta, ``A {QUBO}
  formulation of minimum multicut problem instances in trees for {D-Wave}
  quantum annealers,'' \emph{Scientific reports}, vol.~9, no.~1, pp. 1--12,
  2019.

\bibitem{boros2014quadratization}
E.~Boros and A.~Gruber, ``On quadratization of {pseudo-Boolean} functions,''
  \emph{arXiv preprint arXiv:1404.6538}, 2014.

\bibitem{dattani2019quadratization}
N.~Dattani, ``Quadratization in discrete optimization and quantum mechanics,''
  \emph{arXiv preprint arXiv:1901.04405}, 2019.

\bibitem{Ishikawa09conf}
H.~Ishikawa, ``Higher-order clique reduction in binary graph cut,'' in
  \emph{IEEE Conference on Computer Vision and Pattern Recognition}, 2009, pp.
  2993--3000.

\bibitem{chancellor2016direct}
N.~Chancellor, S.~Zohren, P.~A. Warburton, S.~C. Benjamin, and S.~Roberts, ``A
  direct mapping of max k-{SAT} and high order parity checks to a {Chimera}
  graph,'' \emph{Scientific reports}, vol.~6, no.~1, pp. 1--9, 2016.

\bibitem{dattani2019all}
N.~Dattani and H.~T. Chau, ``All 4-variable functions can be perfectly
  quadratized with only 1 auxiliary variable,'' \emph{arXiv preprint
  arXiv:1910.13583}, 2019.

\bibitem{boros2020compact}
E.~Boros, Y.~Crama, and E.~Rodríguez-Heck, ``Compact quadratizations for
  pseudo-{Boolean} functions,'' \emph{Journal of Combinatorial Optimization},
  vol.~39, no.~3, p. 687–707, 2020.

\bibitem{tanburn2015reducingp1}
R.~Tanburn, E.~Okada, and N.~Dattani, ``Reducing multi-qubit interactions in
  adiabatic quantum computation without adding auxiliary qubits. part 1: The"
  deduc-reduc" method and its application to quantum factorization of
  numbers,'' \emph{arXiv preprint arXiv:1508.04816}, 2015.

\bibitem{okada2015reducingp2}
E.~Okada, R.~Tanburn, and N.~S. Dattani, ``Reducing multi-qubit interactions in
  adiabatic quantum computation without adding auxiliary qubits. part 2: The"
  split-reduc" method and its application to quantum determination of ramsey
  numbers,'' \emph{arXiv preprint arXiv:1508.07190}, 2015.

\bibitem{dridi2017prime}
R.~Dridi and H.~Alghassi, ``Prime factorization using quantum annealing and
  computational algebraic geometry,'' \emph{Scientific reports}, vol.~7, no.~1,
  pp. 1--10, 2017.

\bibitem{Ishikawa14conf}
H.~Ishikawa, ``Higher-order clique reduction without auxiliary variables,'' in
  \emph{IEEE Conference on Computer Vision and Pattern Recognition}, 2014, pp.
  1362--1369.

\bibitem{rosenberg1975reduction}
I.~G. Rosenberg, ``Reduction of bivalent maximization to the quadratic case,''
  in \emph{Cahiers du Centre d’Etudes de Recherche Operationnelle 17}, 1975,
  p. 71–74.

\bibitem{boros2018quadratizations}
E.~Boros, Y.~Crama, and E.~Rodr{\'\i}guez-Heck, ``Quadratizations of symmetric
  pseudo-{Boolean} functions: Sub-linear bounds on the number of auxiliary
  variables,'' in \emph{International Symposium on Artificial Intelligence and
  Mathematics (ISAIM)}, 2018.

\bibitem{Kohli94SIAM}
R.~Kohli, R.~Krishnamurti, and P.~Mirchandani, ``The minimum satisfiability
  problem,'' \emph{SIAM J. Discret. Math.}, vol.~7, no.~2, p. 275–283, May
  1994.

\bibitem{Umair20SIP}
U.~Arif, R.~Benkoczi, D.~R. Gaur, and R.~Krishnamurti, ``On the minimum
  satisfiability problem,'' in \emph{Algorithms and Discrete Applied
  Mathematics}, M.~Changat and S.~Das, Eds.\hskip 1em plus 0.5em minus
  0.4em\relax Cham: Springer International Publishing, 2020, pp. 269--281.

\bibitem{LARROSA2008AI}
J.~Larrosa, F.~Heras, and S.~{de Givry}, ``A logical approach to efficient
  max-sat solving,'' \emph{Artificial Intelligence}, vol. 172, no.~2, pp.
  204--233, 2008.

\bibitem{Fine2017book}
B.~Fine, A.~Gaglione, A.~Moldenhauer, G.~Rosenberger, and D.~Spellman,
  \emph{Algebra and Number Theory: A Selection of Highlights}.\hskip 1em plus
  0.5em minus 0.4em\relax De Gruyter, 2017.

\bibitem{Hadfield19Alg}
S.~Hadfield, Z.~Wang, B.~O’Gorman, E.~G. Rieffel, D.~Venturelli, and
  R.~Biswas, ``From the quantum approximate optimization algorithm to a quantum
  alternating operator ansatz,'' \emph{Algorithms}, vol.~12, no.~2, 2019.

\bibitem{recommendation2012detailed}
I.~Recommendation, ``Detailed specifications of the terrestrial radio
  interfaces of international mobile telecommunications advanced
  {(IMT-Advanced)},'' \emph{ITU-R, Tech. Rep. M}, 2012.

\bibitem{Smith75TCOM}
J.~Smith, ``Odd-bit quadrature amplitude-shift keying,'' \emph{{IEEE} Trans.
  Commun.}, vol.~23, no.~3, pp. 385--389, 1975.

\bibitem{Wesel01IT}
R.~Wesel, X.~Liu, J.~Cioffi, and C.~Komninakis, ``Constellation labeling for
  linear encoders,'' \emph{{IEEE} Trans. Inf. Theory}, vol.~47, no.~6, pp.
  2417--2431, 2001.

\bibitem{Vitthaladevuni05TWC}
P.~Vitthaladevuni, M.-S. Alouini, and J.~Kieffer, ``Exact {BER} computation for
  cross {QAM} constellations,'' \emph{{IEEE} Trans. Wireless Commun.}, vol.~4,
  no.~6, pp. 3039--3050, 2005.

\bibitem{Qiskit}
H.~Abraham \emph{et~al.}, ``Qiskit: An open-source framework for quantum
  computing,'' doi: 10.5281/zenodo.2562110, 2019.

\bibitem{powell2007view}
M.~J. Powell, ``A view of algorithms for optimization without derivatives,''
  \emph{Mathematics Today-Bulletin of the Institute of Mathematics and its
  Applications}, vol.~43, no.~5, pp. 170--174, 2007.

\bibitem{Lee85tcom}
P.~Lee, ``Computation of the bit error rate of coherent {M-ary PSK} with {Gray}
  code bit mapping,'' \emph{{IEEE} Trans. Commun.}, vol.~34, no.~5, pp.
  488--491, 1986.

\end{thebibliography}
}
\end{document}